# Symbiotic causal network of seagrass-bacteria-algae-diatoms interactions

Hirokuni Miyamoto[1,2,3,4,5,6]*, Kenta Suzuki[7,8], Shigeharu Moriya[9], Makiko Matsuura[3,5], Naoko Tsuji[5], Teruno Nakaguma[3,6], Chitose Ishii[1,5], Takayuki Nagatsuka[5,6], Takashi Satoh[10], Wataru Suda[1], Tamotsu Kato[1,3], Chie Shindo[1], Atsushi Kurotani[1,11], Hiroaki Kodama[3], Hiroshi Masuya[7], Satoshi Wada[8], Nobuhiro Kawachi[12], Hisashi Miyamoto[5,6,13], Yukinari Tsuruda[14], Yohei Shimasaki[14], Shouzo Ogizo[15], Nobuo Suzuki[15], Tomoharu Yuge[16], Toshio Takahashi[16], Tomohito Ojima[17], Toshio Furota[17], Akio Sakamoto[17], Keiichi Takimoto[18], Kozo Kugimiya[19], Takehiro Tanaka[20], Takashi Kimura[2], Yuuji Oshima[13], Jun Kikuchi[2,4,21]*, and Hiroshi Ohno[1]*

Affiliations:
[1]RIKEN Center for Integrative Medical Sciences, Yokohama, Kanagawa 230-0045, Japan
[2]NPO Umibezukurikenkyukai, Yokohama, Kanagawa, 220-0023, Japan
[3]Graduate School of Horticulture, Chiba University, Matsudo, Chiba 271-8501, Japan
[4]Graduate School of Medical Life Science, Yokohama City University, Yokohama, Kanagawa 230-0045, Japan
[5]Sermas Co., Ltd., Ichikawa, Chiba 272-0033, Japan
[6]Japan Eco-science (Nikkan Kagaku) Co., Ltd., Chiba, Chiba 263-8522, Japan
[7]RIKEN, BioResource Research Center, Tsukuba, Ibaraki 305-0074, Japan
[8]Institute for Multidisciplinary Sciences, Yokohama National University, Yokohama, Kanagawa 240-8501, Japan.
[9]RIKEN Center for Advanced Photonics, Wako, Saitama 351-0198, Japan
[10]Division of Hematology, Kitasato University School of Allied Health Sciences, Sagamihara, Kanagawa 252-0373, Japan
[11]Research Center for Agricultural Information Technology, National Agriculture and Food Research Organization, Tsukuba, Ibaraki 305-0856, Japan
[12]Kawachi Suisan Ltd., Saiki, Oita, 876-2302, Japan
[13]Miroku Co. Ltd., Kitsuki, Oita 873-0021, Japan
[14]Laboratory of Marine Environmental Science, Graduate School of Bioresource and Bioenvironmental Sciences, Kyushu University, Fukuoka, 819-0395, Japan
[15]Institute of Nature and Environmental Technology, Kanazawa University, Kanazawa, 920-1192, Japan
[16]Kaneda Fisheries Cooperative Association (Kaneda Gyokyou), Kisarazu, Chiba, 292-0008, Japan
[17]The beautification society (Umiotsukurukai voluntary association), Yokohama, Kanagawa 221-0852, Japan
[18]Hinase Fisheries Cooperative Association (Hinase Gyokyou), Hinase, Okayama, 701-3204, Japan
[19]Sumiyoshihama Resort Park, Hakuai-kai Social Welfare Corporation, Kitsuki, Oita, 873-0033, Japan
[20]NPO Satoumi Research Institute (Satoumizukuri Kenkyukaigi), Okayama, 704-8194, Japan
[21]RIKEN Center for Sustainable Resource Science, Yokohama, Kanagawa, 230-0045, Japan

* Co-corresponding authors
Correspondence: Hirokuni Miyamoto Ph.D., RIKEN IMS, Umibezukurikenkyukai Non-Profit Organization, Chiba University, Yokohama City University, Sermas Co., Ltd., and Japan Eco-science Co., Ltd.
Tel: +81-43-290-3947
E-mail: hirokuni.miyamoto@riken.jp, h-miyamoto@faculty.chiba-u.jp

Correspondence: Jun Kikuchi Ph.D., RIKEN CSRS, Umibezukurikenkyukai Non-Profit Organization, Yokohama City University
Tel: +81-43-290-3942
E-mail: jun.kikuchi@riken.jp

Correspondence: Hiroshi Ohno Ph.D. & M.D., RIKEN IMS
Tel: +81-45-503-7031
E-mail: hiroshi.ohno@riken.jp

## Abstract

Seagrass meadows contribute to the conservation of marine ecosystems, reduction in global warming impacts and pathogen controls. However, the decline in seagrass habitats due to environmental loads has become an urgent global issue. One way to address this issue is to better understand healthy seagrass habitats. Here, we estimate the structural characteristics of symbiotic and metabolic systems in sediments from eight coastal regions of Japan, with each region containing both seagrass-covered areas and adjacent unvegetated areas. Notably, seagrasses commonly maintain a balanced symbiotic relationship characterized by a positive association with cable bacteria (Desulfobulbaceae), nitrogen-cycling bacteria (Hyphomonadaceae), and coral algae (Corallinophycidae) and a negative association with diatoms (Diatomea). Furthermore, seagrass growth conditions influence metabolic pathways by activating nitrogen-related metabolism while attenuating methanogenesis. Our findings highlight the crucial roles of marine plants and their symbiotic systems in ensuring environmental conservation within the context of blue carbon storage across environmental gradients.

*Abbreviations*: DID, Difference-in-differences; ML, machine learning; AA, association analysis; ROC, receiver operating characteristic; AUC, the area under the curve; ELA, energy landscape analysis; EFA, exploratory factor analysis; CFA, confirmatory factor analysis; SEM, structural equation modeling; CI, causal inference; CMA, causal mediation analysis

## Introduction

Seagrasses play an important role in ensuring the conservation of marine ecosystems by generating dissolved oxygen in the ocean and serving as spawning sites for marine organisms[1]. Environmental research on seagrasses has revealed the need for efficient cycling of nitrogen and phosphorus[2-4] by seagrasses and sediments to control the negative effects of carbon dioxide[5]. As part of the blue carbon system[6], seagrass is involved in widespread environmental conservation efforts beyond the ocean, and its potential to suppress greenhouse gases is approximately 40 times greater than the potential of terrestrial forest soils to sequester organic carbon[7-8]. Additionally, seagrass meadows reduce the generation of methane gas, which is also a greenhouse gas[9]. In this way, seagrass has important implications for the conservation and management of coastal ecosystems and is necessary for the conservation of the entire global environment. However, important seagrass bed areas are being lost at comparable rates to those for artificial forests, coral reefs, and tropical rainforests, and a trend towards extinction has been confirmed[10-11]. Carbon dioxide emissions have increased due to seagrass loss [12]. Global warming and warming-related ocean acidification are the main triggers of these losses[5], while water pollution resulting from anthropogenic activities (e.g., domestic wastewater and aquaculture) [13-15] is also suspected to be a relevant factor. In contrast, evaluating the characteristics of natural sediments and sediments from areas where seagrass abundance is maintained via transplantation is an extremely important for identifying potential data sources for future environmental conservation.

It has been suggested that seagrass plays important roles in not only ensuring the conservation of marine ecosystems and the reduction in the environmental impacts (greenhouse gas generation) by seagrass itself but also in controlling the symbiotic microbiome. Although such analyses are more limited than those of terrestrial plants, several important reports exist[4,16-19]. Seagrass meadows reduce bacterial pathogens that affect fish, invertebrates, and humans[16]. In addition, nitrogen-fixing bacteria[7] and cable bacteria[18] involved in the metabolism of nitrogen and sulfur coexist. It is reported that seagrass symbionts inhibit the growth of harmful algae *Chatonella antiqua*[20], *Heterosigma akashiwo*[21] and *Alexandrium tamarense*[21], which can also promote the occurrence of red tides. Thus, understanding the symbiotic microbiota of seagrass as well as the role of seagrass itself in shaping the system is very important.

Here, we comprehensively investigated the bacterial and eukaryotic biomes of marine sediment from eight seagrass-growing areas in Japan. We identified metabolic network structures in addition to the major groups of bacteria-algae that are common across regional characteristics via causal network analyses. Gaining insight into the symbiotic microbes and metabolic processes in the seagrass sediment may offer novel perspectives on environmental conservation strategies.

**Results**

*Overview of the survey region in this study and their analysis procedure*

Here, the microbial and chemical characteristics of sediments from seagrass-rich areas along eight coastlines in Japan were investigated (Fig. 1a). These regions were categorized as two regions (Tokyo-HK and Tokyo-KC) facing the Pacific Ocean, three regions (Kitsuki, Saiki, and Okayama) located in the Seto Inland Sea, and two regions (Hakata and Noto) located in the Sea of Japan. The sea temperatures near these regions have been confirmed to have increased over the past decade according to data from the Japan Meteorological Agency (Fig. 1a), and sampling was conducted during the high temperature period on the basis of annual water temperature assessments from 2022 to 2023 (Figs.S1 and 1a). In these regions, the sediment in areas where seagrass grew and that in areas where seagrass did not grow were chosen as the subjects of analysis (Fig. 1b). Across regions, seagrass growth varied (Fig. S2), but genetically, the same genus occurred (the genus Zostera, Amamo as the Japanese name) (Table S1). Sediments were collected from each region with only a scoop or via custom methods (Fig. S3). The microbial (bacterial and eukaryotic biome) and chemical datasets derived from the regions were computationally analyzed, and an essential component group for seagrass growth in the marine sediments was estimated via feature selection and latent structure analyses as shown in Fig. 1c.

*Seagrasses modulate the microbial and chemical diversity of the sediments*

The diversity of bacteria and eukaryotes in the marine sediments significantly differed in terms of regional dependence and in the presence or absence of seagrass (Figs. 2, S4, and S5 and Table S2). Three indicators of α diversity for bacteria (Fig. S4) and eukaryotes (Fig. S5) were analysed. Significant differences were not necessarily observed between the seagrass and non-seagrass areas, although differences depending on the regions were confirmed (Table S2). Similarly, a similar trend in nonmetric multidimensional scaling (NMDS) for β diversity was observed (Fig. 2 and Table S3).

The population of individual bacteria was also recognized as exhibiting regional characteristics (Figs. 3a, S6, and S7), but the relative abundance of the families Desulfobulbaceae (coefficient=0.57, q=0.013), Marinilabiliaceae (coefficient=0.82, q=0.0063), and the Unknown family (coefficient=0.82, q=0.00041) belonging to the class Gammaproteobacteria, which depends on the presence or absence of seagrass beyond regionality are significantly different (Fig. S8), and the family Marinilabiliaceae was the only one whose abundance increased in the non-seagrass area. The population of individual eukaryotes also had regional characteristics (Figs. 3b, S9, and S10), but the relative abundance of the clade Deatomea (coefficient=0.66, q=0.021) and the Novel Clade Gran1 (coefficient=1.79, q=0.0074), belonging to the phylum Cercozoa, differed and increased in the seagrass-free area (Fig. S11). Furthermore, no significant differences in the characteristics of chemical components were observed depending on the presence or absence of seagrass within any seagrass growth area (Fig. S12). Thus, these difference-in-differences (DID) partly selected feature components depending on the presence or absence of seagrasses.

*Seagrasses affect the stabilities of the ML-selected symbiotic microbiota*

As seen in Fig. S13, thirteen bacteria (bacterium_enrichment_culture_clone, Crocinitomicaceae, Clostridiaceae_1, D3_BD7_8_other, Defluviitaleaceae, Desulfobulbaceae, Erysipelotrichaceae, Flavobacteriaceae, Hyphomonadaceae, Kiloniellaceae, Marinilabiliaceae, Pirellulaceae, and uncultured_Aminicenantes_bacterium) and four eukaryotes (Corallinophycidae, Intramacronucleata, Novel_Apicomplexa_Class_1, and Rhodymeniophycidae) were selected as characteristic microbial components that may be present or absent in seagrass sediments based on association analysis (AA), one of unsupervised machine learning (ML) algorithms. These components could be ranked via the use of other ML algorithms such as random forest, extreme gradient boosting (XGBoost), and light gradient boosting (LightGBM). As a result, twelve components with an increasing trend and five components with a decreasing trend were selected for seagrass flourishing (Figs. 4a and S13b). Furthermore, when energy landscape analysis (ELA) was performed based on the dominance rate as the abundance ratio, eleven positive factors (Rhodymeniophycidae, Novel_Apicomplexa_Class_1, Corallinophycidae, uncultured_Aminicenantes_bacterium, bacterium_enrichment_culture_clone, Hyphomonadaceae, Erysipelotrichaceae, Desulfobulbaceae, Defluviitaleaceae, Crocinitomicaceae, and Clostridiaceae_1) and six negative factors (Intramacronucleata, Pirellulaceae, Marinilabiliaceae, Kiloniellaceae, Flavobacteriaceae, and D3_BD7_8_other) were selected for seagrass abundance (Fig. 4b). The ELA-interaction network revealed that these factors exhibit complex relationships in seagrass bottom sediments (Fig. S14). Furthermore, at least the positive relationship among the cable bacteria, the family Desulfobulbaceae, and bacterium_enrichment_culture_clone belonging to the phylum Proteobacteria remained stable at the energy level under seagrass overgrowth conditions (Fig. 4c) as shown in Fig. 4d. In contrast, the families Flavobacteriaceae, Kiloniellaceae, Marinilabiliaceae, and D3_BD7_8_other remained stable under conditions without seagrass (Fig. 4c).

*Seagrasses balance the symbiotic microbial causal interactions*

Even in seagrass-rich areas, there are parts where seagrass is growing well and parts where it is not. The balance of the relationships between microorganisms was evaluated for the feature components (bacterium_enrichment_culture_clone, Crocinitomicaceae, Clostridiaceae_1, D3_BD7_8_other, Defluviitaleaceae, Desulfobulbaceae, Erysipelotrichaceae, Flavobacteriaceae, Hyphomonadaceae, Kiloniellaceae, Marinilabiliaceae, Pirellulaceae, uncultured_Aminicenantes_bacterium, Corallinophycidae, Intramacronucleata, Novel_Apicomplexa_Class_1, Rhodymeniophycidae, Diatomea, Unknown_Family, and Novel_Clade_Gran_1) selected by DID and ML algorithms. The relationships among the feature components were revealed via causal learn in python, except eukaryotes in the subclass Rhodymeniophycidae (Fig. 5a). Next, since the correlations of the selected component group included components with positive or negative correlations (Fig. S15), the function 'promax', an oblique rotation that allows components to be correlated, was chosen as the calculation function. The systemic tree of the feature component candidates was calculated using the minimum residual (MR) method as the default method. The number of MR factors for EFA was set to 4 according to the very simple structure (VSS) criterion rather than Kaiser–Guttman criterion (Fig. S16a). The components selected for EFA (Table S4) were based on the following criteria: KMO (Kaiser-Meyer-Olkin>0.5) and factor loadings (|loading|>0.3). As the result, nine bacteria (bacterium_enrichment_culture_clone, Crocinitomicaceae, D3_BD7_8_other, Desulfobulbaceae, Flavobacteriaceae, Hyphomonadaceae, Kiloniellaceae, Marinilabiliaceae, and Unknown_Family) and three eukaryotes (Corallinophycidae, Novel_Clade_Gran_1, and Diatomea) were clustered (Fig. S16b). The finding suggested that these factors are strongly related. Furthermore, five bacteria and three eukaryotes were selected via ROC curve analysis as positive and negative component groups, respectively, associated with the presence or absence of seagrass (Fig.5b and Table S4). In addition, the components that showed strong positive relationship with seagrass (AUC>0.6) were the bacterial families Desulfobulbaceae and Hyphomonadaceae, and the eukaryotic subclass Corallinophycidae. The components with a negative relationship (AUC<0.4) were the bacterial families Pirellulaceae, Kiloniellaceae, BD7_8_other bacteria, and the eukaryotic clade Diatomea and the subphylum Intramacronucleata. Structural equation modeling (SEM) for confirmatory factor analysis (CFA) was performed of these components. Because most of the feature components comprised non-Gaussian distribution data ($p<0.05$, according to the Anderson-Darling test, Table S4), the calculation was performed using the robust estimation function (estimator = "MLR"). The components shown in Fig.S17 constructed a structural equation model showing the optimal value ranges (Chisq p-value>0.05; CFI.robust ≒ 1; TLI.robust ≒ 1; RMSEA<0.07; SRMR<0.08; GFI>0.95; and AGFI>0.95) (Fig. 5c and Table S5). Structural equation modeling indicated that two causal directions between seagrass and the family Desulfobulbaceae were statistically supported: (No.1_b in Table S5) seagrass influencing Desulfobulbaceae abundance, and (No.1_a in Table S5). Both models showed equivalent fit indices (ex., $\chi^2$ = 0.113, p = 0.737, CFI.robust= 1.000, RMSEA=0.000, SRMR = 0.008), suggesting that these relationships are tightly linked. In other words, the relationship between seagrass and the family Desulfobulbaceae statistically appear bidirectional. The mediating effect was statistically evaluated using these structural equation models. The family Hyphomonadaceae had a significant negative mediating effect (38.3%) on seagrass via the family Desulfobulbaceae (Prop. Mediated = -0.383, p = 0.002), whereas a positive mediating effect (16.8%) via the eukaryotic Corallinophycidae on seagrass (Prop. Mediated = 0.168, p = 0.002) was confirmed. In addition, the mediating effect on seagrass from the eukaryote Corallinophycidae mediated 20.7% via the family Desulfobulbaceae (Prop. Mediated = -0.207, p = 0.02). The complexity among these components was suggested to be important as a group (Table S6).

These results revealed that the presence of seagrass was positively related to the bacterial families Desulfobulbaceae and Hyphomonadaceae and the eukaryotic Corallinophycidae and negatively related to the eukaryotic Diatomea. The selected microbial groups not only provide an optimal fit in the SEM, indicating their collective importance in explaining interactions with seagrass, but also show significant direct and indirect effects in the causal mediation analysis. Together, these results highlight their functional relevance in the seagrass-associated microbial community.

*Seagrasses balance the symbiotic metabolic causal interaction*

On the basis of the dataset obtained via PICRUST2 analysis, an evaluation was conducted by the limma-voom method. The value of the coefficient of determination for the smoothed line (LOESS $R^2$) was greater than 0.9 (LOESS_$R^2$=0.9253) (Fig. S18a), and the value (coefficient value=-0.9538) of Spearman's correlation was close to the absolute value of 1 (Fig. S18a). Therefore, statistical evaluation could be carried out. As shown in Fig. 6a and S18b, seven pathways were selected as MetaCyc pathway IDs by LogFC and FDR, of which five pathways were significantly activated, and two pathways were inhibited. Furthermore, causal learning for these selected pathways revealed that there was a causal relationship except for PWY-6167 (flavin biosynthesis) was exhibited (Fig. 6b). Amino acids metabolisms (PWY-6339, a superpathway of phenylethylamine degradation; and PWY-2941, L-lysine biosynthesis), nitrogen metabolism (PWY-4722, creatinine degradation), and peptidoglycan biosynthesis (PWY-6470 and PWY-6471) and an essential factor for methanogens (P241-PWY, coenzyme B biosynthesis) were related to seagrass flourishing conditions. Furthermore, ROC curve analysis selected three positive pathways (PWY-6471, peptidoglycan biosynthesis; PWY-4722, creatinine degradation; and PWY-2941, L-lysine biosynthesis) (AUC>0.6) and two negative pathways (PWY-6167, flavin biosynthesis and P241-PWY, coenzyme B biosynthesis) (AUC<0.4) related to seagrass (Figs.S18c and S19). Among these, an optimal structural equation model was constructed as a group with PWY-2941, PWY-6471, PWY-4722, and P241-PWY (Chisq p-value>0.05; CFI.robust≒1；TLI.robust ≒ 1；RMSEA<0.07; SRMR<0.08; GFI>0.95; and AGFI>0.95) (Figs. 6c and 6d, and Table S6). In this model, only P241-PWY indicated a negative relationship. The presence or absence of seagrass affects two pathways (PWY-2941 and PWY-4722), and PWY-6471 and P241-PWY may also influence these pathways. Although no statistically significant mediation effects were observed, these pathways, together with the presence or absence of seagrass, was considered to be important as a collective group, supported by the optimal fit of the structural equation model (Table S7).

**Discussion**

Here, common symbiotic microbial interaction in seagrass sediments were identified across an environmental gradient. The growth of seagrass as a blue carbon sink is restricted not only overseas but also in Japan due to environmental deterioration. In this study, we found universal characteristics of the symbiotic microbiota in the eight places where seagrass occurs, even under such degraded environmental conditions, from the perspectives of bacteria and eukaryotes. First, there were region-dependent changes in the microbiota, but in contrast, there was a general trend towards differences in the stable state of the dominant microbial group between the seagrass-free plots and the seagrass plots (Fig.4c). For example, the abundance of pathogenic bacteria increases when no seagrasses flourish[16]; in fact, the phylum Flavobacteria, a type of pathogenic bacteria, was selected as a key component in regions without seagrass (non-Seagrass C1 in Fig. 4c). In addition, the presence of cable bacteria such as Desulfobulbaceae is essential (Seagrass C10 in Fig. 4c). The observations also contributed to understanding the stable state of seagrass growth. The abundance of the order Flavobacteriales, which contains bacteria that are pathogenic to fish[22], can decrease around seagrass[16,18]. Cable bacteria Desulfobulbaceae[18] dominate in the seagrass sediments where seagrass grows. Desulfobulbaceae is a crucial component for seagrass growth, even in aquaculture facilities, where seagrasses generally do not flourish[18]. These reports do not contradict the results of this study. Further investigation revealed that the balance may be maintained depending on the presence or absence of seagrass (Figs.5a and S16b). Among other findings, a positive association with the cable bacteria Desulfobulbaceae, the nitrogen-cycling bacteria Hyphomonadaceae, and the coral algae Corallinophycidae, alongside a negative association with the diatoms Diatomea were confirmed as an optimal core structures (Figs.5c and 5d). As a result of mediation analysis for the model, the importance of these components as a group was statistically confirmed (Table S6 and Fig. S20a). In addition, the structural model was also reviewed via the use of other causal inference algorithms, and the analysis of the degree of influence in the structural equation model also suggested that the presence of seagrass itself may affect the symbiotic microorganisms in the optimal structure model (Fig. 5c). Regardless, the importance of these components as a group was confirmed in terms of the calculation (Fig. S20b). Next, a structural equation model was constructed in which the metabolic pathways involved in nitrogen metabolism[23], the biosynthesis of peptidoglycan[23] involved in bacterial growth, and the biosynthesis of coenzyme B [24], an essential factor for methane production[23], were confirmed to be related to seagrass (Fig.6c). Furthermore, the degree of influence of these metabolic pathways in the structural equation model suggested that the presence of seagrass itself was important (Fig. S21).

These results indicate that these microbial structures as well as metabolic structures are universally essential in the waters near Japan. Interestingly, these structures do not point to the importance of a single component but are computationally important for the group as a whole (Figs. 5c, 5d, 6c, and 6d, and Tables S5, S6, and S7). It is also notable that the role of individual factors is based on the previous reports, and several other reports exist.

For example, studies of seagrass models (*Halophila ovalis* and *Zostera muelleri*) have shown root-to-root oxygen loss (ROL)[25] results in the formation of thin oxidation zones around the roots. There are bacteria that use this redox boundary to transport electrons and oxidize sulfides. Specifically, in seagrass beds, it is known that the roots and rhizosphere are susceptible to reducing stresses of sulfide, and sulfate-reducing bacteria (SRBs) [26] are important at the shade/rhizosphere interface of the sediment where the oxygen supply is limited. The cable bacteraia Desulfobulbaceae [27], which is one of seagrass symbiont SRBs and can reduce sulfate to sulfide and oxidize sulfide to sulfur [28-31], has been shown to be involved in the nitrogen cycle [32-33]. In addition, *Desulfobulbaceae* has been confirmed in the sediment of two seagrasses, *Thalassia hemprichii* [34] and *Halophila ovalis* [35], in different regions. In the structural equation model in this study, on the basis of the relationships among the components that were positively related to seagrass, it is assumed that the sulfuric acid reduction reaction may occur on the surface of the seagrass sediment and in the surrounding hypoxic environment. SRBs have been reported to possess a calcification support mechanism, such as decomposing organic matter and reducing sulfuric acid to hydrogen sulfide ($H_2S$), which promotes calcium carbonate deposition[36]. However, the direct relationship between the coral algae Corallinophycidae and the cable bacteraia Desulfobulbaceae is unknown. The bacterial family Hyphomonadaceae often live in the marine environments and have been referred to as biofilm-formers, for example, in algae and marine surface microbial communities[37]. In fact, in this study, the biosynthesis of peptidoglycan and amino acids was activated (Fig. 6) and was not thought to be inconsistent in the formation of biofilms. In the presence of the coral algae Corallinophycidae, the abundance of the phylum Proteobacteria including the family Hyphomonadaceae increases, while its abundance also increases under warming plus acidification conditions [38]. In contrast, Diatomea, which exhibits a negative relationship in a seagrass flourishing environment, has been confirmed to coexist with SRBs[39]. These results may provide a bird's-eye view of the change in the balance between the constituents, rather than simply their increase or decrease.

In terms of red tides, which are an example of a detrimental environmental burdens, the dinoflagellates involved with red tides may increase in the next step of diatom increase[40-44]. In this study, the effect of seagrass on dinoflagellates is not known because dinoflagellates were not selected as an important factor; however, the presence of seagrass decreased the proportion of diatoms, and it was structurally negatively related. Therefore, the presence of seagrass may reduce the amount of diatoms in the bottom sediments or may be adsorbed, resulting in a reduction in floating diatoms. As a result, it is possible that the balance of the abundance ratio of diatoms in the early stages of red tide occurrence is maintained and seagrass acts in a restrained manner against red tide occurrence. Notably, it is interesting to note that among the eight areas surveyed, Hinase (Okayama_G and Okayama_K) of Okayama Prefecture, where aquaculture facilities are located, is known to have few occurrences of red tides despite the presence of aquaculture facilities.

From a metabolic pathway perspective, the biosynthesis of peptidoglycan and lysine during the formation of symbiotic biofilms with seagrasses may be associated with the proliferation of symbiotic bacteria. Additionally, the accelerated breakdown of creatinine is reminiscent of the treatment of biological waste in symbiosis with seagrasses. Conversely, the biosynthesis of Coenzyme B, which is negatively correlated with seagrass, suggests a reduction in methane-producing bacteria. Since methanogens are generally anaerobic, the generation of oxygen from seagrasses may inhibit them. On the basis of these results, seagrass flourishing is expected to contribute to the maintenance of marine ecosystems at the metabolic level. Furthermore, the biosynthetic pathway of lysine (https://metacyc.org/pathway?orgid=META&id=PWY-2941), which is derived from the family Bacillaceae, is expected to be altered by external factors. In fact, previous studies[18] revealed that at an aquaculture facility using a fermented feed containing the thermophilic Bacillaceae[45-49] for many years, seagrass growth was observed after 2004 when use of the fermented feed began, even though no seagrass growth was seen at nearby aquaculture facilities. These facts underscore a potential relationship between the low abundance of the family Bacillaceae and seagrass growth. Additionally, *Caldibacillus hisashii* (synonym: *Bacillus hisashii*)[50-51], a thermophilic Bacillaceae isolated from the fermented feed, decrease methanogen levels in the intestines of calves[52]. In chicken, it efficiently digests crude protein in feed and reduces the concentration of crude protein and urate in excrement[53]. Although the microbiota of marine sediment and anaerobic gut microbiota differ, an interesting connection has been observed. It may be necessary to better understand how microorganisms interact within ecosystem cycles from the perspective of One Health. Our findings also suggest that controlling symbiotic microorganisms may be important for supporting the cultivation of seagrass through an environmental conservation strategy. Initially, transplants were conducted in Tokyo_HK from Tokyo_CK and other areas, but rooting required many years [54-55]. One reason for this is thought to be the need for a colonization period for symbiotic microorganisms, and conditions that increase their abundance. Thus, our findings in this study provide an essential clue on how seagrass growth can be enhanced for ensuring sustainable ecosystems from a broad perspective, such as by conserving marine ecosystems, reducing greenhouse warming and controlling pathogens.

## Methods
### *Surveyed regions and sampling methods*
Sampling of marine sediments was carried out in the following eight regions where seagrass grew: Hakata (N33.596115, E130.269691) in Imazu Bay of Hakata Bay in Fukuoka Prefecture (https://maps.gsi.go.jp/vector/#10/33.596115/130.269691/&ls=photo&disp=1&d=l); Kitsuki (N33.409902, E131.675306) in Morie Bay of Beppu Bay in Oita Prefecture (https://maps.gsi.go.jp/vector/#10/33.41026/131.670816/&ls=photo&disp=1&d=l); Noto (N37.308803, E137.234492; N37.18253, E137.14124) in Tsukumo Bay of Noto peninsula in Ishikawa Prefecture (https://maps.gsi.go.jp/vector/#10/37.308803/137.234492/&ls=photo&disp=1&d=l); Okayama_G (N34.711492, E134.285321) in Genji Bay of Hinase in Okayama Prefecture (https://maps.gsi.go.jp/vector/#10/34.711492/134.285321/&ls=photo&disp=1&d=l); Okayama_K (N34.719837, E134.284655) in Yonago Bay of Hinase in Okayama Prefecture (https://maps.gsi.go.jp/vector/#10/34.719837/134.284655/&ls=photo&disp=1&d=l); Saiki (N32.844180, E131.976601; N32.850582, E131.966180; N32.844678, E131.977572) in Nishinoura Bay of Saiki in Oita Prefecture (https://maps.gsi.go.jp/vector/#10/32.84418/131.976601/&ls=photo&disp=1&d=l; https://maps.gsi.go.jp/vector/#10/32.850582/131.96618/&ls=photo&disp=1&d=l; https://maps.gsi.go.jp/vector/#10/32.84418/131.976601/&ls=photo&disp=1&d=l; Tokyo-HK (N35.338525, E139.637851) Umi-no-kouen in Yokohama (Hakkeijima) of Kanagawa Prefecture (https://maps.gsi.go.jp/vector/#10/35.338525/139.637851/&ls=photo&disp=1&d=l); Tokyo-KC (N35.432861, E139.905639) Kaneda gyokou in Kisarazu of Chiba Prefecture (https://maps.gsi.go.jp/vector/#10/35.432537/139.90555/&ls=photo&disp=1&d=l). Each of these regions was close to the following ocean and inland sea in Japan: regions facing the Pacific Ocean (Tokyo-HK and Tokyo-KC); regions located in the Seto Inland Sea (Kitsuki, Saiki, and Okayama); and regions facing the Sea of Japan (Hakata and Noto). The sea temperatures near these regions were summarized on the basis of the data from the Japan Meteorological Agency (https://www.data.jma.go.jp/kaiyou/data/db/kaikyo/series/engan/engan.html). Marine sediment was sampled in the summer to coincide with high tides and was collected with a shovel.

Noto and Tokyo_HK were sampled by diving. Samples for the Okayama_G and Okayama_K were collected by an Ekman-Birge bottom sampler. Custom methods that differs from ordinary shovel collection were shown in Fig. S3.

*Detection of bacteria and eukaryotes*

Total DNA was obtained according to previously described methods[18]. Bacterial 16S rRNA gene analysis was carried out via PCR amplification with the barcoded forward primer 341F (5′-CCTACGGGNGGCWGCAG-3′) and reverse primer 805R (5′- GACTACHVGGGTATCTAATCC-3′) primers targeting the hypervariable regions (V3-V4) of the 16S rRNA gene according to a previous report[56]. Eukaryote 18S rRNA gene analysis was carried out via PCR amplification with the barcoded forward primer Euk454F (5′-CCAGCASCYGCGGTAATTCC-3′) and reverse primer Euk964R (5′- ACTTTCGTTCTTGATYRA-3′) primers targeting the hypervariable regions (V4) of the 18S rRNA gene[57]. Multiplexed amplicon sequences for the samples were sequenced via the MiSeq platform according to the manufacturer's instructions. The obtained FASTAQ format files were analysed using QIIME2 (https://qiime2.org) [58]. In addition, the sequences were trimmed via maxEE (default: inf) of the DADA2 plugin, and the trimmed sequences were taxonomically classified via vsearch (using the SILVA database). All rRNA sequence datasets were deposited in the GenBank Sequence Read Archive database as described in the "Data and materials availability" section.

*Microbial diversity*

The values of α-diversity obtained by Qiime2 were assessed by multiple comparisons as described in the "Statistical analysis" section. The statistical analysis of nonmetric multidimensional scaling (NMDS) values for β-diversity was performed using the "vegan" (distance=bray as the calculation condition) and "pairwiseAdonis" libraries in R software. These were visualized by the following libraries and modules in Python (version 3.10.8) in Mac OS Sequoia (version 15.3) on arm64: "numpy", "pandas", "seaborn", "matplotlib.pyplot", "statsmodels.formula.api", "statsmodels.stats.multitest", "matplotlib.patches", "matplotlib.lines".

*Physicochemical analyses*

The following analyses were performed as previously described[18]. In brief, the contents of Cu, Fe, Mn, and Zn were determined using an iCE3000 atomic absorption spectrophotometer (Thermo Scientific Co., Ltd.). The contents of CaO, $P_2O_5$, $K_2O$, MgO, $NH_4$-N, $NO_3$-N, phosphate absorption coefficient, CEC (cation exchange capacity), and humus were determined using a soil and plant clinical analyzer SFP-4j (Fujidaira, Co., Ltd.). Total nitrogen and total carbon were measured using NC-220F (Sumika Chemical Analysis Service, Ltd.). The electric conductivity (EC) and pH values were measured using an MM-60R multi-water quality meter (Towa DKK Co., Ltd.).

*Feature selection*

Here, feature selection using difference-in-differences (DID) and ML algorithms was performed respectively. On the DID, the region-dependent microbiota and chemical composition were assessed. The R-package "MaAsLin3" as Microbiome Multivariable Association with Linear Models was used to analyze the differences between comparator groups. Next, feature component candidates that could not be selected by DID were extracted using ML algorithsm. Four machine learning (ML) algorithms were employed to identify feature component candidates independent of DID. Association analysis (AA), an elementary method of unsupervised ML, was applied [59-60]. In brief, the omics data were classified on the basis of the median value (M) of the data themselves and sorted as 0 (< M) or 1 (>M). The library packages "arules" and "arulesViz" of the R software were used to conduct association rule mining of these datasets. To extract feature components via ML algorithms, random forest models [61], one of ML algorithms with bagging (bootstrap aggregation), and XGBoost (eXtreme Gradient Boost)[62], one of ML algorithms with extreme gradient boosting[63], and LightGBM (light gradient boosting machine), one of ML algorithms for a highly efficient gradient boosting decision tree[64], were used. These ML algorithms can be used for not only for predicting test data on the basis of training data, but also for selecting features to classify between groups (feature selection) [63,65-66]. These indicators were MeanDecreaseGini (random forest), importance (XGBoost and LightGBM), and **SHapley Additive exPlanations (**SHAP) value [67-69] (randomforest, XGBoost, and LightGBM). The following libraries and modules in Python (version 3.10.8) were used in Mac OS Sequoia (version 15.3) on arm64: "xgboost", "RandomForestClassifier", "lightgbm", "train_test_split" (from sklearn.model_selection), "accuracy_score" (from sklearn.metrics), "numpy", "pandas", "re", and "shap". These selected feature components were visualized as a bubble chart using the "matplotlib" (https://matplotlib.org/stable/index.html).

In order to evaluate the diagnostic accuracy of the feature components, receiver operating characteristic (ROC) curve analysis was performed using the R library "Epi". The conditions administering the compost extract were set to 1. The calculated area under the curve (AUC) were used as the classification indicator.

*Energy landscape analysis*

Energy landscape analysis (ELA) was performed as previously described[18,70]. ELA is a data-driven method for constructing landscapes that explain the stability of the community composition across environmental gradients. The probability of observing a microbial composition $\sigma^{(k)}$ in an environment $\epsilon$ is determined by microbial interactions, the effect of presence or absence of seagrass, site-specific effects, and other underlying regularities. In summary, the ELA was based on an extended pairwise maximum entropy model that explains the probability of the occurrence of the ecological state of a sample $k$, $\sigma^{(k)}$ given the environmental condition $\epsilon^{(k)}$; as the ecological state, we combined the presence/absence status of selected taxa and the levels of microbial factors:

$\sigma^{(k)}: \sigma^{(k)} = \left(\sigma_1^{(k)}, \sigma_2^{(k)}, \ldots \sigma_N^{(k)}\right)$

where N=$n_b$ + $n_e$ represents the total number of taxa, with $n_b$ and $n_e$ denoting bacterial and eukaryotic taxa, respectively. As the environmental condition, we introduced a binary vector consisting of nine elements: one environmental factor representing the effect of seagrass presence or absence (absence=0, presence=1) and eight factors reprenting site-specific effects (absence=0, presence=1): $\epsilon^{(k)} = (\epsilon_{seagrass}, \epsilon_{site1}, \epsilon_{site2}, \epsilon_{site3}, \epsilon_{site4}, \epsilon_{site5}, \epsilon_{site6}, \epsilon_{site7}, \epsilon_{site8})$. Here, the site-specific factors serve as dummy variables, where exactly on variable takes the value 1 and all others are 0 for each sampling point. The model can be expressed as follows:

(I) $P(\sigma^{(k)}|\epsilon^{(k)}) = \frac{e^{-E(\sigma^{(k)}|\epsilon^{(k)})}}{\Sigma_l e^{-E(\sigma^{(k)}|\epsilon^{(k)})}}$,

(II) $E(\sigma^{(k)}|\epsilon^{(k)}) = -(\Sigma_i \Sigma_j J_{ij} \sigma_i^{(k)} \sigma_j^{(k)})/2 + \Sigma_i g_i^{seagrass} \epsilon_{seagrass}^{(k)} \sigma_i^{(k)} + \Sigma_i g_i^{site1} \epsilon_{site1}^{(k)} \sigma_i^{(k)} + \cdots + \Sigma_i h_i \sigma_i^{(k)})$.

where, $P(\sigma^{(k)}|\epsilon^{(k)})$ is the probability of the occurrence of an ecological state $\sigma^{(k)}$ given environmental factor $\epsilon^{(k)}$. Eq. (I) indicates that the probability is high when the energy $E(\sigma^{(k)}|\epsilon^{(k)})$ is low and *vice versa*. In eq. (II), $E(\sigma^{(k)}|\epsilon^{(k)})$ is defined as the sum of the effects of interactions among components, seagrass flourishing, and the net effect of unobserved environmental factors. The parameters in eq. (II), namely, $J_{ij}$, $g_i^{factors}$ ($factors \in \{seagrass, site1, site2, site3, site4, site4, site5, site6, site7, site8\}$) and $h_i$, indicate the effects of the relationships among components ($J_{ij} > 0$ favours and $J_{ij} < 0$ disfavours the co-occurrence of components $i$ and $j$), the effect of seagrass on component $i$ (the presence of seagrass positively ($g_i^{seagrass} > 0$) or negatively ($g_i^{seagrass} < 0$) affects the occurrence of component $i$), and how likely component $i$ is when the other factors are equal. All the components in $\sigma^{(k)}$ were converted to the range 0-1 as follows. We defined the relative abundance of each component in the sediment samples) as the presence (1>median value) or absence (0<median value) of each type of bacterial and eukaryotic data. Consequently, we obtained a set of explanatory variables $\sigma^{(k)}$ with components that accompany the environmental condition $\epsilon^{(k)}$, which represents the status of the presence of seagrasses and surveyed regions.

*Causal learning*
To infer the causal structure among variables, we applied the constraint-based PC algorithm to estimate a causal directed acyclic graph (DAG). For continuous variables, normality was assessed using the Shapiro–Wilk test. When one or more variables were identified as nonnormally distributed (p < 0.05), a nonparametric independence test (KCIT, kernel-based conditional independence test) was employed. If discrete variables were detected, then the chi-square test was used; otherwise, the Fisher-z test was applied for normally distributed data. The resulting causal graphs were visualized using the following Python libraries: "numpy", "pandas", "network" (from causallearn.search.ConstraintBased.PC), "kci" (from causallearn.utils.cit), "shapiro" (from scipy.stats), and "matplotlib.pyplot". These selected feature factors were visualized as a bubble chart (https://matplotlib.org/stable/index.html).

*Exploratory factor analysis*
In brief, we conducted a search for important factors via exploratory factor analysis (EFA) [71-72] conducted with the library packages "psych" and "GPArotation" in R software. The analysis codes were obtained from the website (https://cran.r-project.org/web/packages/psych/index.html). The number of factors and their values were calculated by the function 'VSS' for a very simple structure (VSS). Here, EFA was performed with 'fm=minres' as a condition, i.e., on the basis of the minimum residual (MR) method. In addition, the values of Spearman's rank correlation coefficients were used to determine whether an oblique rotation or an orthogonal rotation should be applied in calculation. The other parameters for EFA were as follows: $h^2$, communality score; $u^2$, uniqueness score; and com, complexity, an information score that is generally related to uniqueness. To select the components for EFA, the values of Kaiser-Meyer-Olkin (KMO) and factor loadings were used (the minimum requirements to implement an EFA: KMO>0.5 and |loading|>0.3). The selected components calculated by the function 'fa' were visualized via the library package "heatmaply" of R software.

*Confirmatory factor analysis*
To assess structural universality across the animal species, structure equation modeling (SEM) for confirmatory factor analysis (CFA) was performed for a group of components with many directed acyclic graphs (DAGs), as previously reported [73-75]. For the selected components, SEM was conducted using the library package "lavaan" [76] in R software. Analysis codes are based on those on the website (https://lavaan.ugent.be). As CFA requires a hypothesis, the components (AUC>0.6 and AUC<0.4) selected via ROC curve analysis after ML application were utilized as factors for a latent construct of metabolites and the microbiota. The distribution of the data was assessed using the Shapiro–Wilk test in the R library "MVN" to determine whether it was Gaussian or non-Gaussian. Here, the hypothesized models were statistically estimated via the robust maximum likelihood parameter estimator 'MLR' (for non-Gaussian data distribution) as a computational condition for the R libraries "lavaan" and "sem". Model fit was assessed by the chi-square p value (p>0.05, nonsignificant), robust comparative fit index (CFI.rbust/cfi.rbust) (>0.95), robust Tucker–Lewis index (TLI.rbust/tli.rbust) (>0.95), root mean square error of approximation (RMSEA/rmsea) (<0.07), standardized root mean residual (SRMR/srmr) (< 0.08), goodness-of-fit index (GFI/gfi) (>0.95), and adjusted goodness-of-fit index (AGFI/agfi) (>0.95) as indices of good model fit [77]. In addition, standardized residuals (SRs) (|SR|<2.58) were complementarily confirmed. To select the optimal structural equation, when the above conditions were satisfied, an equation with a negligible difference between the GFI and AGFI values was adopted. Next, the model with a low value of the Akaike information criterion (AIC) was adopted from among many candidate models. Standard errors were calculated via 1000 bootstrap draws via the maximum likelihood method. The path diagrams of the good model were constructed with 'layout=tree' using the package "semPlot" of R software [78].

*Pathway analysis*
The pathway data of the microbiota were obtained using the analysis method of PICRUST2 in Qiime2. These count raw data were used to visualize mean-variance trends as well as volcano plots with false discovery rate (FDR) and LogFC (fold change). The R packages "edgeR", "limma", "ggplot2", "EnhancedVolcano", "pheatmap", "RColorBrewer", "vegan", and "dplyr" were employed, and the values of LOESS_R2 (0.09 or higher is preferable), RMSE_smoothness (small numbers are preferable), and Spearman_correlation coefficient (absolute value close to 1), DEGs_voom_adj_0.05 (number of genes with FDR<0.05), and DEGs_raw_adjP_0.05 (number of DEGs before normalization) were quantified. Furthermore, the raw data were transformed by centered Log-ratio and thereafter used for structural equation models and causal inference.

*Causal mediation analysis*
To investigate whether the effect of the treatment on the outcome was mediated by a given variable, causal mediation analysis (CMA)[79] was performed using the packages "mediation", "DiagrammeR", "DiagrammeRsvg", and "rsvg" in R software partly referred to a tutorial website. Each regression model with "~" for the optimal structural equation model was assessed using the "lm" function. Subsequently, the values of the causal relationships of treatment-mediator-outcome were evaluated. The estimated average causal mediation effect (ACME), average direct effect (ADE), and proportion of the total effect through mediation were calculated by nonparametric bootstrapping ('boot=TRUE' as a command) and a quasi-Bayesian approximation ('boot=FALSE' as a command) with 'sims=1000' as the number of iterative calculations.

*Causal discovery*
The BayesLiNGAM method [80], a Bayesian score-based approach, was applied for causal structural inference among the components in the optimal model as previously described [74-75]. The BayesLiNGAM method was established by the "fastICA" package (https://cran.r-project.org/web/packages/fastICA) of R software. On the basis of information on the website (https://www.cs.helsinki.fi/group/neuroinf/lingam/bayeslingam/), the percentage data calculated by BayesLiNGAM were visualized by the R library package "igraph" as previously described [74-75].

*Statistical analyses*
Data for frequentist equivalence testing were analysed as follows: the Shapiro–Wilk test, Henze-Zirkler test, and/or the Anderson-Darling test was used to evaluate the Gaussian distribution and select parametric and nonparametric analyses. The Shapiro–Wilk test, Henze-Zirkler test, and/or the Anderson-Darling test was performed using the R software function "shapiro. test" and the R library "MVN", and/or python library "shapiro". The F test was used to evaluate equal variances and to select the unpaired t test or the Welch's t test for a parametric analysis. The Wilcoxon rank-sum test and Wilcoxon signed-rank test were used for nonparametric analysis. In addition, ANOVA followed by Tukey's HSD test and the Kruskal–Wallis test followed by the Dunn's test, as appropriate methods dependent upon datasets, were performed. The Dunn's test was performed using "dunn.test" library of R software, respectively. Significance was declared at p<0.05, and a tendency was assumed at 0.05≤p<0.10. Spearman's rank correlation coefficients were visualized using the library "corrplot" ("method=spearman") function of R software. The data were calculated via R software (version 4.3.3). Microsoft Office (version 16.66.1) were also used. The data are presented as the means ± SDs. Data collection and analyses were not performed in a blinded manner.

## Acknowledgments
We are grateful to Ms. Yasuharu Katte (Kyushu University) for help with the preparation of the samples. Special thanks for Ms. Risa Monna (Reso Co., Ltd.), Ms. Akiyo Muto and Mr. Takashi Kobayashi (Agrokanesho Co., Ltd.) for special technical support, and Ms. Makiko Hoshi, and Mr. Hiroshi Hirakawa for help with preparation of field photograph.

## Author contributions
Hirokuni Miyamoto, Hisashi Miyamoto, and Jun Kikuchi conceptualised the experiments; Hirokuni Miyamoto, Nobuhiro Kawachi, Yukinari Tsuruda, Yohei Shimasaki, Shouzo Ogizo, Nobuo Suzuki, Tomoharu Yuge, Toshio Takahashi, Tomohito Ojima, Toshio Furota, Akio Sakamoto, Keiichi Takimoto, Kozo Kugimiya, Takehiro Tanaka, Takashi Kimura, Yuuji Oshima, and Jun Kikuchi performed the sampling of the marine sediments and/or the adjustments; Hirokuni Miyamoto, Nobuhiro Kawachi, Makiko Matsuura, Naoko Tsuji, Teruno Nakaguma, Chitose Ishii performed the preparation of the sample to analyze; Shigeharu Moriya, Wataru Suda, Tamotsu Kato, Chie Shindo performed the NGS experiments; Hirokuni Miyamoto, Kenta Suzuki, Shigeharu Moriya, Makiko Matsuura, Naoko Tsuji, Teruno Nakaguma, and Jun Kikuchi analysed the raw data; Hirokuni Miyamoto, Kenta Suzuki, Shigeharu Moriya, Atsushi Kurotani, Hiroaki Kodama, Hiroshi Masuya, Satoshi Wada, Jun Kikuchi, Hiroshi Ohno provided analytical methods and/or equipments; Hirokuni Miyamoto, Jun Kikuchi, and Hiroshi Ohno supervised the study; Hirokuni Miyamoto, Kenta Suzuki, Shigeharu Moriya, Wataru Suda, Jun Kikuchi, and Hiroshi Ohno wrote and reviewed the manuscript. All the authors have read and approved the final manuscript.

## Data availability
Raw files of the bacterial 16S rRNA (V3-V4) data and eukaryotic 18S rRNA (V4) data were deposited in the DNA Data Bank of Japan (DDBJ) under NCBI BioProject accession number PRJDB043178 (BioSample Accession no. SAMD1689534-SAMD1689999). All the data were stored in a source data file ("Data_seagrass.xlsx"). The source commands will be provided on the GitHub sites. Please contact the corresponding authors to obtain any additional information. If you need the information not listed here, please contact the corresponding authors.

## Funding
The authors received no specific funding for this work.

## Competing interests
The authors declare no competing interests.

# References

1. Eger, A. M., Best, R. J. & Baum, J. K. Dominance determines fish community biomass in a temperate seagrass ecosystem. *Ecol Evol* **11**, 10489-10501 (2021).
2. David T., W. Nitrogen fixation in seagrass meadows: Regulation plant bacteria interactions. *Ecology Letters* **3**, 58-71 (2000).
3. Fraser, M. W., Gleeson, D. B., Grierson, P. F., Laverock, B. & Kendrick, G. A. Metagenomic Evidence of Microbial Community Responsiveness to Phosphorus and Salinity Gradients in Seagrass Sediments. *Front Microbiol* **9**, 1703 (2018).
4. Mohr, W. et al. Terrestrial-type nitrogen-fixing symbiosis between seagrass and a marine bacterium. *Nature* **600**, 105-109 (2021).
5. Ricart, A. M. et al. Coast-wide evidence of low pH amelioration by seagrass ecosystems. *Glob Chang Biol* **27**, 2580-2591 (2021).
6. Oreska, M. P. J. et al. The greenhouse gas offset potential from seagrass restoration. *Sci Rep* **10**, 7325 (2020).
7. McLeod, E. et al. A blueprint for blue carbon: toward an improved understanding of the role of vegetated coastal habitats in sequestering CO2. *Frontiers in Ecology and the Environment* **9**, 552-560 (2011).
8. Fourqurean, J. W. et al. Seagrass ecosystems as a globally significant carbon stock. *Nature Geoscience* **5**, 505-509 (2012).
9. Hall, N. et al. Coastal methane emissions driven by aerotolerant methanogens using seaweed and seagrass metabolites. *Nat Geosci* **18**, 854-861 (2025).
10. Waycott, M. et al. Accelerating loss of seagrasses across the globe threatens coastal ecosystems. *Proc Natl Acad Sci U S A* **106**, 12377-12381 (2009).
11. Strydom, S. et al. Too hot to handle: Unprecedented seagrass death driven by marine heatwave in a World Heritage Area. *Glob Chang Biol* **26**, 3525-3538 (2020).
12. Salinas, C. et al. Seagrass losses since mid-20th century fuelled CO2 emissions from soil carbon stocks. *Glob Chang Biol* **26**, 4772-4784 (2020).
13. Liu, S. et al. Nutrient loading decreases blue carbon by mediating fungi activities within seagrass meadows. *Environ Res* **212**, 113280 (2022).
14. Thomsen, E., Herbeck, L. S. & Jennerjahn, T. C. The end of resilience: Surpassed nitrogen thresholds in coastal waters led to severe seagrass loss after decades of exposure to aquaculture effluents. *Marine Environmental Research* **160** (2020).
15. Murata, H., Hara, M., Yonezawa, C. & Komatsu, T. Monitoring oyster culture rafts and seagrass meadows in Nagatsura-ura Lagoon, Sanriku Coast, Japan before and after the 2011 tsunami by remote sensing: their recoveries implying the sustainable development of coastal waters. *PeerJ* **9**, e10727 (2021).
16. Lamb, J. B. et al. Seagrass ecosystems reduce exposure to bacterial pathogens of humans, fishes, and invertebrates. *Science* **355**, 731-733 (2017).
17. Imai, I., Inaba, N. & Yamamoto, K. Harmful algal blooms and environmentally friendly control strategies in Japan. *Fisheries Science* **87**, 437-464 (2021).
18. Miyamoto, H. et al. Computational estimation of sediment symbiotic bacterial structures of seagrasses overgrowing downstream of onshore aquaculture. *Environ Res* **219**, 115130 (2022).
19. Sun, H. et al. Novel insights into the rhizosphere and seawater microbiome of Zostera marina in diverse mariculture zones. *Microbiome* **12**, 27 (2024).
20. Inaba, N. et al. Dynamics of seagrass bed microbial communities in artificial Chattonella blooms: A laboratory microcosm study. *Harmful Algae* **84**, 139-150 (2019).
21. Inaba, N. et al. Algicidal and growth-inhibiting bacteria associated with seagrass and macroalgae beds in Puget Sound, WA, USA. *Harmful Algae* **62**, 136-147 (2017).
22. Loch, T. P. & Faisal, M. Emerging flavobacterial infections in fish: A review. *J Adv Res* **6**, 283-300 (2015).
23. Kastenmuller, G., Schenk, M. E., Gasteiger, J. & Mewes, H. W. Uncovering metabolic pathways relevant to phenotypic traits of microbial genomes. *Genome Biol* **10**, R28 (2009).
24. Kirschning, A. On the evolution of coenzyme biosynthesis. *Nat Prod Rep* **39**, 2175-2199 (2022).
25. Martin, B. C. et al. Oxygen loss from seagrass roots coincides with colonisation of sulphide-oxidising cable bacteria and reduces sulphide stress. *ISME J* **13**, 707-719 (2019).
26. Gao, P. et al. Genomic insight of sulfate reducing bacterial genus Desulfofaba reveals their metabolic versatility in biogeochemical cycling. *BMC Genomics* **24**, 209 (2023).
27. Burdorf, L. D. W. et al. Long-distance electron transport occurs globally in marine sediments. *Biogeosciences* **14**, 683-701 (2017).
28. Nielsen, L. P., Risgaard-Petersen, N., Fossing, H., Christensen, P. B. & Sayama, M. Electric currents couple spatially separated biogeochemical processes in marine sediment. *Nature* **463**, 1071-1074 (2010).
29. Pfeffer, C. et al. Filamentous bacteria transport electrons over centimetre distances. *Nature* **491**, 218-221 (2012).
30. Reguera, G. Bacterial power cords. *Nature* **491**, 201-202 (2012).
31. Larsen, S., Nielsen, L. P. & Schramm, A. Cable bacteria associated with long-distance electron transport in New England salt marsh sediment. *Environ Microbiol Rep* **7**, 175-179 (2015).
32. Marzocchi, U. et al. Electric coupling between distant nitrate reduction and sulfide oxidation in marine sediment. *ISME J* **8**, 1682-1690 (2014).
33. Risgaard-Petersen, N., Damgaard, L. R., Revil, A. & Nielsen, L. P. Mapping electron sources and sinks in a marine biogeobattery. *Journal of Geophysical Research: Biogeosciences* **119**, 1475-1486 (2014).
34. Cai, Z. et al. Bacterial epiphyte and endophyte communities of seagrass Thalassia hemprichii: the impact of feed extract solution. *Environ Microbiol Rep* **13**, 757-772 (2021).
35. Tarquinio, F., Attlan, O., Vanderklift, M. A., Berry, O. & Bissett, A. Distinct Endophytic Bacterial Communities Inhabiting Seagrass Seeds. *Front Microbiol* **12**, 703014 (2021).
36. Spring, S. et al. Sulfate-Reducing Bacteria That Produce Exopolymers Thrive in the Calcifying Zone of a Hypersaline Cyanobacterial Mat. *Front Microbiol* **10**, 862 (2019).
37. Nelson, C. E. et al. Coral and macroalgal exudates vary in neutral sugar composition and differentially enrich reef bacterioplankton lineages. *ISME J* **7**, 962-979 (2013).
38. Lemay, M. A. et al. Morphological complexity affects the diversity of marine microbiomes. *ISME J* **15**, 1372-1386 (2021).
39. Siebers, R. et al. Marine particle microbiomes during a spring diatom bloom contain active sulfate-reducing bacteria. *FEMS Microbiol Ecol* **100** (2024).
40. Zhang, Y. et al. Metatranscriptomic Signatures Associated With Phytoplankton Regime Shift From Diatom Dominance to a Dinoflagellate Bloom. *Front Microbiol* **10**, 590 (2019).
41. Sun, X. et al. Effects of Algal Blooms on Phytoplankton Composition and Hypoxia in Coastal Waters of the Northern Yellow Sea, China. *Frontiers in Marine Science* **9** (2022).
42. Liang, Y. et al. Nutrient-limitation induced diatom-dinoflagellate shift of spring phytoplankton community in an offshore shellfish farming area. *Mar Pollut Bull* **141**, 1-8 (2019).
43. Ollison, G. A. et al. Physiology governing diatom vs. dinoflagellate bloom and decline in coastal Santa Monica Bay. *Front Microbiol* **14**, 1287326 (2023).
44. Chai, X. et al. Blooms of diatom and dinoflagellate associated with nutrient imbalance driven by cycling of nitrogen and phosphorus in anaerobic sediments in Johor Strait (Malaysia). *Mar Environ Res* **169**, 105398 (2021).
45. Tanaka, R. et al. Feed additives with thermophile-fermented compost enhance concentrations of free amino acids in the muscle of the flatfish Paralichthys olivaceus. *J Gen Appl Microbiol* **56**, 61-65 (2010).
46. Miyamoto, H. et al. The oral administration of thermophile-fermented compost extract and its influence on stillbirths and growth rate of pre-weaning piglets. *Res Vet Sci* **93**, 137-142 (2012).
47. Satoh, T. et al. Impact of oral administration of compost extract on gene expression in the rat gastrointestinal tract. *J Biosci Bioeng* **114**, 500-505 (2012).
48. Miyamoto, H. et al. Thermophile-fermented compost as a possible scavenging feed additive to prevent peroxidation. *J Biosci Bioeng* **116**, 203-208 (2013).
49. Tanaka, R. et al. Thermophile-fermented compost as a fish feed additive modulates lipid peroxidation and free amino acid contents in the muscle of the carp, Cyprinus carpio. *J Biosci Bioeng* **121**, 530-535 (2016).
50. Miyamoto, H. et al. Potential probiotic thermophiles isolated from

51. Nishida, A. et al. Bacillus hisashii sp. nov., isolated from the caeca of gnotobiotic mice fed with thermophile-fermented compost. *Int J Syst Evol Microbiol* **65**, 3944-3949 (2015).
52. Inabu, Y. et al. Development of a novel feeding method for Japanese black calves with thermophile probiotics at postweaning. *J Appl Microbiol* **132**, 3870-3882 (2022).
53. Kikusato, M. & Namai, F. Effect of Bacillus hisashii Supplementation on the Growth Performance, Nitrogen Excretion, and Intestinal Microbiota of Broiler Chickens Fed Low Crude Protein Diet. *Anim Sci J* **96**, e70088 (2025).
54. Lee, K. S. & Park, J. I. An effective transplanting technique using shells for restoration of Zostera marina habitats. *Mar Pollut Bull* **56**, 1015-1021 (2008).
55. Oliveira, V. H. et al. Transplantation of seagrass (Zostera noltei) as a potential nature-based solution for the restoration of historically contaminated mudflats. *Sci Total Environ* **959**, 178257 (2025).
56. Klindworth, A. et al. Evaluation of general 16S ribosomal RNA gene PCR primers for classical and next-generation sequencing-based diversity studies. *Nucleic Acids Res* **41**, e1 (2013).
57. Stoeck, T. et al. Multiple marker parallel tag environmental DNA sequencing reveals a highly complex eukaryotic community in marine anoxic water. *Mol Ecol* **19 Suppl 1**, 21-31 (2010).
58. Bolyen, E. et al. Reproducible, interactive, scalable and extensible microbiome data science using QIIME 2. *Nat. Biotechnol.* **37**, 852–857 (2019).
59. Shiokawa, Y., Misawa, T., Date, Y. & Kikuchi, J. Application of market basket analysis for the visualization of transaction data based on human lifestyle and spectroscopic measurements. *Anal. Chem.* **88**, 2714–2719 (2016).
60. Wei, F., Sakata, K., Asakura, T., Date, Y. & Kikuchi, J. Systemic homeostasis in metabolome, ionome, and microbiome of wild yellowfin goby in estuarine ecosystem. *Sci. Rep.* **8**, 3478 (2018).
61. Breiman, L. Random Forests. *Machine Learning* **45**, 5-32 (2001).
62. Chen, T. & Guestrin, C. in Proceedings of the 22nd ACM SIGKDD International Conference on Knowledge Discovery and Data Mining 785-794 (2016).
63. Miyamoto, H. & Kikuchi, J. An evaluation of homeostatic plasticity for ecosystems using an analytical data science approach. *Comput Struct Biotechnol J* **21**, 869-878 (2023).
64. Ke, G. et al. LightGBM: A highly efficient gradient boosting decision tree. *Proceedings of the 31st Annual Conference on Neural Information Processing Systems, Long Beach, CA, USA, 4–9 December 2017*, pp. 3146-3154 (2017).
65. Yu, Z., Ma, J., Qu, Y., Pan, L. & Wan, S. PM(2.5) extended-range forecast based on MJO and S2S using LightGBM. *Sci Total Environ* **880**, 163358 (2023).
66. Taguchi, Y. et al. Causal estimation of maternal-offspring gut commensal bacterial associations under livestock grazing management conditions. *Computational and Structural Biotechnology Reports* **1** (2024).
67. Karwowska, Z., Aasmets, O., Estonian Biobank research, t., Kosciolek, T. & Org, E. Effects of data transformation and model selection on feature importance in microbiome classification data. *Microbiome* **13**, 2 (2025).
68. Cao, S. & Hu, Y. Interpretable machine learning framework to predict gout associated with dietary fiber and triglyceride-glucose index. *Nutr Metab (Lond)* **21**, 25 (2024).
69. Myers, T. et al. Chronological age estimation from human microbiomes with transformer-based Robust Principal Component Analysis. *Commun Biol* **8**, 1159 (2025).
70. Suzuki, K., Nakaoka, S., Fukuda, S. & Masuya, H. Energy landscape analysis elucidates the multistability of ecological communities. *Ecological Monographs* **91**, e01469 (2021).
71. Schmitt, T. A. Current methodological considerations in exploratory and confirmatory factor analysis. *J. Psychoeduc. Assess.* **29**, 304–321 (2011).
72. Peprah, S. et al. *Inverse association of falciparum* positivity with endemic Burkitt lymphoma is robust in analyses adjusting for pre-enrollment malaria in the EMBLEM case-control study. *Infect. Agent Cancer* **16**, 40 (2021).
73. Miyamoto, H. et al. An agroecological structure model of compost-soil-plant interactions for sustainable organic farming. *ISME Commun.* **3**, 28 (2023).
74. Miyamoto, H. et al. Computational estimation of sediment symbiotic bacterial structures of seagrasses overgrowing downstream of onshore aquaculture. *Environ. Res.* **219**, 115130 (2023).
75. Miyamoto, H. et al. A potential network structure of symbiotic bacteria involved in carbon and nitrogen metabolism of wood-utilizing insect larvae. *Sci. Total Environ.* **836**, 155520 (2022).
76. Rosseel, Y. lavaan: an R package for structural equation modeling. *J. Stat. Softw.* **48**, 1–36 (2012).
77. Hooper, D., Coughlan, J. & Mullen, M. R. Structural equation modelling: guidelines for determining model fit. *Electronic Journal of Business Research Methods* **6**, 53–60 (2008).
78. Epskamp, S., Stuber, S., Nak, J., Veenman, M. & Jorgensen, T. D. Path diagrams and visual analysis of various SEM packages' output. R package version 1.1.2 (2019).
79. Tingley, D., Yamamoto, T., Hirose, K., Keele, L. & Imai, K. mediation: R Package for Causal Mediation Analysis. *Journal of Statistical Software* **59**, 1-38 (2014).
80. Hoyer, P. O. & Hyttinen, A. in Proceedings of the Twenty-Fifth Conference on Uncertainty in Artificial Intelligence (UAI '09) 240–248 (AUAI Press, 2009).

# Figure legends

## Fig. 1 Survey regions with overgrown seagrass in Japan and research overview of this study

(a) Study regions shown on the maps of Japan. The eight areas shown in the figure were investigated. The red arrows in the maps indicate the sampling locations. The abbreviations are as follows: Hakata, a bay in Fukuoka Prefecture. K and G, bays in Hinase of Okayama Prefecture, hereinafter described as Okayama_K and Okayama_G, respectively; Noto, Tukumo Bay in Noto of Ishikawa Prefecture; Kitsuki, Morie Bay in Beppu Bay in Kitsuki City of Oita Prefecture; Saeki, Nishinoura Bay in Kamae of Saeki city of Oita Prefecture; Tokyo_HK, near the sea park (Umi-no-koen) in Yokohama (Hakkeijima) of Kanagawa Prefecture (a part of Tokyo bay); Tokyo_KC, near Kaneda Fishing Port in Kisarazu city of Chiba Prefecture (a part of Tokyo bay). The line graph shows the change in sea temperature in the sea near the survey regions. The arrows indicate the sampling years. (b) Image of sampling portions for seagrass *Zostera* sediment and non-seagrass sediment. The sampling portions for seagrass and non-seagrass sediments are shown in green and pink parentheses, respectively. (c) The workflow of this study. The abbreviations are as follows: DID, difference-in-differences; ROC, receiver operating characteristic; ELA, energy landscape analysis; EFA, exploratory factor analysis; CI, causal inference; SEM, structural equation modeling; SCN, symbiotic causal network.

## Fig. 2 Symbiotic microbiome diversity in the marine sediments.

Characteristics of the microbiome of marine sediment classified via nonmetric multidimensional scaling (NMDS). The NMDS data of the (a) bacterial and (b) eukaryotic populations are visualized to evaluate of β-diversity. The circles and squares indicate the seagrass and non-seagrass sediment, respectively, as described in the legend 'CND (marker shape)'. The statistical values calculated using the R library "pairwiseAdonis" are shown in the upper right portion. The sampling sites are distinguished by different colours according to the legend 'Env (region)'. The statistical values for the 'Env(region)' are listed in Table S1.

## Fig.3 Relative abundance of the marine sediment bacterial families.

The relative abundances of (a) bacterial families (D4 category selected by QIIME2 for 16S rRNA sequences) and (b) eukaryotes (D4 category selected by QIIME2 for 18S rRNA sequences) in seagrass and non-seagrass sediment are presented. These bacteria were statistically selected by maasilin3. In addition, statistical comparisons of the data between seagrass and non-seagrass sediment are shown with asterisks. The abbreviations are as follows: sgr, seagrass sediment; non, non-seagrass sediment. The names of the regions are the same as Fig.1. Abbreviations are as follows: B_D4_, bacterial families (D4) classification selected QIIME2 for 16S rRNA sequences; E_D4_, eukaryote D4 classification selected QIIME2 for 18S rRNA sequences; Oka_G, Okayama_G in Fig.1a; Oka_K, Okayama_K in Fig. 1a; Ty_HK, Tokyo_HK in Fig.1a; Ty_KC, Tokyo_KC in Fig.1a; and *, $p<0.05$ in Fig. S3 and S6.

## Fig. 4 Assessment of the symbiotic feature component candidates by machine learning (ML) algorithms.

(a) Bubble chart of the feature component candidates selected via the three types of ML algorithms. Definitions: Random Forest SHAP, a measure of random forest; XGBoost SHAP, a measure of XGBoost; LightGBM SHAP, a measure of LightGBM. (b) Scores for the feature component candidates by energy landscape analysis. (c) The axis formed the energy landscape with compositional energy. The labels "non-Seagrass C1" and "Seagrass C10" represent the labels of stable states for non-seagrass and seagrass conditions, respectively. The vertical axis shows the energy level. (d) Image of the stable state in the energy landscape. The dimples are an example of two groups in a stable state.

## Fig. 5. Causal structural estimation of seagrass sediment symbiotic feature components and a putative model based on the experimental data.

(a) Causal structure of seagrass sediment feature components estimated using causal-learn in Python. The DAGs composed of the components selected by DID, feature selection, and ROC curve analysis are shown. (b) The feature components selected by ROC curve analysis. The seagrass condition was set to 1, and the selected components (green circle, AUC>0.6; and violet circle, AUC<0.4) were visualized. The feature components selected by the DID or ELA stable states (C1 and C10) were also marked as described in the figure. (c) Optimal model estimated via structural equation modeling (SEM) for confirmatory factor analysis (CFA) is shown. The colours indicate the following: green, putative positive interaction; purple, putative negative interaction. The asterisks (*) and (***) indicate $p<0.05$ and $p<0.001$, respectively. Abbreviations are as follows: Sgr, seagrass conditions; Chisq, chi-square $\chi^2$; p, p value for chi-square; CFI.robust, robust comparative fix index; TLI.robust, robust Tucker–Lewis index; RMSEA, root mean square error of approximation; SRMR, standardized root mean residual; GFI, goodness-of-fit index; and AGFI, adjusted goodness-of-fit index. (d) Putative model of seagrass-sediment bacteria-algae-diatoms relationship deduced by this study.

## Fig. 6. Causal structural estimation of seagrass sediment features metabolic pathway candidates and a putative model based on the experimental data.

(a) Causal structure of seagrass sediment features metabolic pathways. The PICRUST2-based pathway candidates selected in Fig. S15 were calculated by causal-learn in Python. The DAGs composed of the components selected by DID, feature selection, and ROC curve analysis are shown. (b) Fold changes in significant feature pathways. This plot displays the log2-fold change (logFC) of selected features significantly altered between experimental conditions (seagrass and non-seagrass). The x-axis shows the magnitude in the seagrass sediment (green colour, positive; blue colour, negative), while the y-axis shows individual functional feature pathways. (c) The optimal model estimated via structural equation modeling (SEM) for confirmatory factor analysis (CFA) is shown. The colours indicate the following: green, putative positive interaction; purple, putative negative interaction. The asterisks (*) and (**) indicate $p<0.05$ and $p<0.01$, respectively. Abbreviations are as follows: Sgr, seagrass conditions; Chisq, chi-square $\chi^2$; p, p value for chi-square; CFI.robust, robust comparative fix index; TLI.robust, robust Tucker–Lewis index; RMSEA, root mean square error of approximation; SRMR, standardized root mean residual; GFI, goodness-of-fit index; and AGFI, adjusted goodness-of-fit index. (d) Putative model of the relationship between seagrass and sediment metabolic pathways deduced in this study.

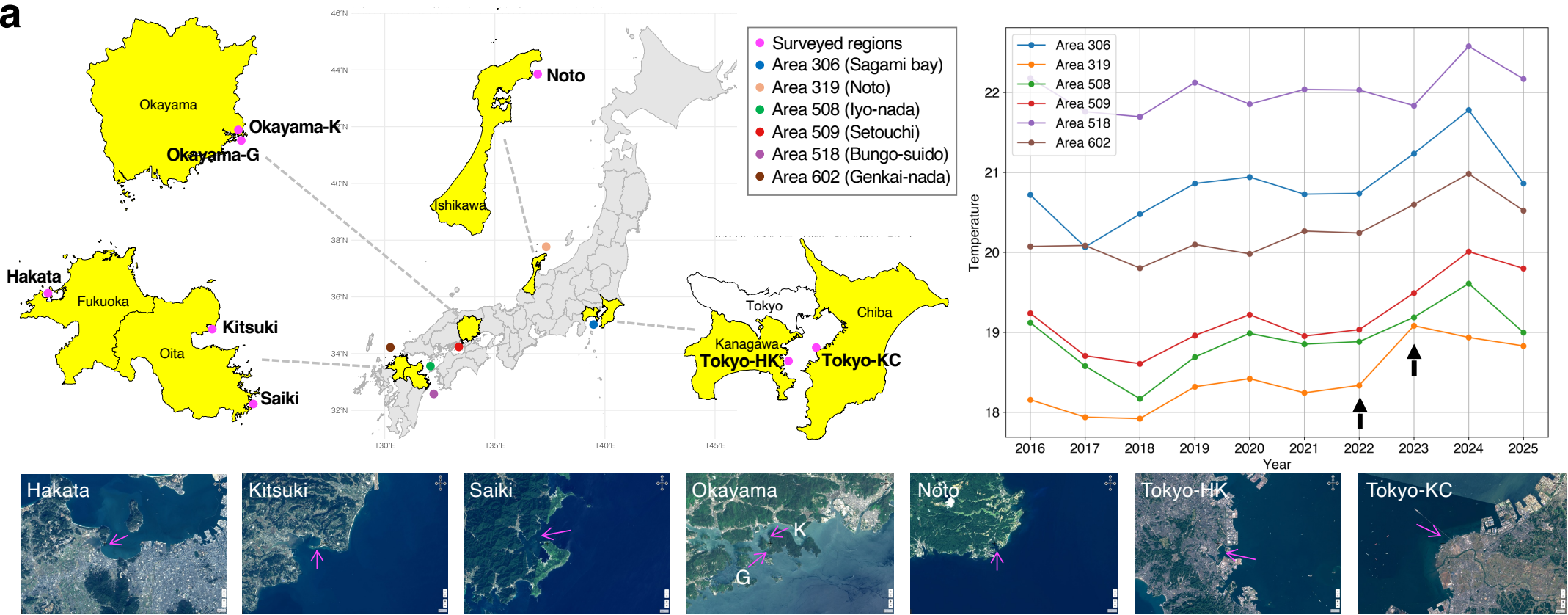
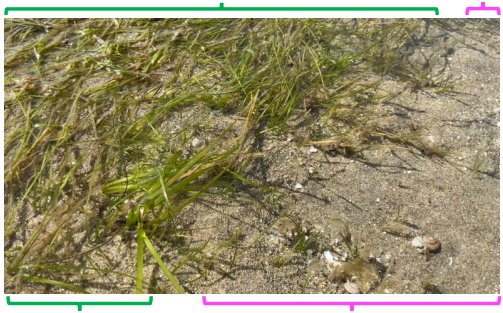
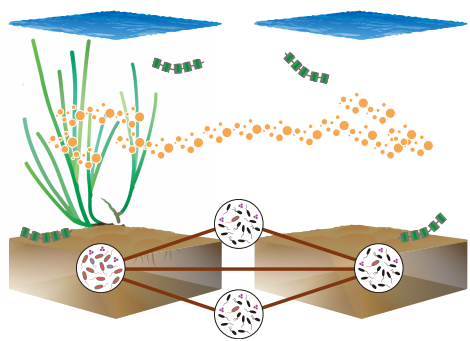
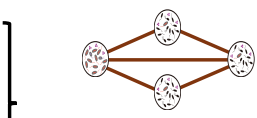

Fig.1

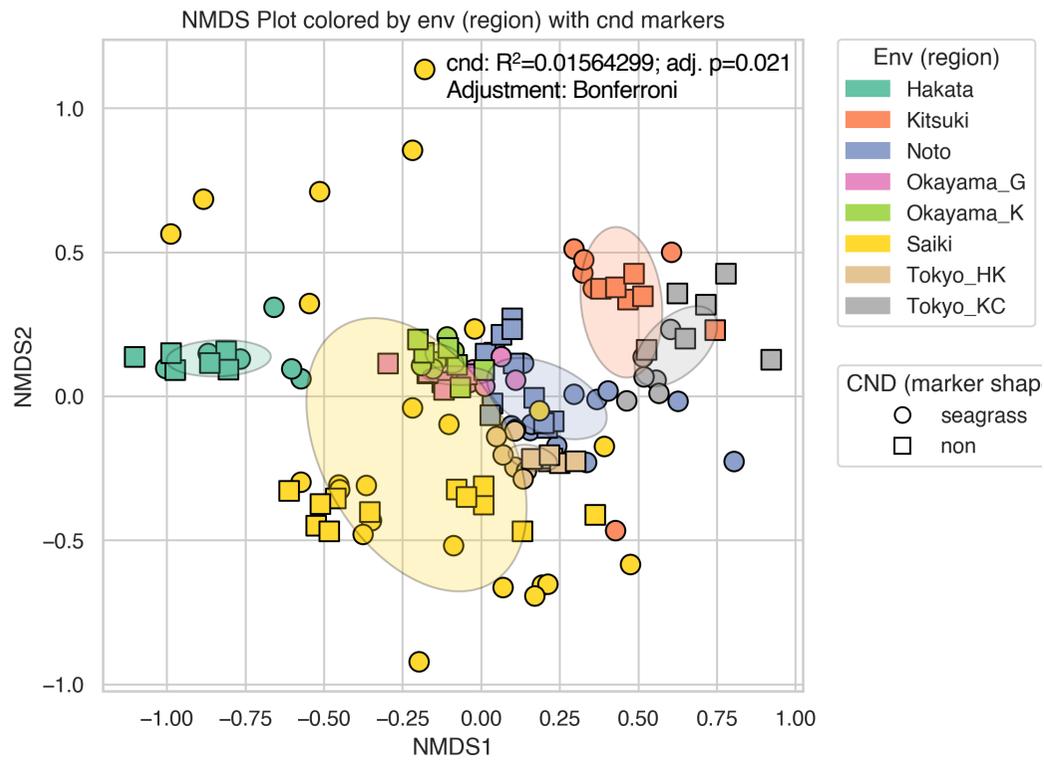 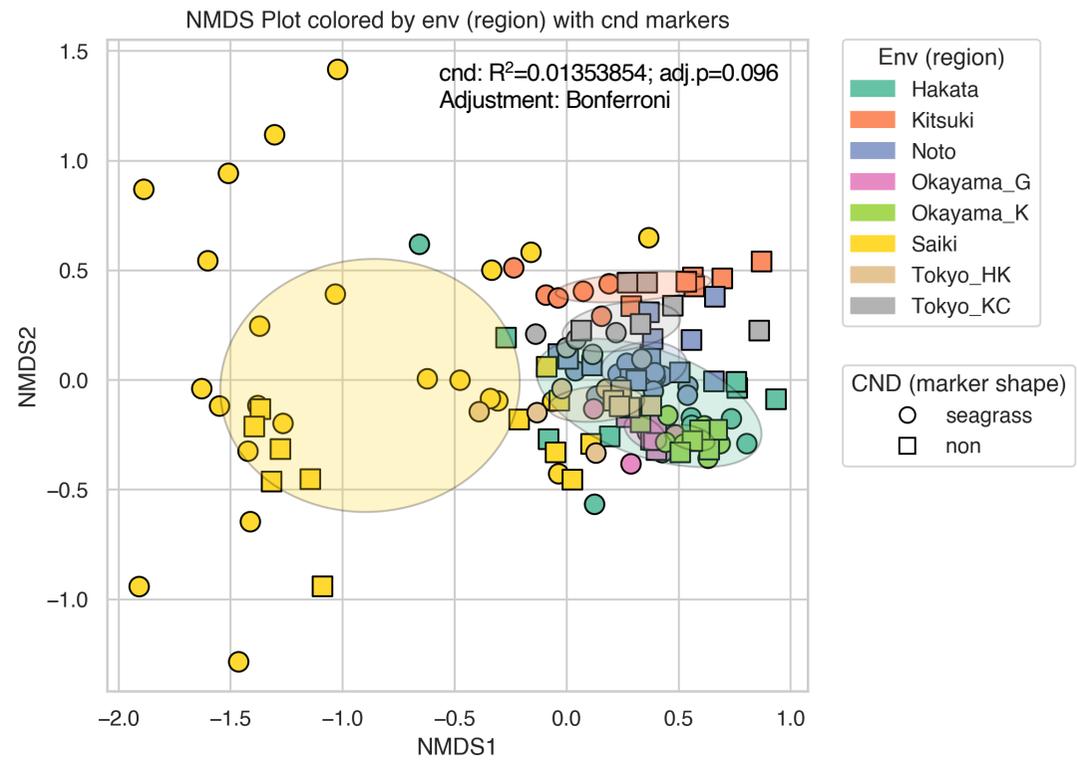

**Fig.2**

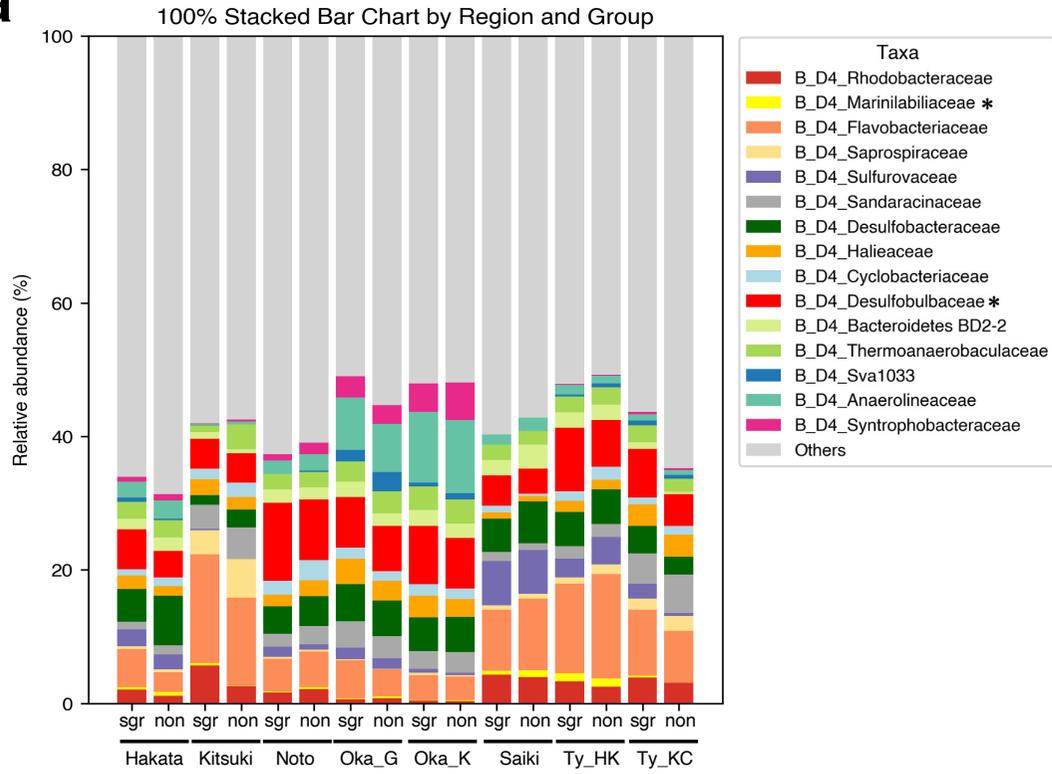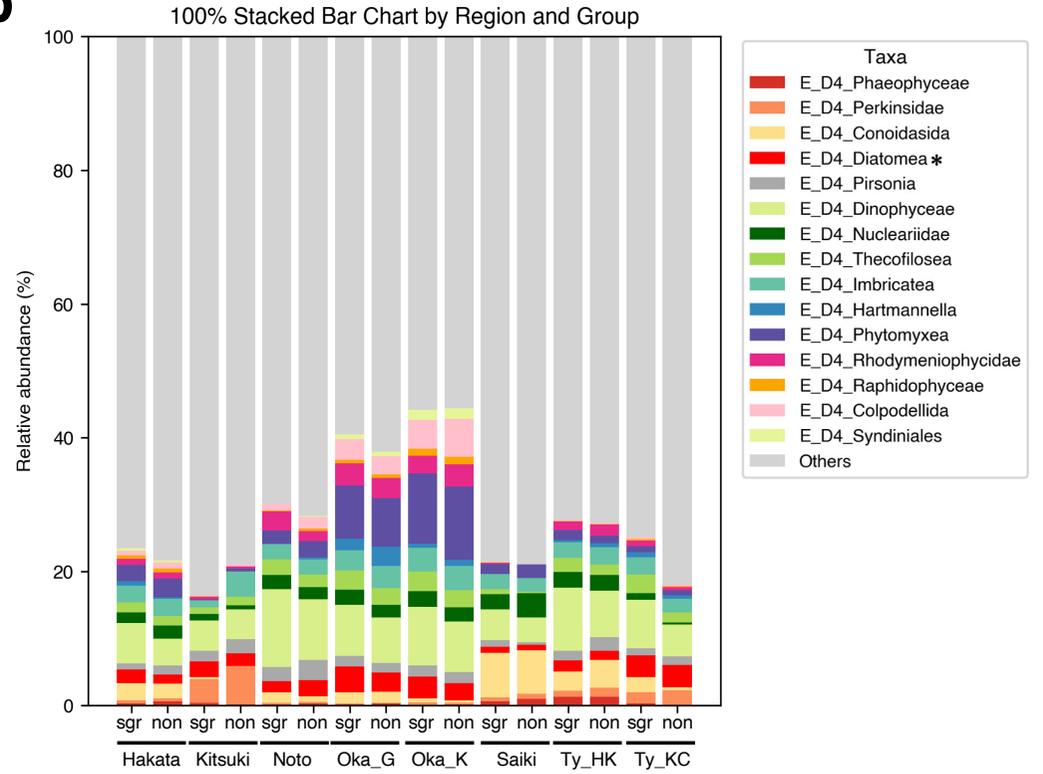

Fig.3

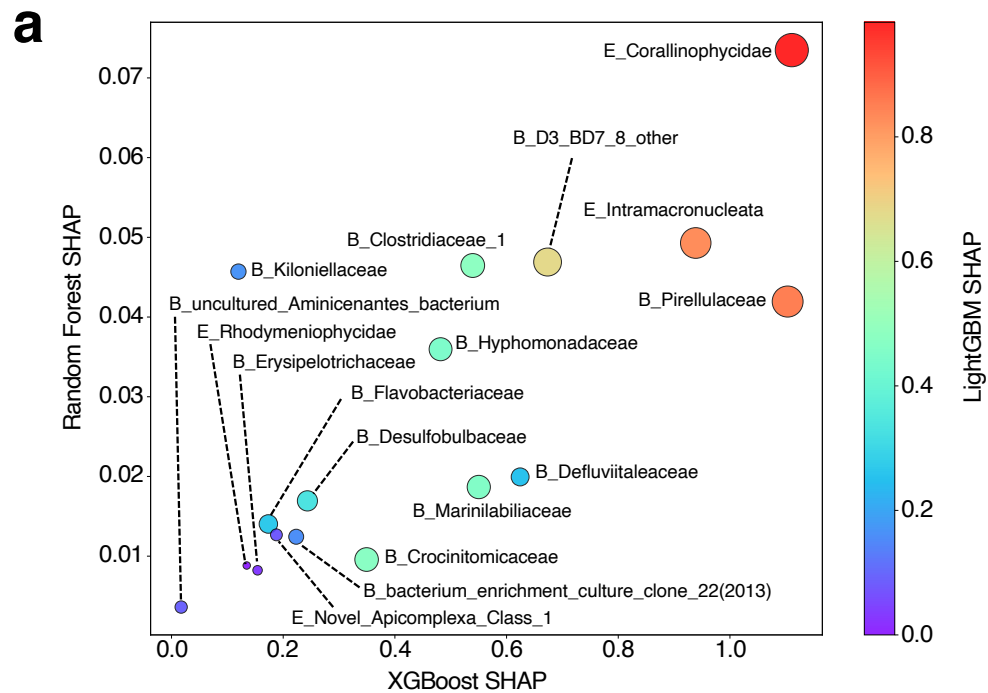
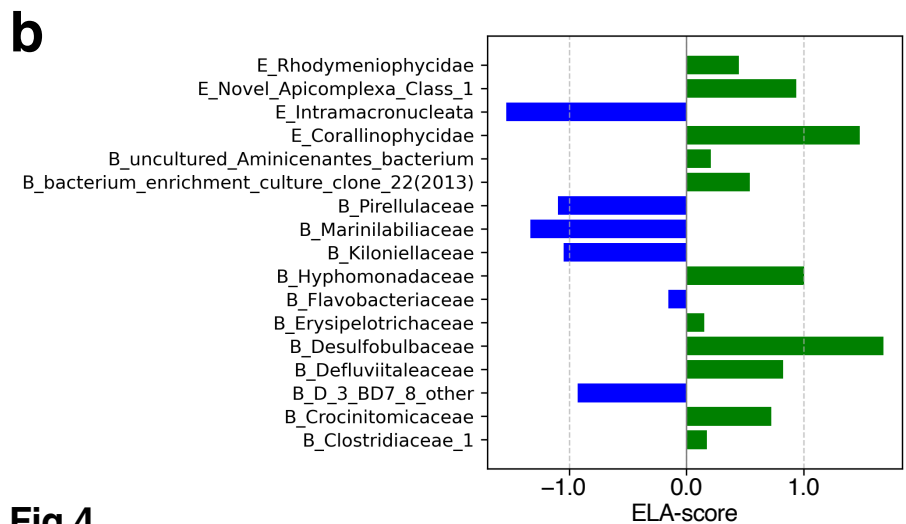
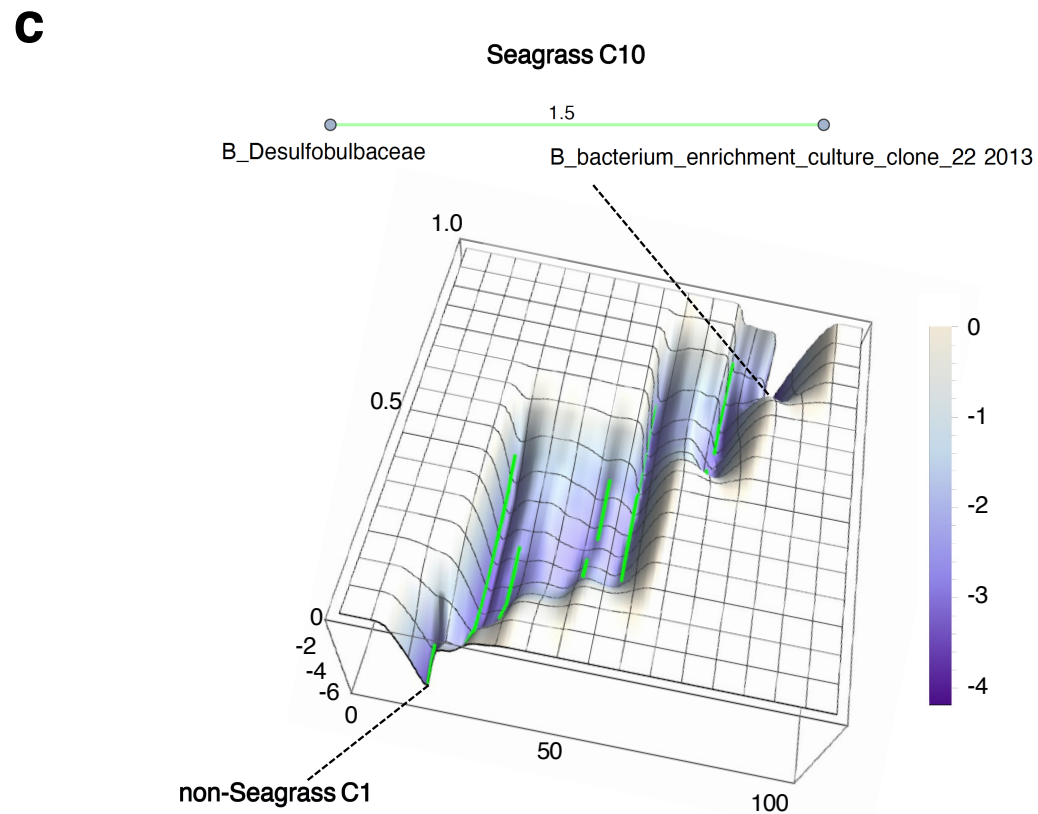
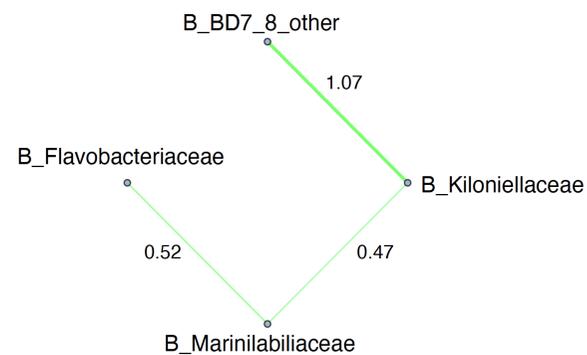
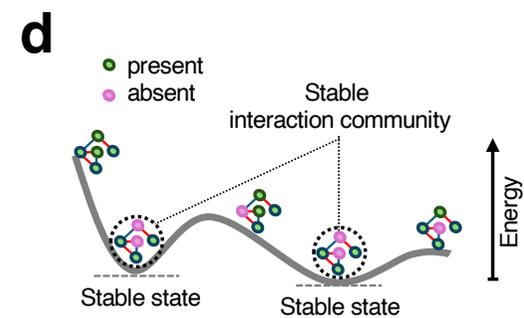

Fig.4

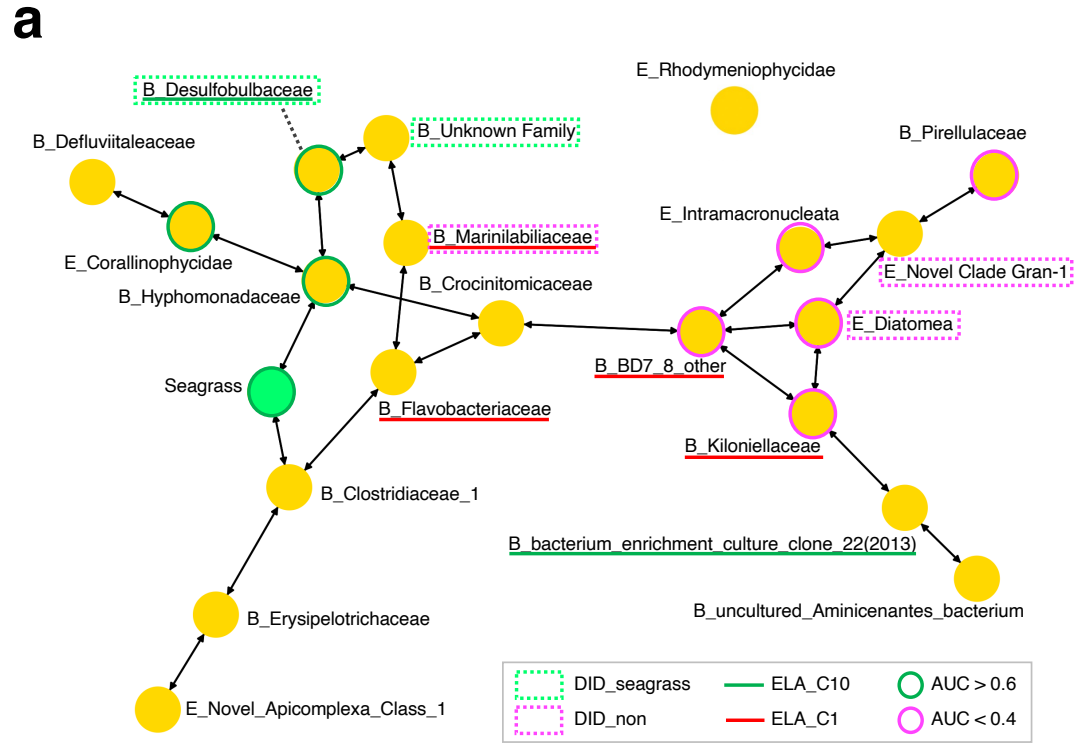
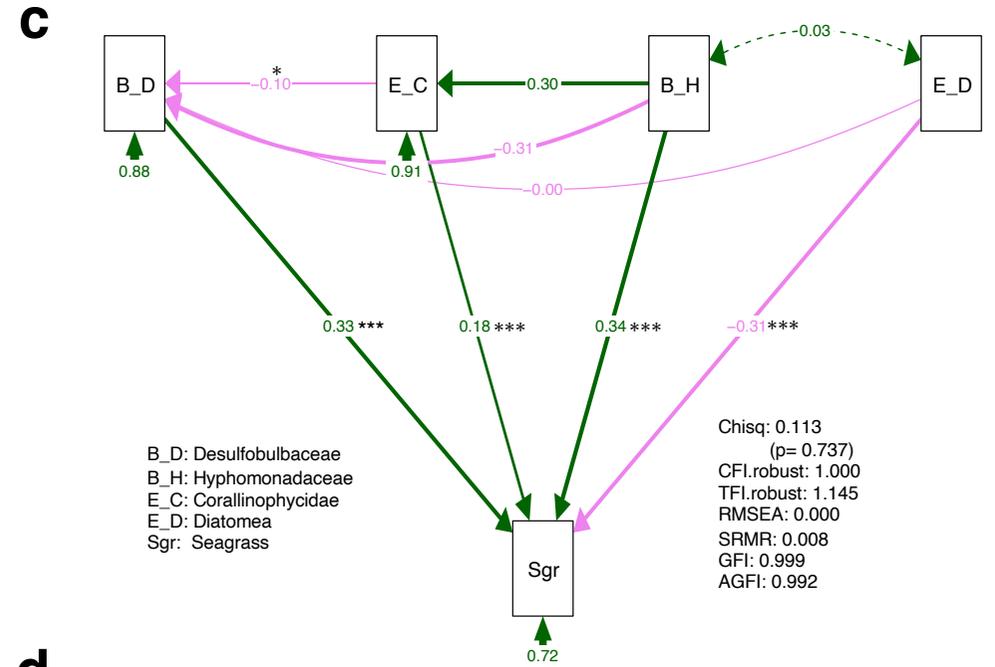
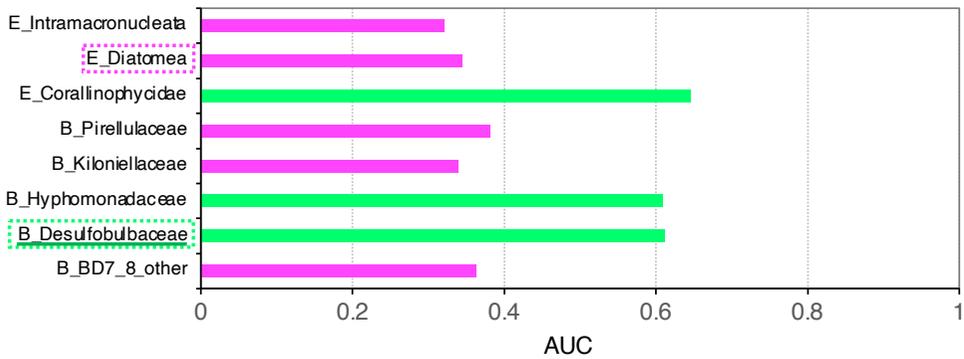
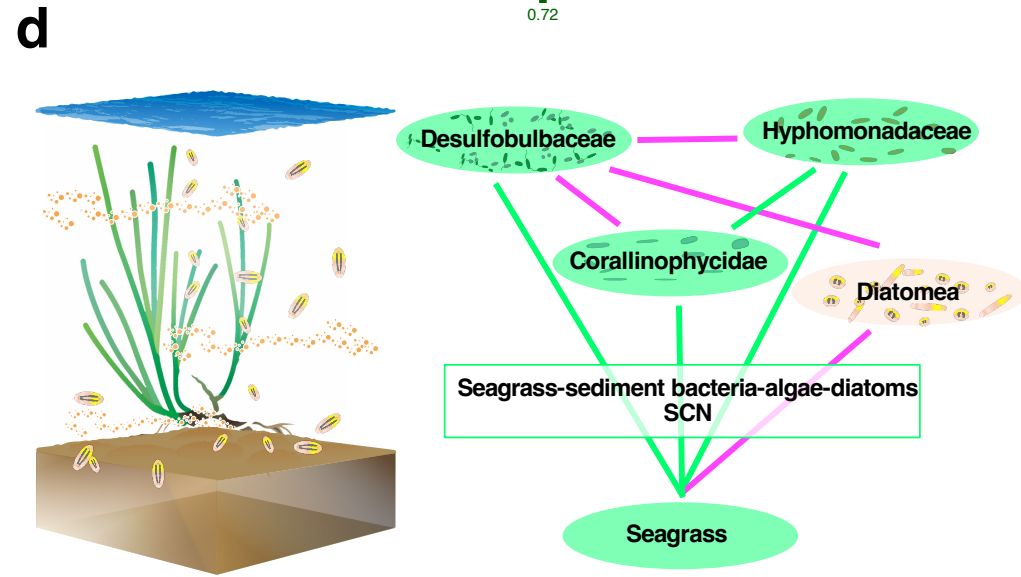

Fig.5

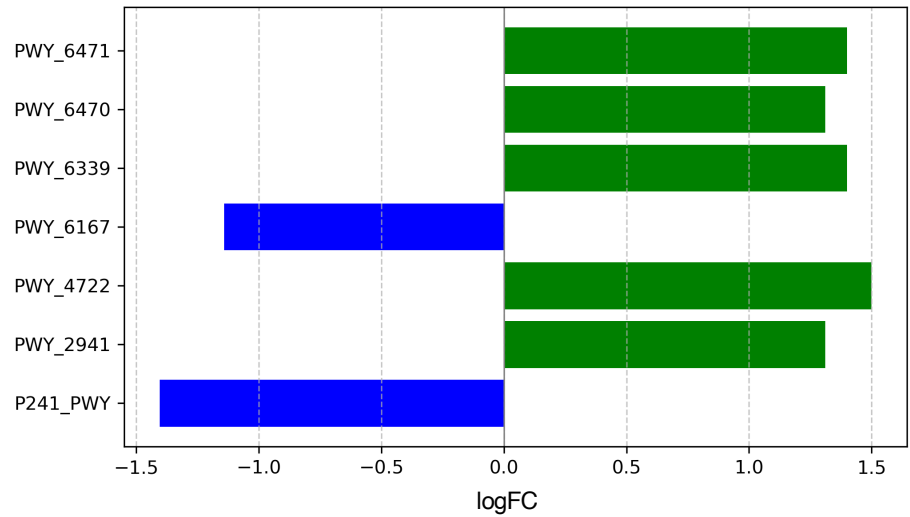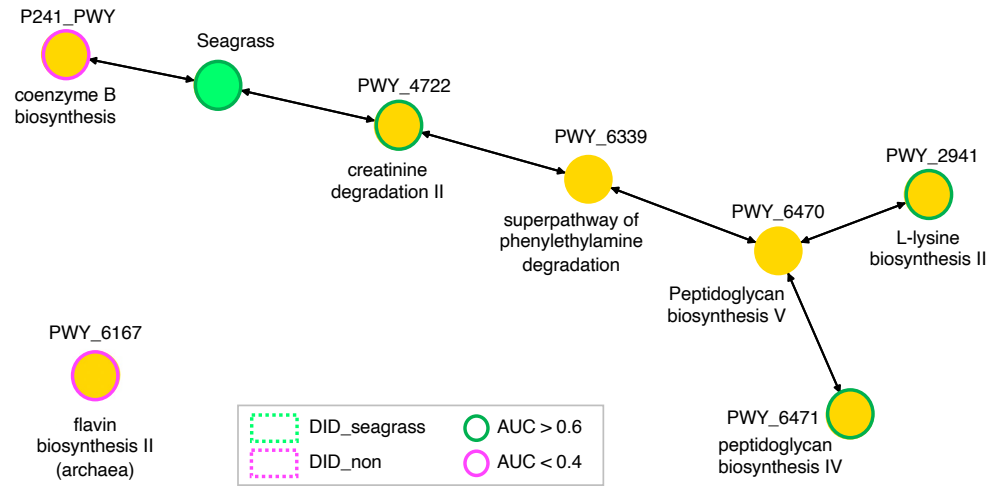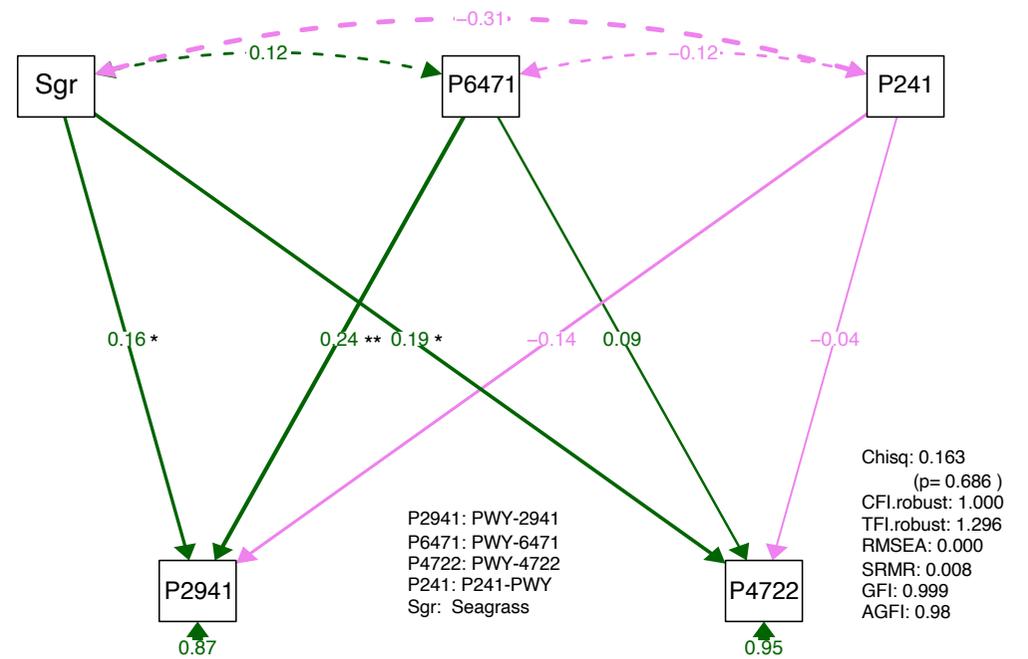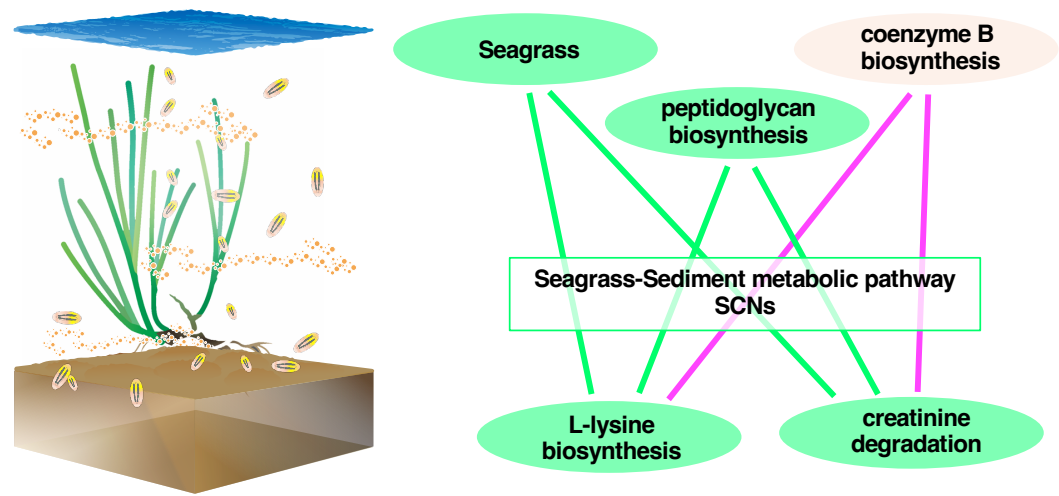

Fig.6

# Supporting Information

# Symbiotic causal network of seagrass-bacteria-algae-diatoms interactions

**\*Corresponding authors**

Hirokuni Miyamoto, Ph.D.,

RIKEN IMS, Umibezukurikenkyukai Non-Profit Organization, Chiba University, Yokohama City University, Sermas Co., Ltd., and Japan Eco-science Co., Ltd.

E-mail: hirokuni.miyamoto@riken.jp, h-miyamoto@faculty.chiba-u.jp

Jun Kikuchi, Ph.D.,

RIKEN CSRS, Umibezukurikenkyukai Non-Profit Organization, Yokohama City University

E-mail: jun.kikuchi@riken.jp

Hiroshi Ohno, M.D. & Ph.D.

RIKEN IMS

E-mail: hiroshi.ohno@riken.jp

# Contents

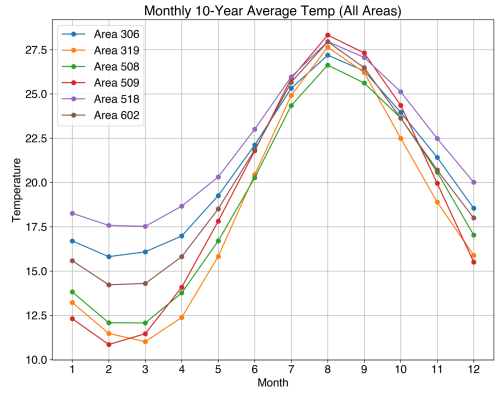
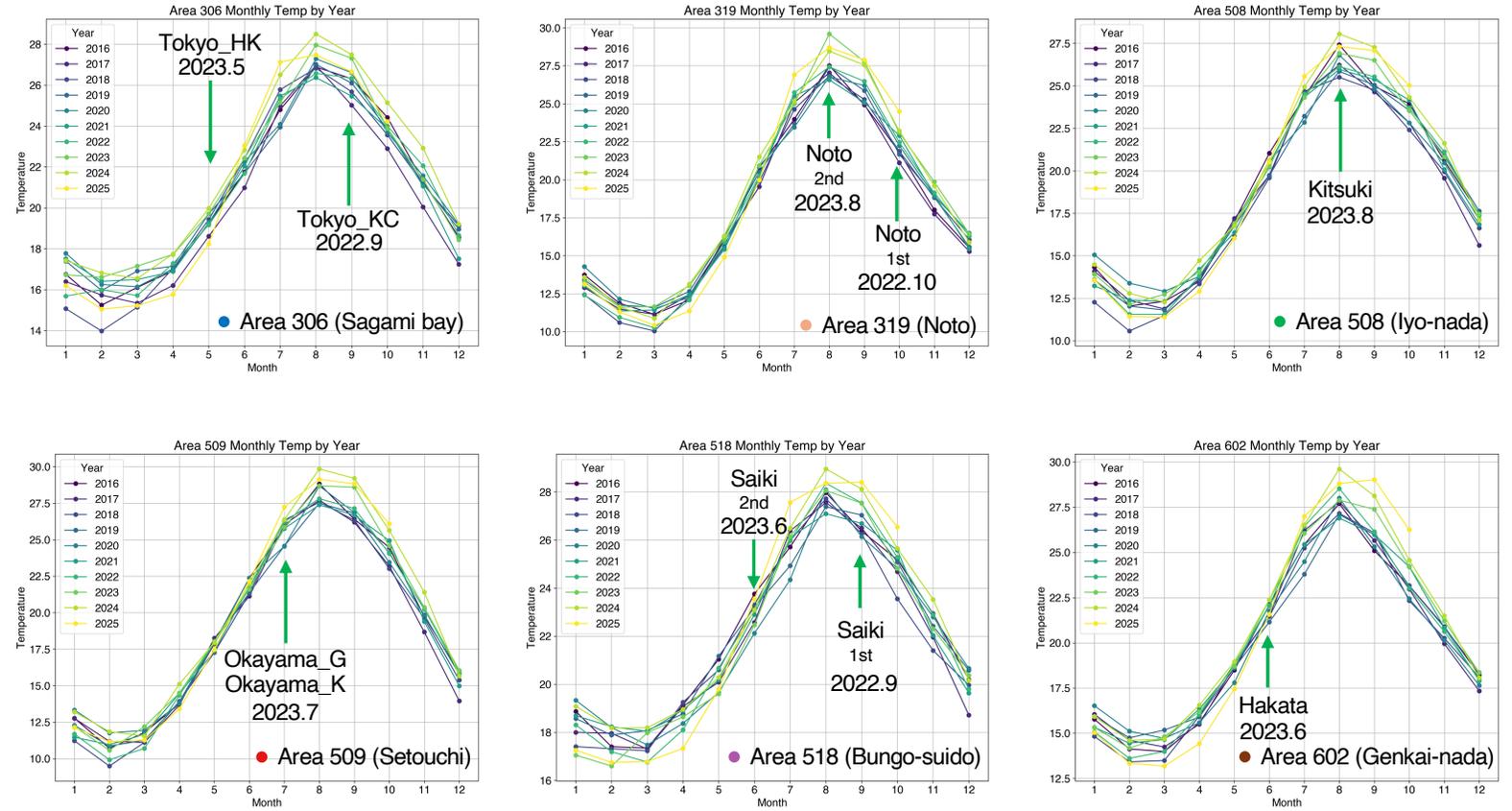

**Fig.S1**
**Temperature of the adjacent waters surveyed in this study.**
(a) Monthly average of water temperature in the sampling area over the past decade. (b) Water temperature changes by year and month in each area. The area codes are identical to those listed in Fig.1a. The green arrows indicate the months sampled in the closely related surveyed area.

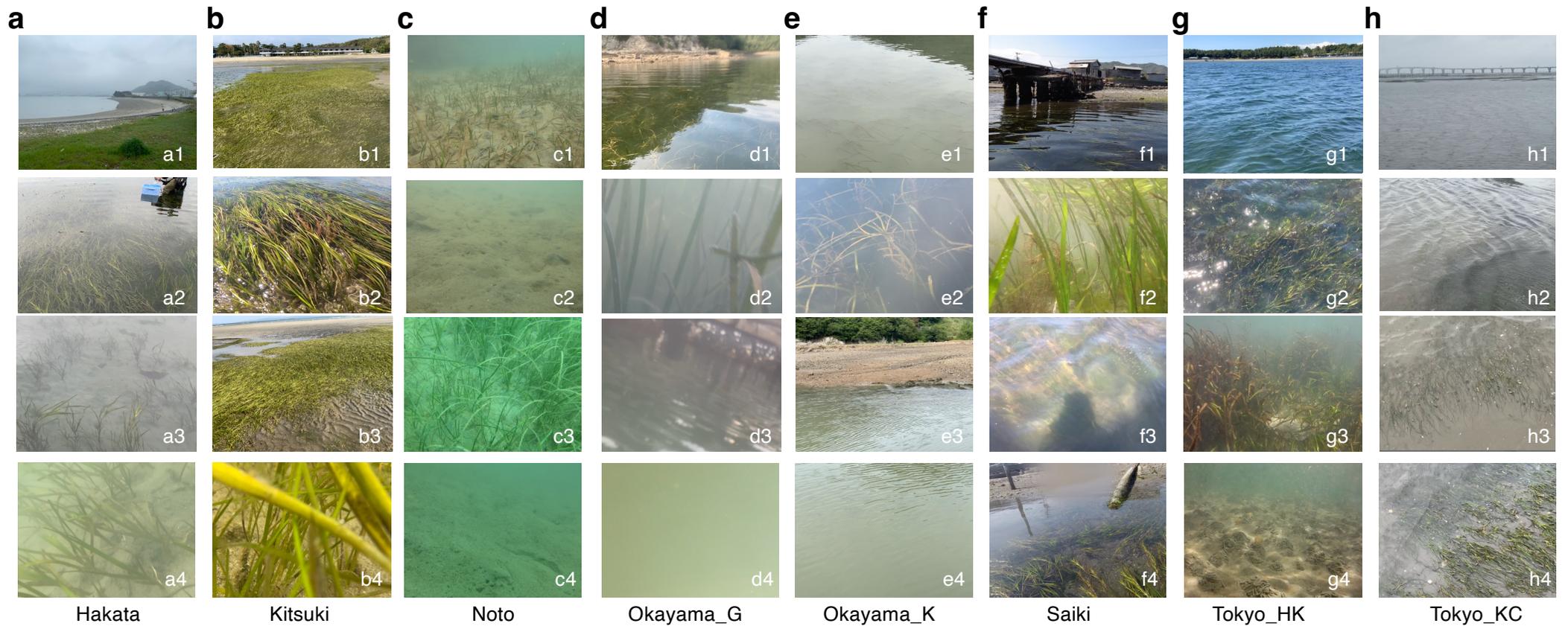

**Fig.S2**
**The marine sediments surveyed in this study.**
Eight of seagrass areas were selected: a) Hakata (Fukuoka Prefecture, Japan Sea side), b) Kitsuki (Oita Prefecture, Seto Inland Sea), c) Noto (Ishikawa Prefecture, Japan Sea side), d) Okayama_G (Okayama Prefecture, Seto Inland Sea), e) Okayama_K (Okayama Prefecture, Seto Inland Sea), f) Saiki (Oita Prefecture, Seto Inland Sea), g) Tokyo _HK (Kanagawa Prefecture, Pacific Ocean), and h) Tokyo _KC (Chiba Prefecture, Pacific Ocean side) The seagrass flourishing area was selected. In a) Hakata Bay, a1) appearance in Hakata Bay, a2) seagrass on the surface of the water, a3) sparsely growing seagrass, and a4) seagrass photographed in seawater. In b) Sumiyoshihama in Kitsuki (Morie Bay in Beppu Bay), b1) overlooking land in Kitsuki, b2) seagrasses with eggs attached, b3) overlooking seagrasses from land, and b4) seagrasses photographed from inside seawater. In c) Tsukumo Bay in Noto, c1) seagrasses photographed from inside seawater (first sampling site), c2) seagrasses-free sediments in the vicinity, c3) seagrasses photographed from inside seawater (second sampling site), and c4) seagrasses-free sediments in the vicinity. In d) Okayama_G (Genji Bay in Hinase), d1) sea surface where seagrasses grow, d2) seagrasses from inside seawater, d3) sea surface where seagrasses do not grow, and d4) underseawater without seagrasses. In e) Okayama_K (Kodama Bay in Hinase), e1) sea surface where seagrasses grow, e2) seagrasses from inside seawater, e3) sea surface where seagrasses do not grow, and e4) underseawater without seagrasses. In f) Saiki (Nishinoura Bay), f1) fish farm and sea surface where seagrasses grow, f2) seagrasses with small fish from inside seawater, f3) the sea surface with small fish where seagrasses grow, and f4) fish farm and sea surface where seagrasses grow. In g) Tokyo_HK (Umi-no-kouen in Hakkeijima of Kanagawa), g1) overlooking land, g2) sea surface where seagrasses grow, g3) seagrasses from under seawater, and g4) underseawater without seagrasses. In h) Tokyo_KC (Kaneda gyokou in Kisarazu of Chiba), h1) overlooking seagrasses from land, h2) sea surface where seagrasses grow, h3) seagrasses growing, and h4) seagrasses exposed from the sea surface.

a

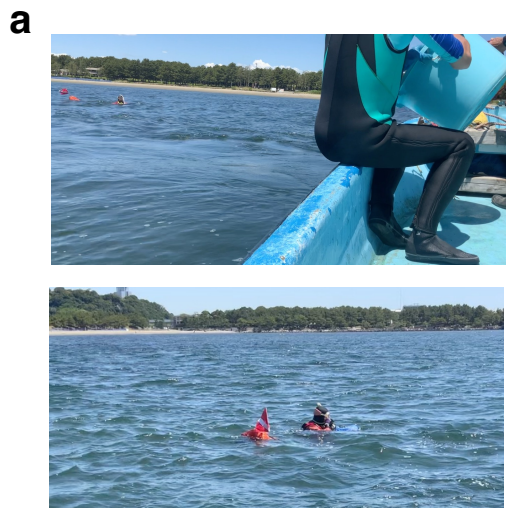

b

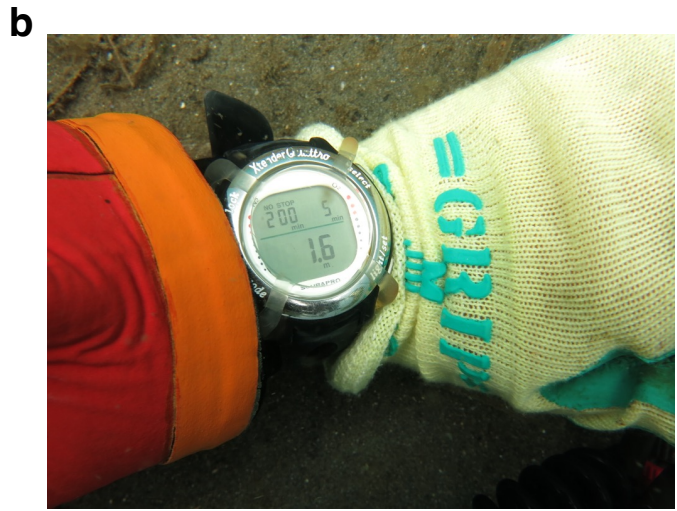

**Fig.S3**
**Custom sampling methods for sampling in this study.**
Custom method that differs from ordinary shovel collection is shown in photos. a) Sampling by divers. Such sampling was conducted in Tokyo_HK and Noto in this study. This photograph shows a sample taken in the Umi-no-kouen in Hakkeijima of Kanagawa. b) Check the water depth. Such sampling was conducted in Noto in this study. c) Sampling by an Ekman-Birge bottom sampler. The photo names c1-c5 show the method of the sampling. The red arrow represents the weight. After the weight moves down the string, the metal fitting (blue arrow) that hooks the automatic shovel is turned off, and the shovel closes, as shown by the green arrow. Marine sediments are collected in this area.

c

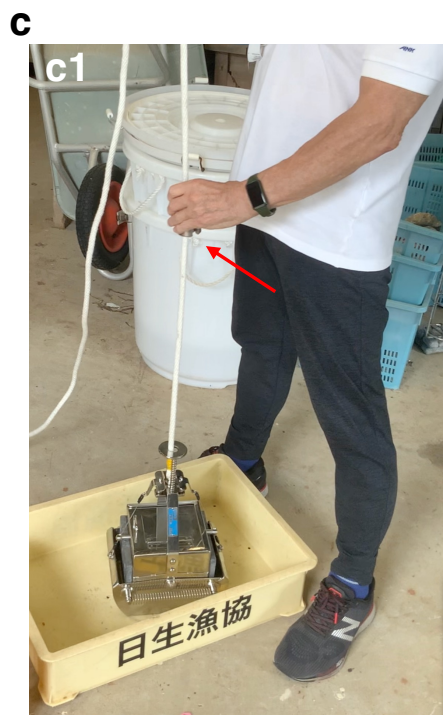 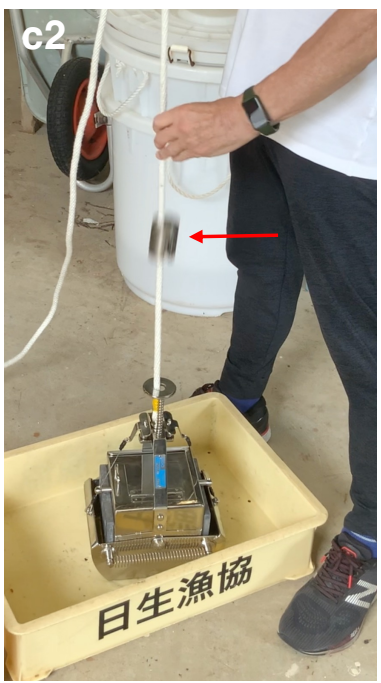 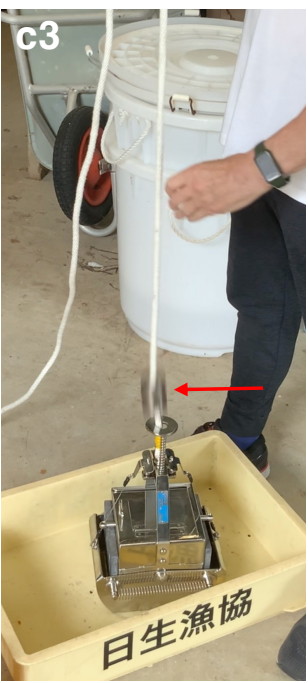 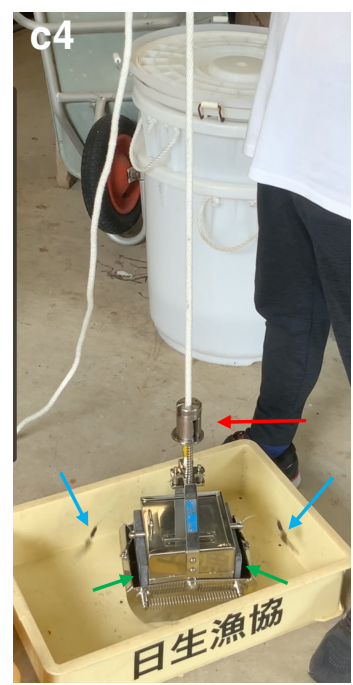 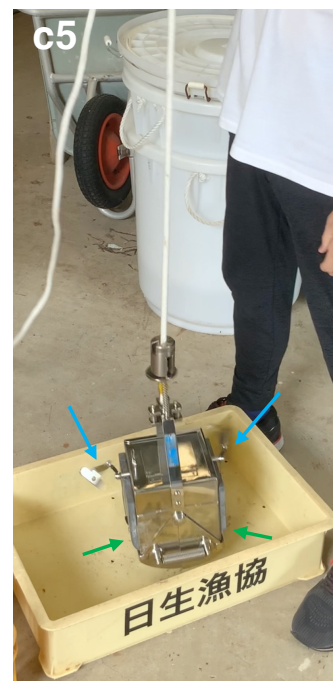

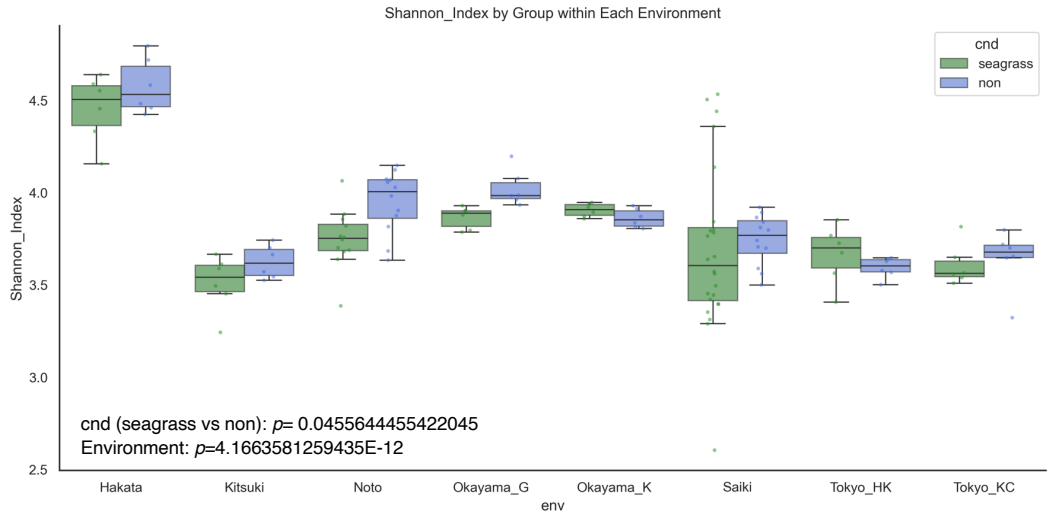

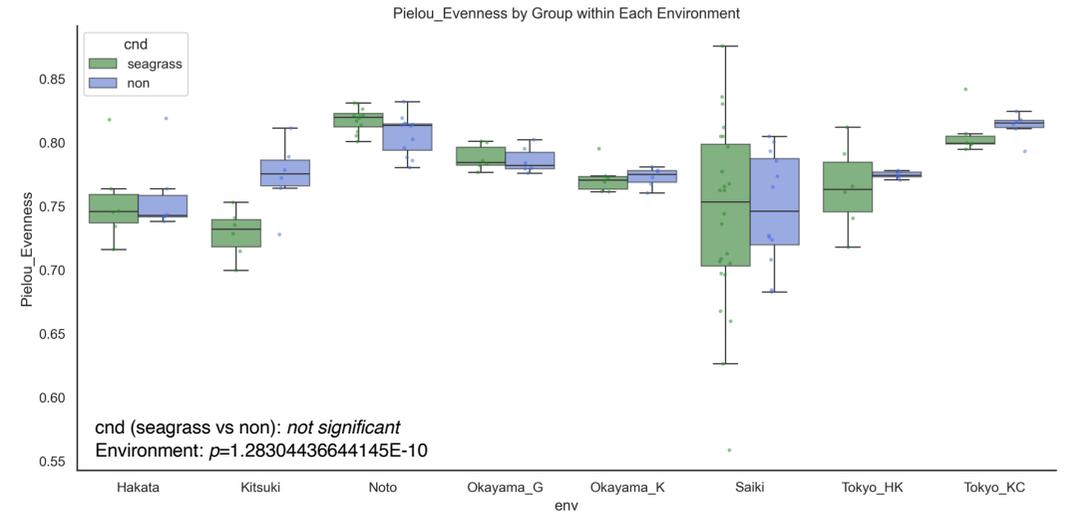

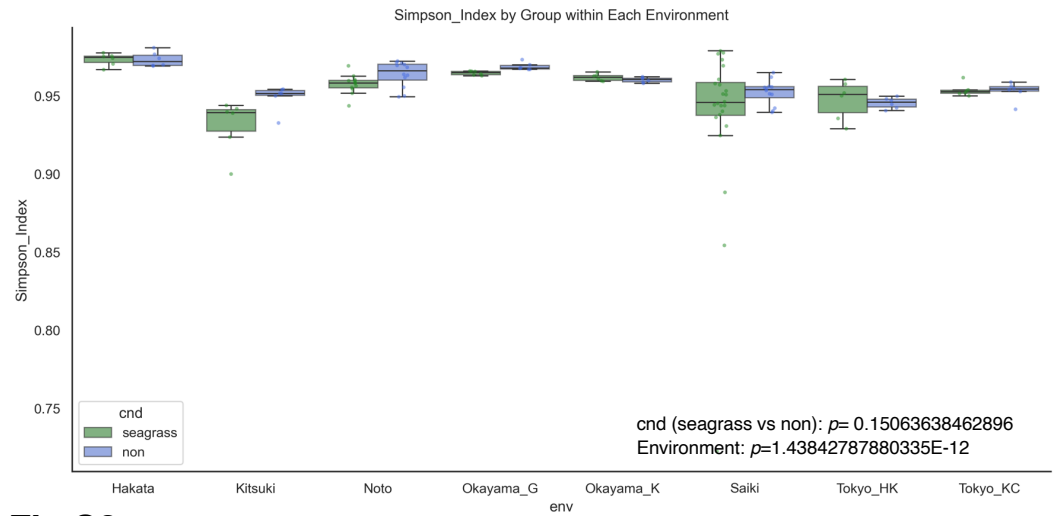

**Fig.S4**

**Symbiotic bacterial diversity in the marine sediment observed in this study.**
(a) The Shannon indices, (b) Pielou evenness values, and (c) Shannon indices, as α-diversity indicators, were calculated on the basis of the bacterial dataset for the sediment calculated via QIIME2. The green and blue indicate the seagrass and non-seagrass sediment, respectively. The sampling sites are shown on the X axis.

Fig.S2

a
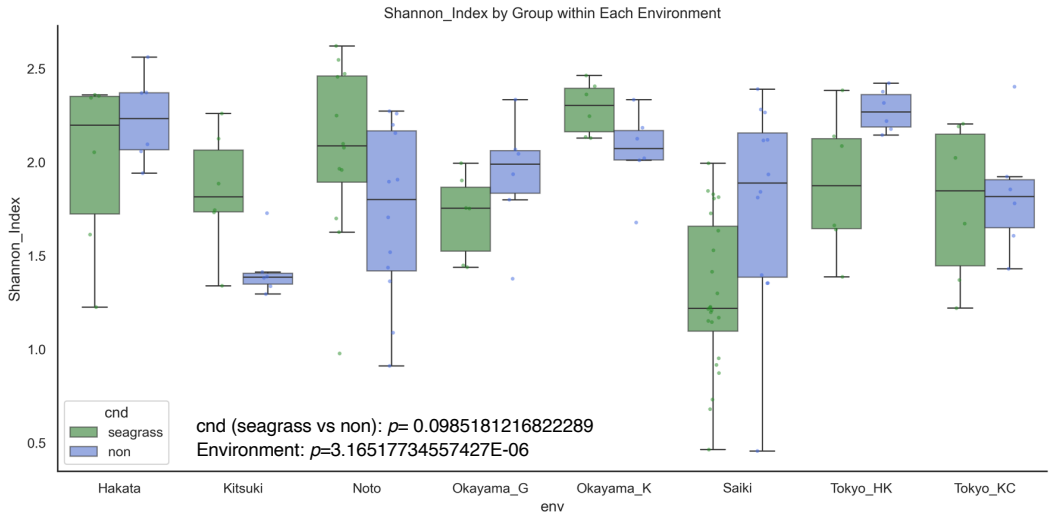

b
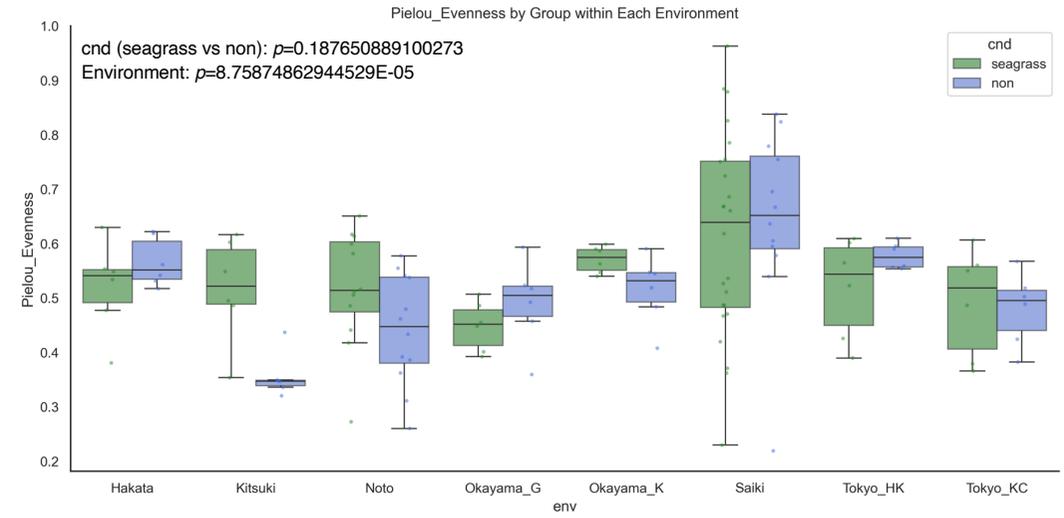

c
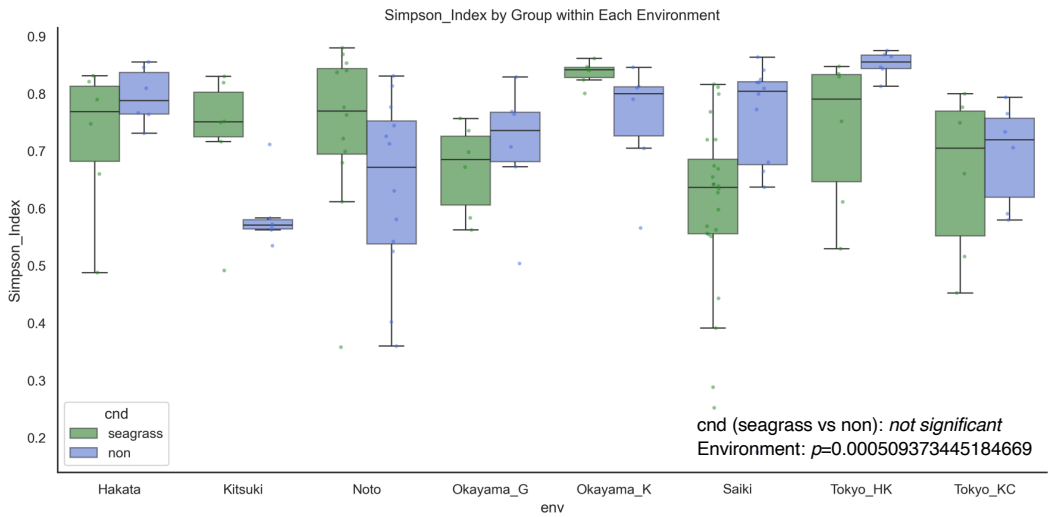

**Fig.S5**
**Symbiotic eukaryotic diversity in the marine sediment observed in this study.**
(a) The Shannon indices, (b) Pielou evenness values, and (c) Shannon indices, as α-diversity indicators, were calculated on the basis of the eukaryotic dataset for the sediment calculated via QIIME2. The green and blue indicate the seagrass and non-seagrass sediment, respectively. The sampling sites are shown on the X axis. The statistical values calculated using the R library "pairwiseAdonis" are shown in the upper right portion.

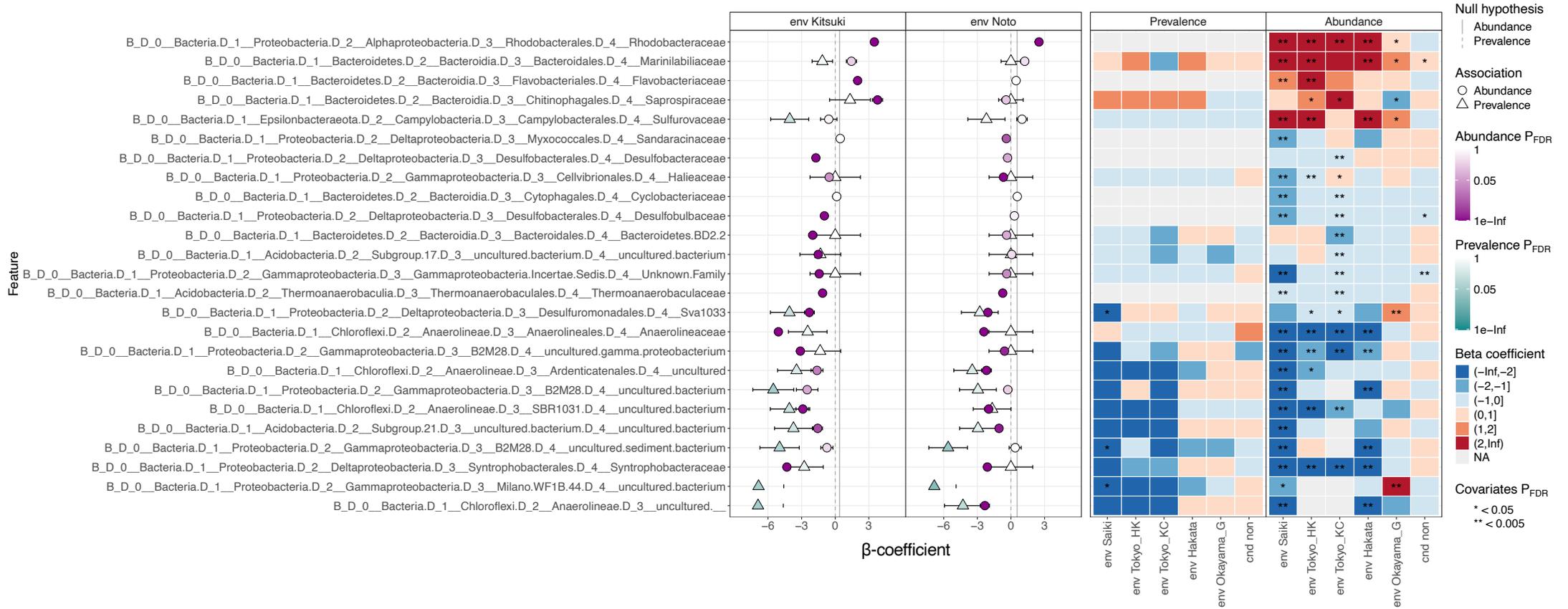

**Fig.S6**
**Difference-in-difference for symbiotic bacterial families (D_4 level) in the marine sediment observed in this study.**
Statistical differences dependent on environmental conditions were evaluated by the R library "MaAsLin3". The categories D_0_ ~ D_4_ are shown based on QIIME2 for the 16S sequencing analysis. Abbreviations are as follows: B_, bacterial; FDR, false discovery rate.

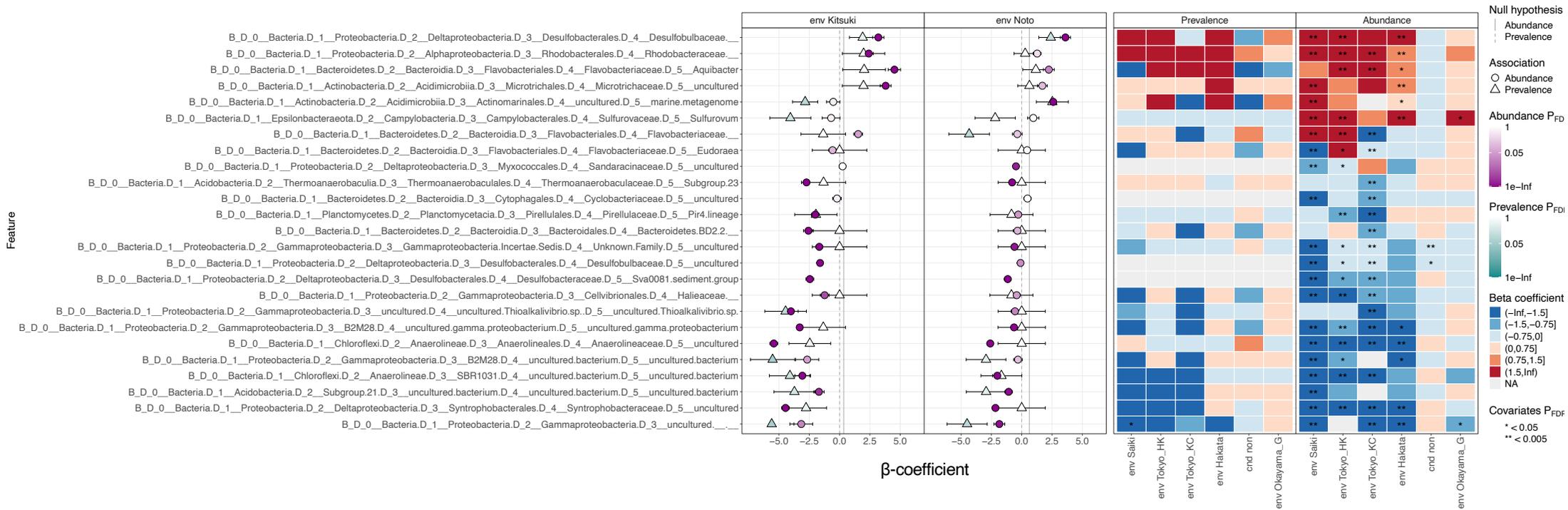

**Fig.S7**
**Difference-in difference for symbiotic bacterial genera (D_5 level) in the marine sediment observed in this study.**
Statistical differences dependent on environmental conditions were evaluated by the R library "MaAsLin3". The categories D_0_ ~ D_5_ are shown based on QIIME2 for the 16S sequencing analysis. Abbreviations are as follows: B_, bacterial; FDR, false discovery rate.

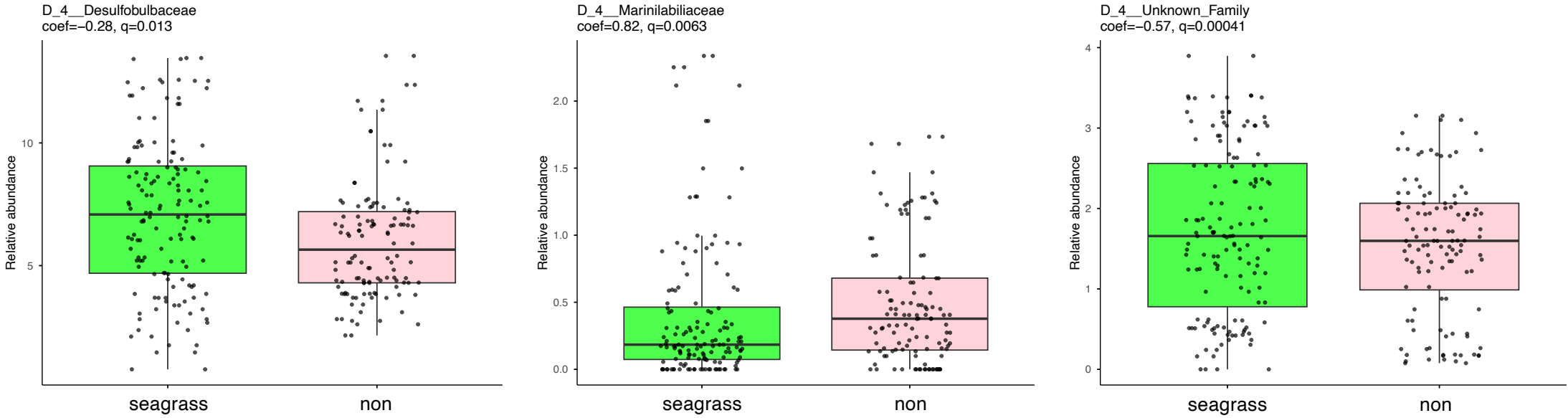

**Fig.S8**
**Significant symbiotic bacterial familes in the marine sediment observed in this study.**
Statistical differences dependent on environmental conditions have been evaluated by the R library "MaAsLin3". The abbreviations are as follows: coef, coefficient; q, FDR-adjusted p-value.

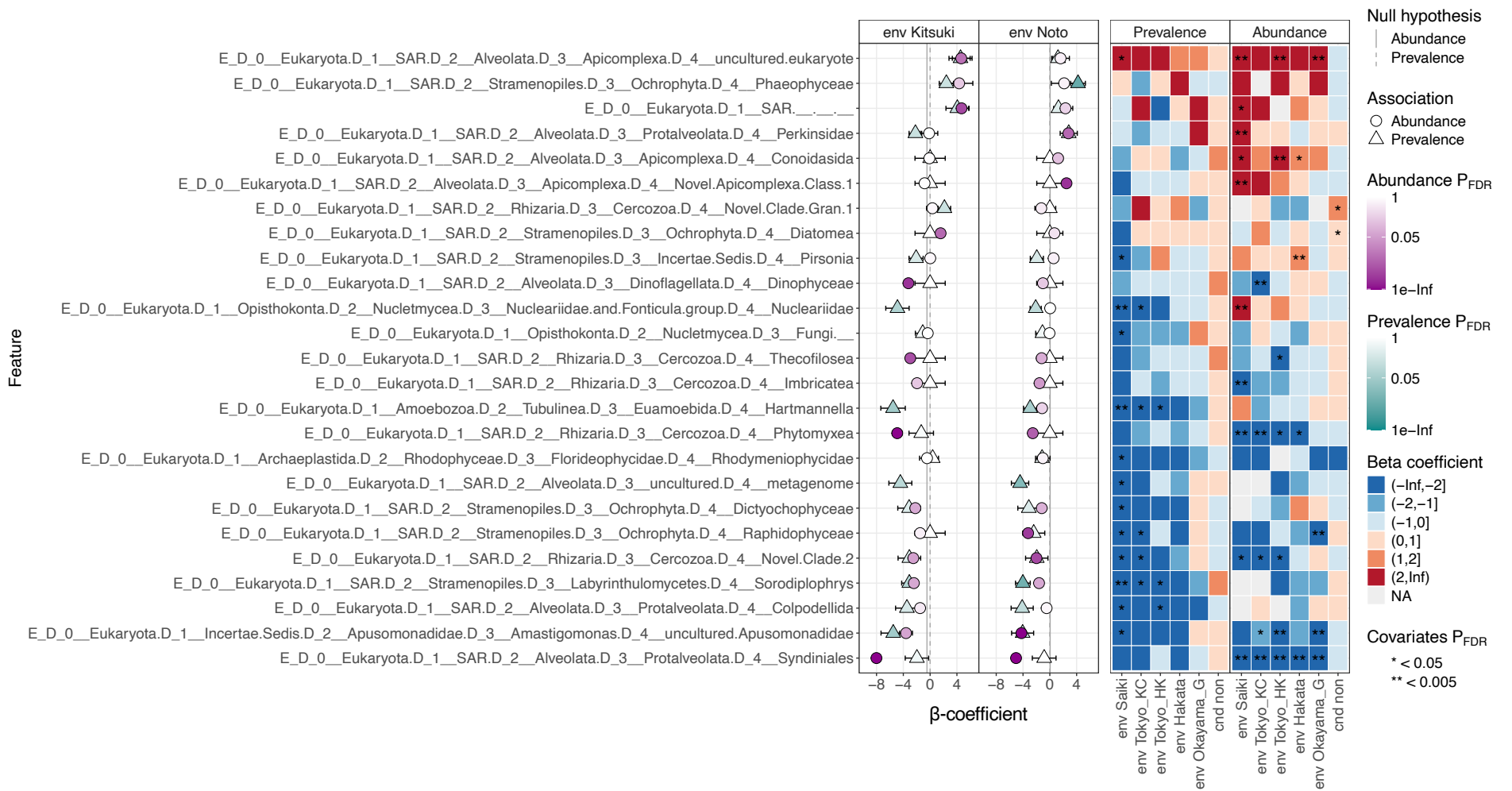

**Fig.S9**
**Difference-in difference for symbiotic eukaryotes (D_4 level) in the marine sediment observed in this study.**
Statistical differences dependent on environmental conditions were evaluated by the R library "MaAsLin3". The categories D_0_ ~ D_4_ are shown based on QIIME2 for the 18S sequencing analysis. Abbreviations are as follows: E_, eukaryotic; FDR, false discovery rate.

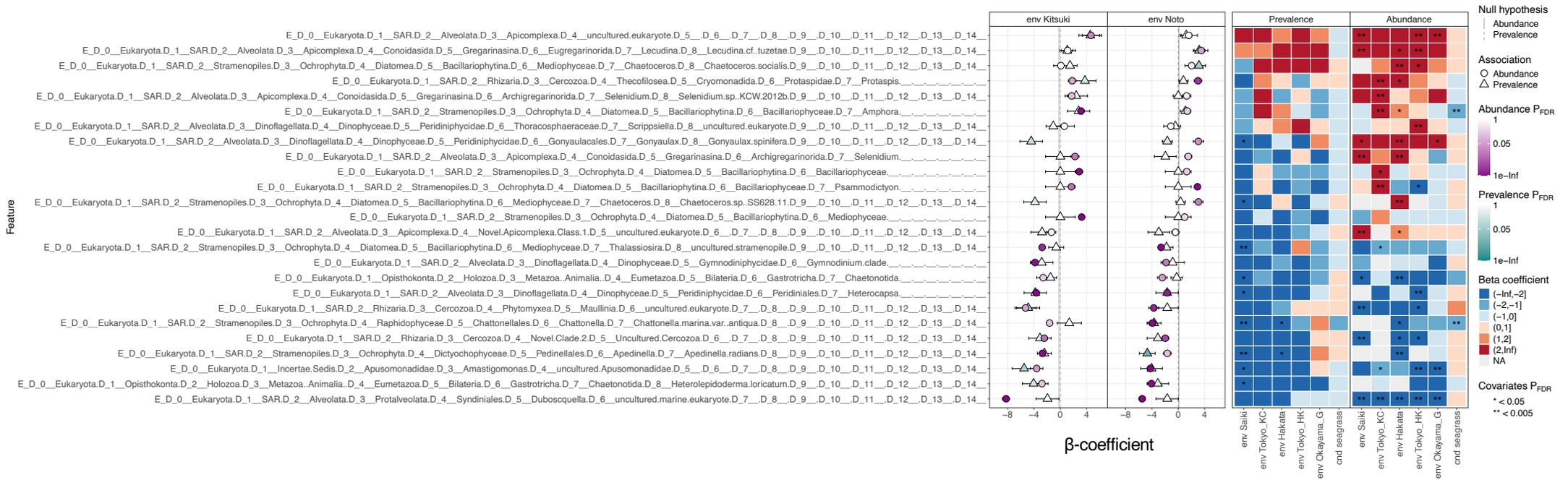

**Fig.S10**
**Difference-in difference for symbiotic eukaryotes (D_14 level) in the marine sediment observed in this study.**
Statistical differences dependent on environmental conditions were evaluated by the R library "MaAsLin3". The categories D_0_ ~ D_14_ are shown based on QIIME2 for the 18S sequencing analysis. Abbreviations are as follows: E_, eukaryotic; FDR, false discovery rate.

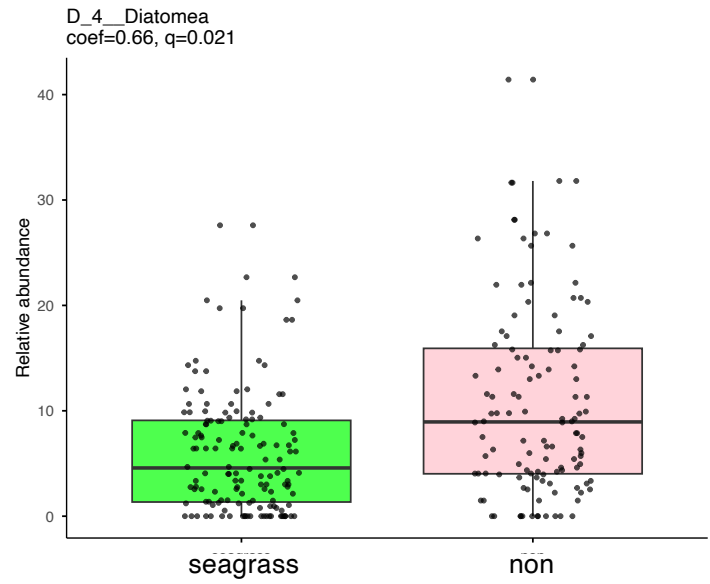 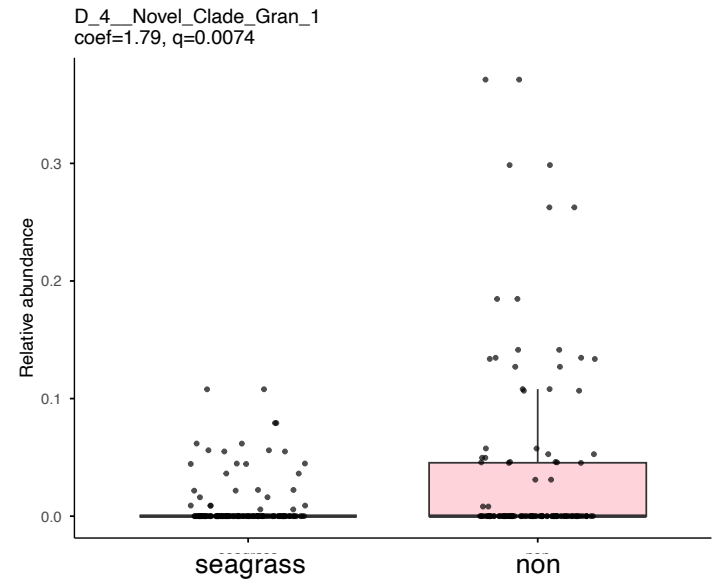

**Fig.S11**
**Significant symbiotic eukaryotes (D_4 level) in the marine sediment observed in this study.**
Statistical differences dependent on environmental conditions were evaluated by the R library "MaAsLin3". The abbreviations are as follows: coef, coefficient; q, FDR-adjusted p-value.

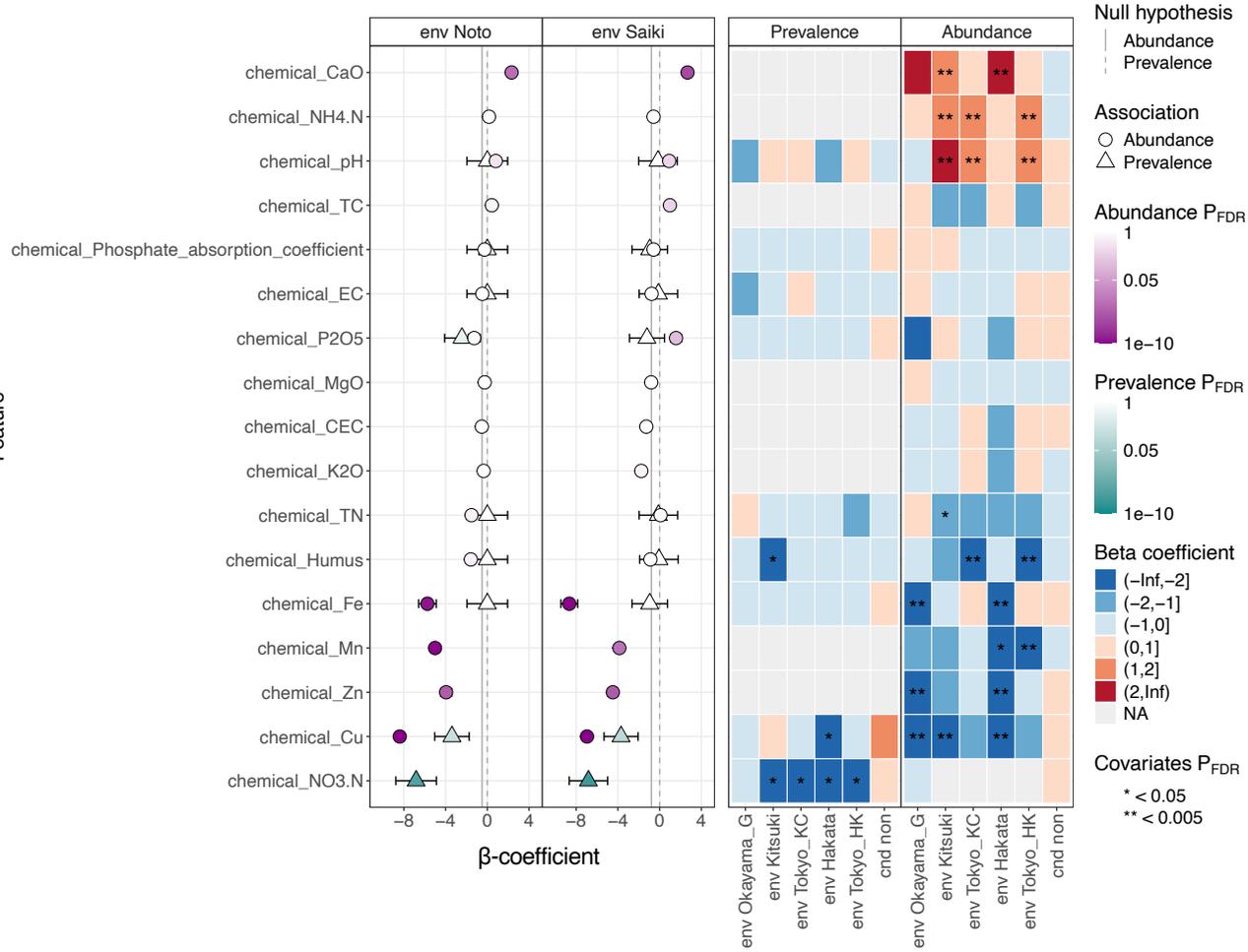

**Fig.S12**
**Difference-in difference for the chemical indices in the marine sediment observed in this study.**
Statistical differences dependent on environmental conditions have been evaluated by the R library "MaAsLin3".

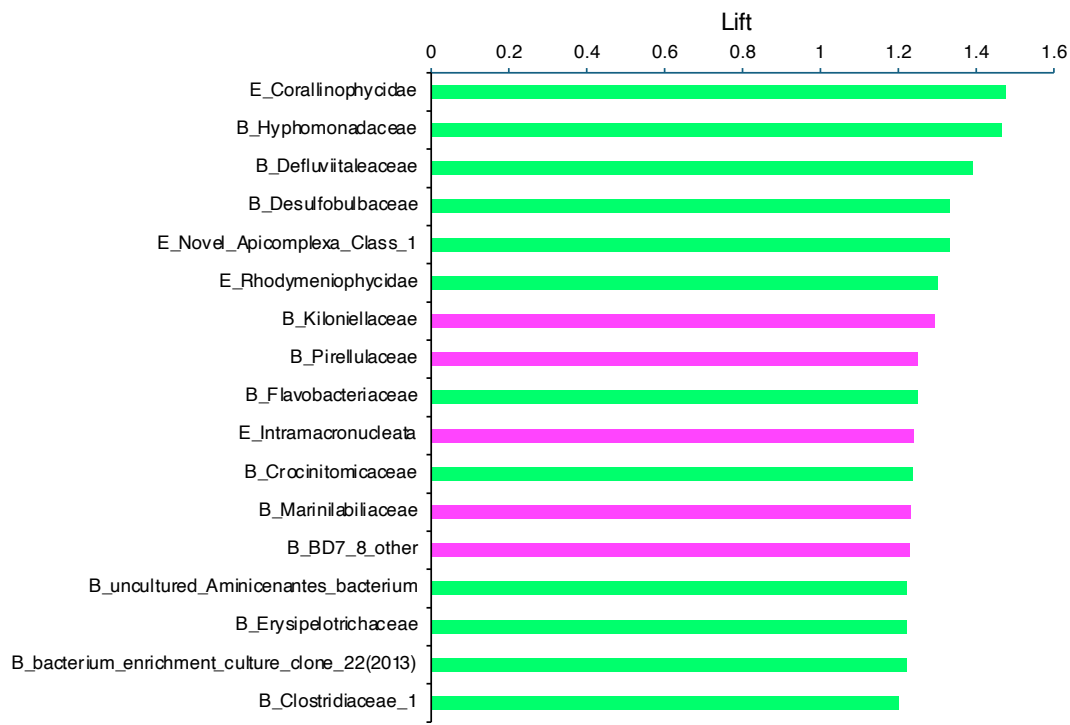 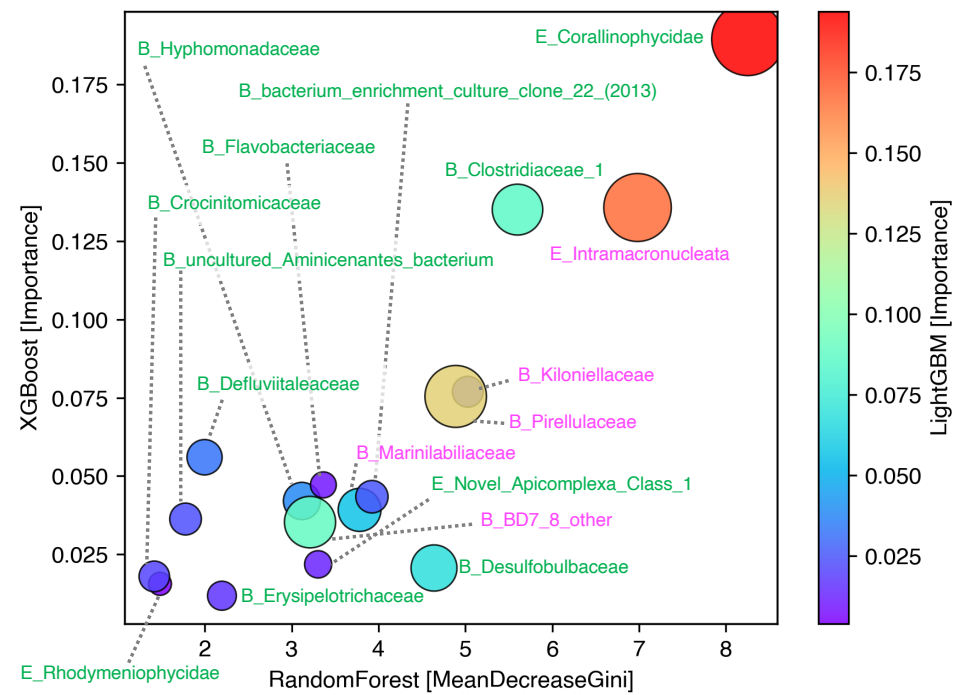

**Fig.S13**
**Values of feature components for the chemical indices in the marine sediment selected via AA, random forest, XGBoost, and LightGBM algorithms.**
(a) Values of feature components selected by association analysis (AA). The seagrass and non-seagrass states were set to 1 and 0, respectively. The colour of the bars was discriminated by the AA (≥0.15 as the support value, ≥0.3 as the confidence values, and ≥1.2 as the lift value under the condition of "seagrass=1"). The green and violet bars indicate positive ("=1" in AA) and negative ("=0" in AA) associations, respectively. (b) Bubble chart of the feature component candidates selected via three types of ML algorithms. Definitions: Random Forest [MeanDecreaseGini], a measure of Random Forest; XGBoost [Importance], a measure of XGBoost; LightGBM [Importance], a measure of LightGBM.

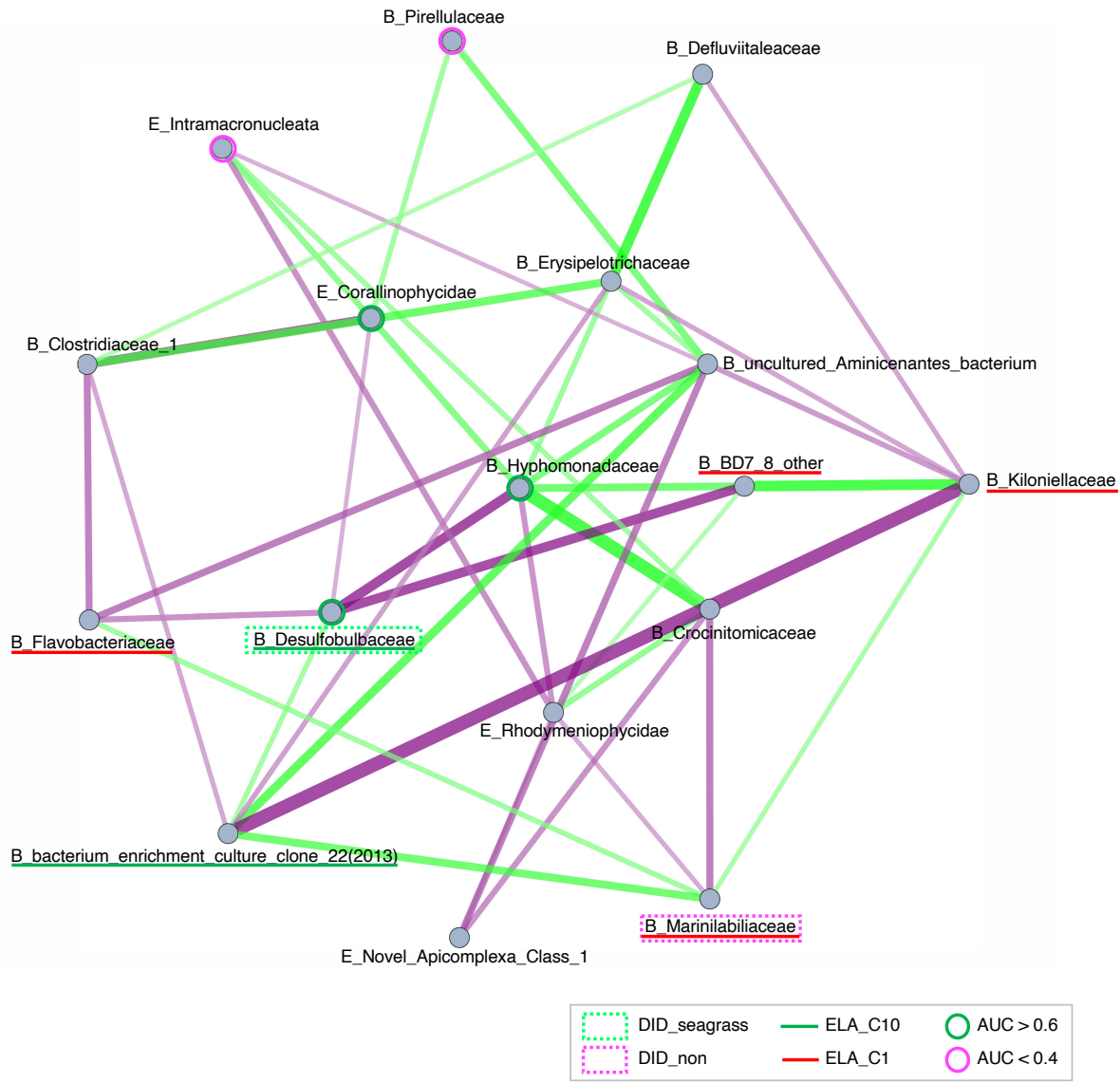

**Fig.S14**

**The ELA interaction network in the extended pairwise maximum entropy model fitted to the observational data.**

Feature components were selected by the ML algorithms (AA, random forest, XGBoost, and LightGBM). The selected feature components were visualized as described in the figure. The feature components selected by the DID or ELA stable states (C1 and C10), and ROC curve analysis are shown in green or violet, respectively. Abbreviations were as follows: DID_seagrass, components significantly increased in the seagrass-rich area; DID_non, components significantly decreased in the seagrass-rich area; ELA_C10, components within seagrass C10 in Fig. 4c; ELA_C1, components within seagrass C1 in Fig. 4c; AUC>0.6, components more than 0.6 as AUC under the seagrass flourishing conditions(1); and AUC>0.4, components less than 0.4 as AUC under the seagrass flourishing conditions(1).

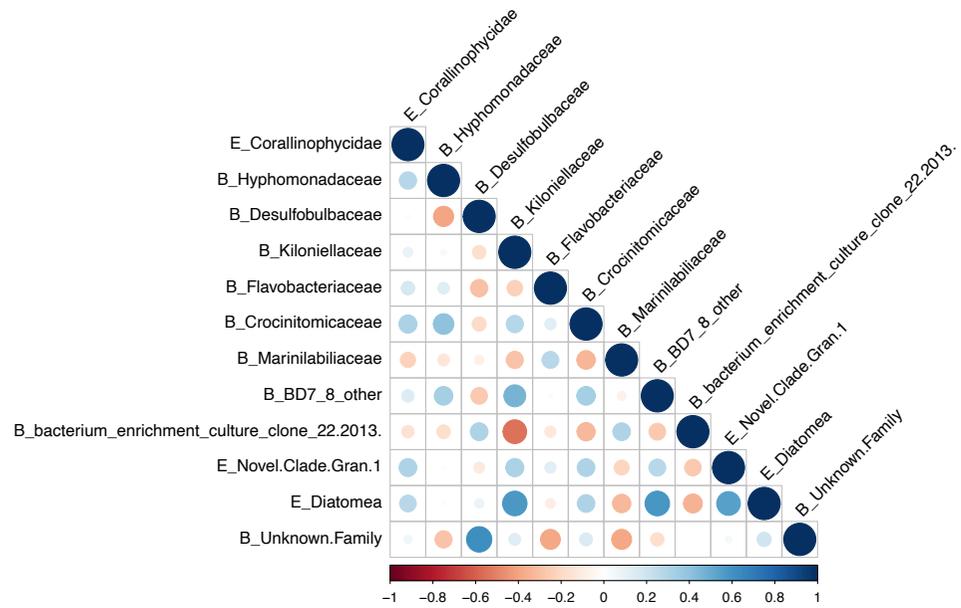

**Fig.S15**
**Correlation heatmap of the feature components in the marine sediment.**
The correlation values of the feature components selected for exploratory factor analysis (EFA) were estimated by the function 'cor' in the R library. The values for the Spearman's rank correlation coefficient were visualized by the R libray "corrplot".

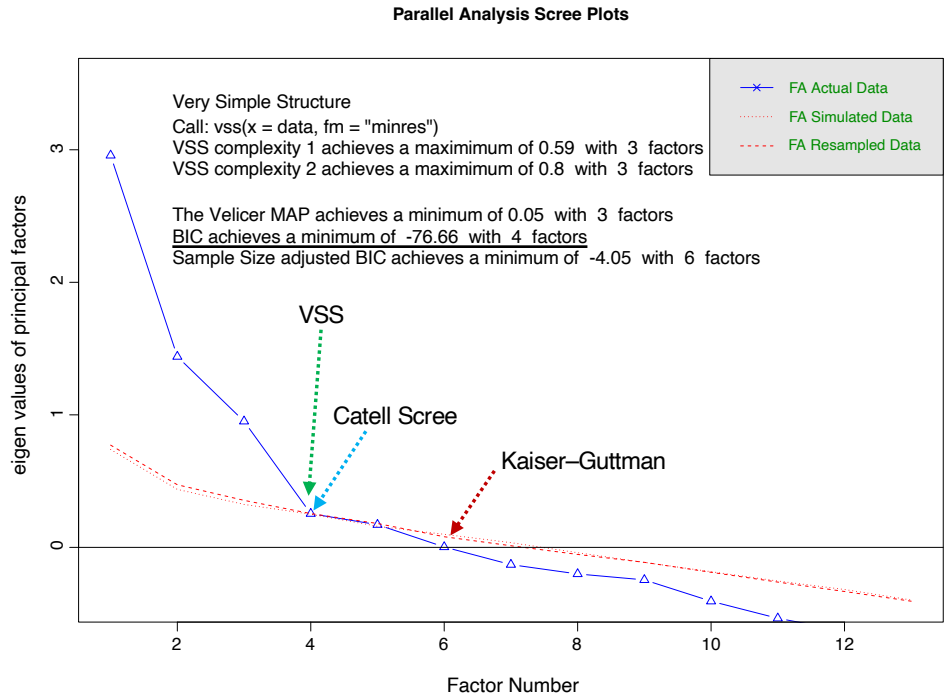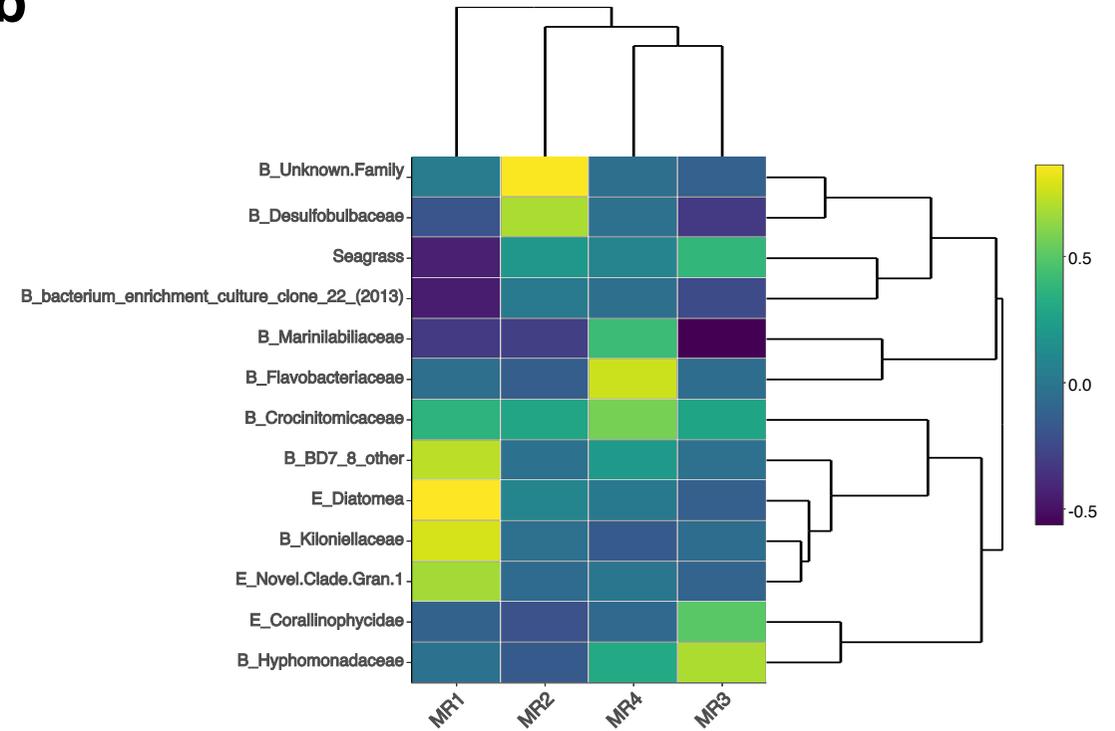

**Fig.S16**
**Evaluation using EFA for feature components in the marine sediment.**
(a) Parallel analysis scree plot for the EFA. The arrow point marked with the Catell Scree (blue broken arrow) indicates the number of factors on the basis of the Catell Scree criteria. The arrow points marked with Kaiser–Guttman (brown broken arrow) indicate the number of factors based on the Kaiser–Guttman criteria. The arrow point marked with 'VSS' (green broken arrow) indicates the number of factors on the basis of the VSS-calculated criteria (underlined criteria). (b) EFA heatmap of the feature components in the marine sediment. The heatmap was constructed with a factor number of 4. The bar on the right indicates the extent of the contribution values. Abbreviations are as follows: VSS, very simple structure; BIC, Bayesian information criterion; MR, minres (minimum residual) method (default method); fa (r = data, nfactors = 4, rotate = "promax", fm = "minres", and use = "complete.obs"), calculation conditions for EFA; B_, bacterial; and E_, eukaryotic.

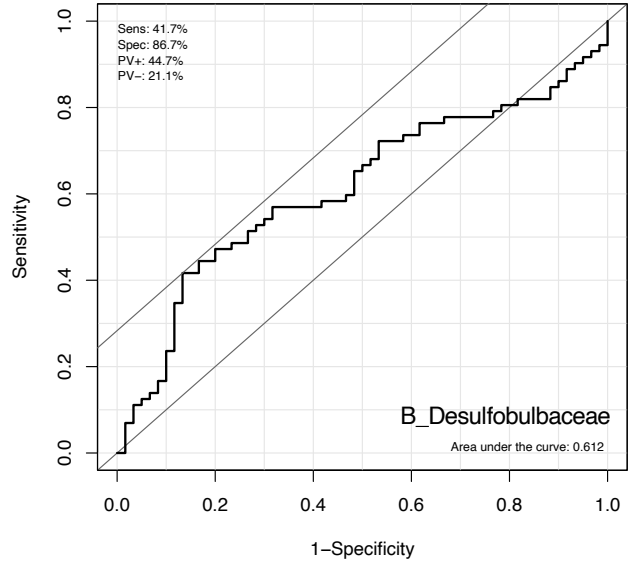
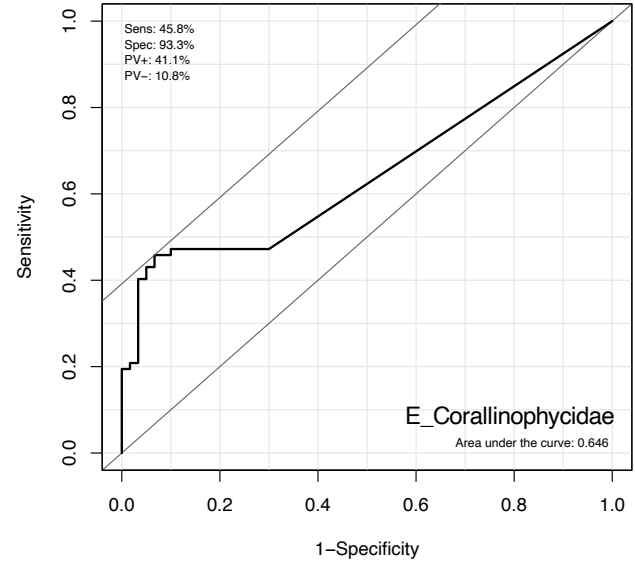
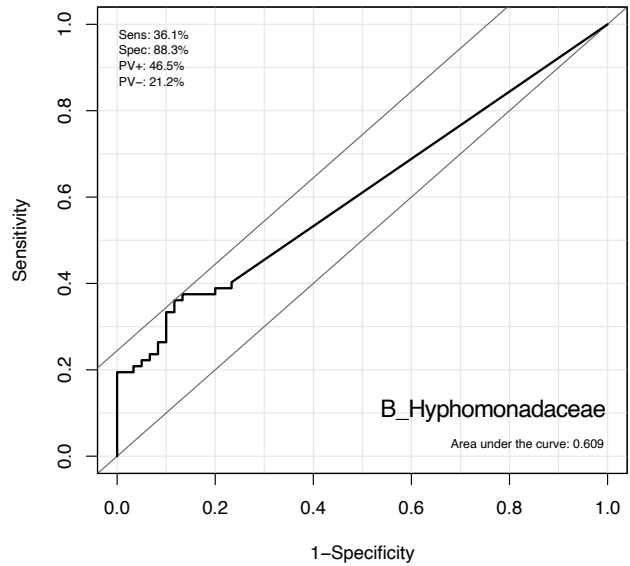
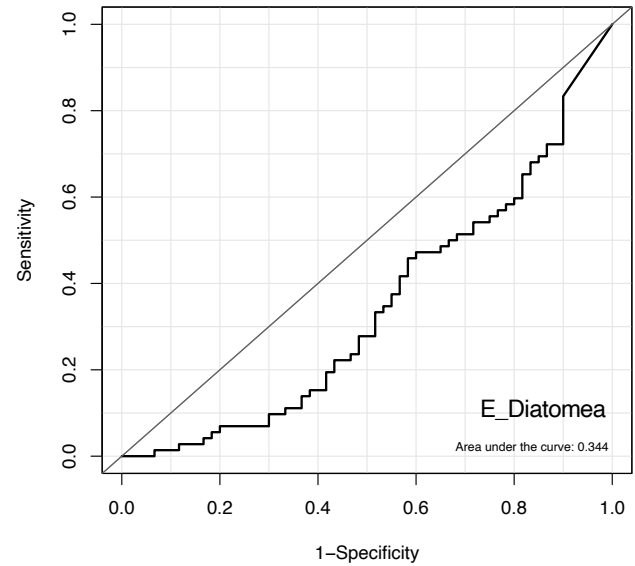

**Fig.S17**

**Thresholds for the selected feature components.**
Receiver operating characteristic (ROC) curve analysis for B_Desulfobulbaceae, B_Hyphomonadaceae, E_Corallinophycidae, and E_Diatomea shown in Fig. 5b. The conditions for seagrass and non-seagrass were assigned as values of 1 and 0, respectively. On the basis of the two-dimensional surface for detection and false-detection rates of abrupt community-compositional changes (abruptness > 0.5), the area under the curve (AUC) was calculated. Abbreviations are as follows: Sens, sensitivity; Spec, specificity; PV+, positive predictive value; PV-, negative predictive value.

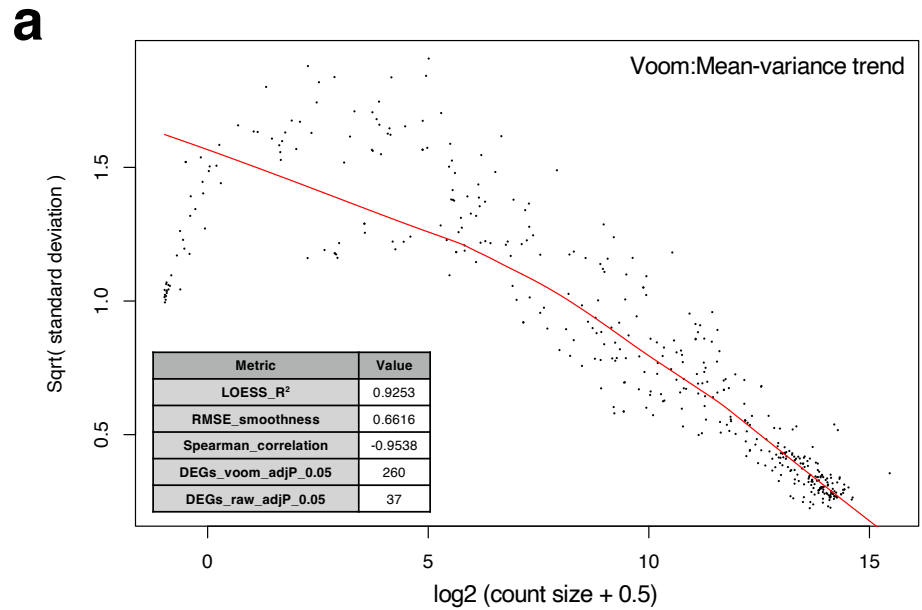
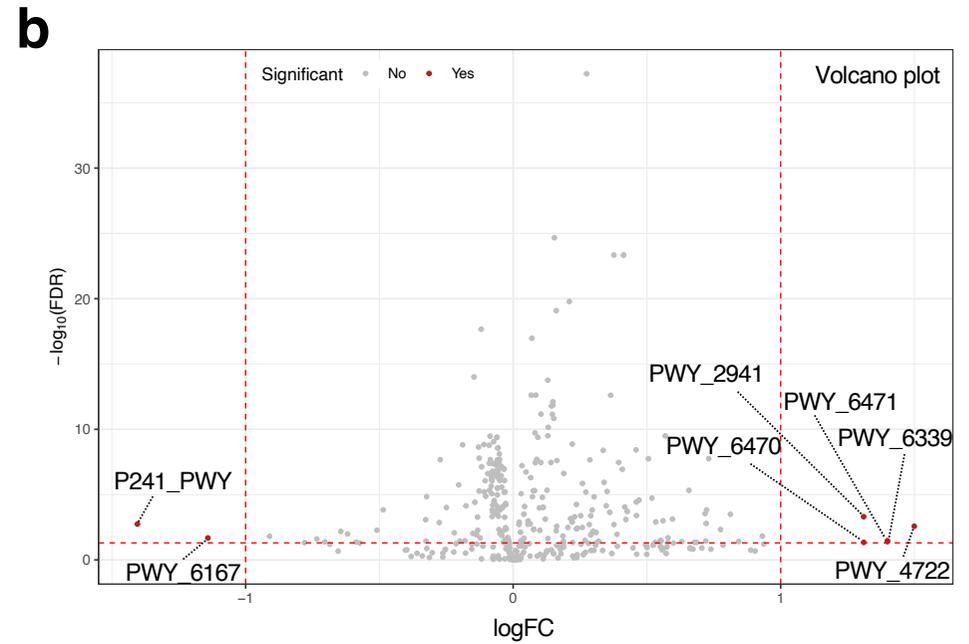
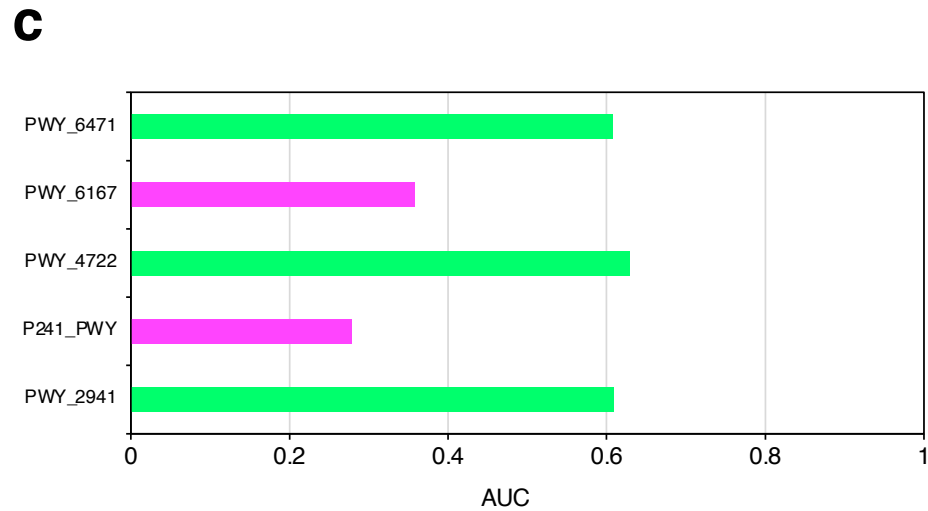

**Fig.S18**

**Differential functional abundance analysis using PICRUST2**

(a) Mean-variance trend estimated by voom transformation for the PICRUST2 dataset in the marine sediment. This plot displays the mean-variance relationship of the log2-transformed feature counts (x-axis: log2(count size + 0.5)) against the square root of the variance (y-axis: sqrt). A loess-smoothed trend line is shown to illustrate the overall variance trend across expression levels. The table within the plot summarizes key metrics for evaluating model fit and data quality. These metrics help assess the performance of voom normalization and the suitability for downstream differential analysis. (b) Volcano plot of the PICRUST2 dataset. The volcano plot shows the relationship between the magnitude of differential abundance (x-axis: log2 fold change, logFC) and statistical significance (y-axis: -log10 FDR-adjusted p-value). Each point represents a single predicted functional feature pathway. Features significantly different between conditions (FDR < 0.05) are highlighted in red, while non-significant features are shown in grey. A red dashed horizontal line indicates the significance threshold (FDR = 0.05 and |logFC|>1.0). This visualization allows the identification of features with both high effect sizes and high statistical confidence. (c) Feature components selected by ROC curve analysis. The seagrass condition was set to 1, and the selected components (AUC>0.6 and AUC<0.4) were visualized. Abbreviations are as follows: FDR, false discovery rate; LOESS $R^2$, coefficient of determination for the smoothed line; RMSE_smoothness, root mean square error of the smoothing; Spearman_correlation, correlation coefficient between mean expression and variance; DEGs_voom_adjP < 0.05, number of differentially expressed features (adjusted p-value < 0.05) based on voom; DEGs_raw_adjP < 0.05, number of DEGs identified without voom transformation.

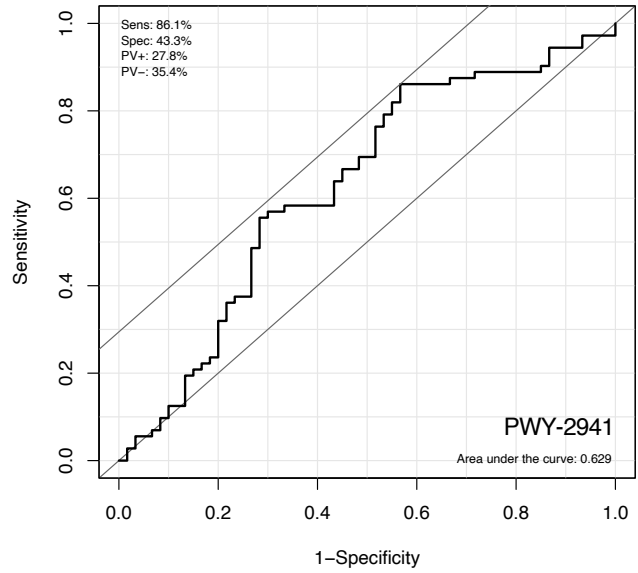
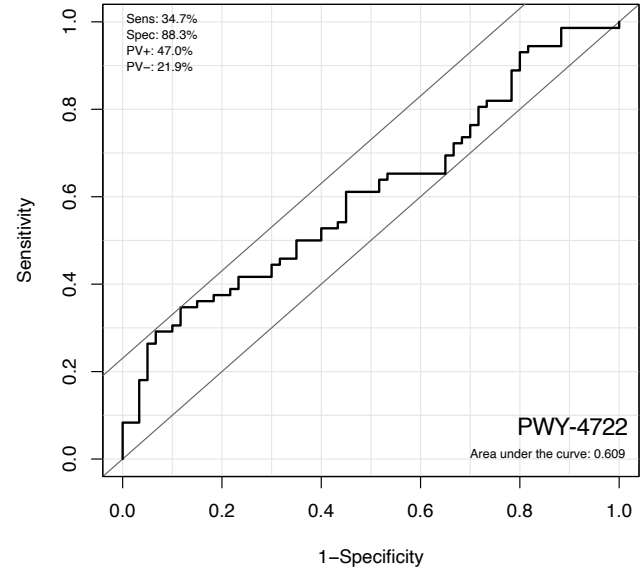
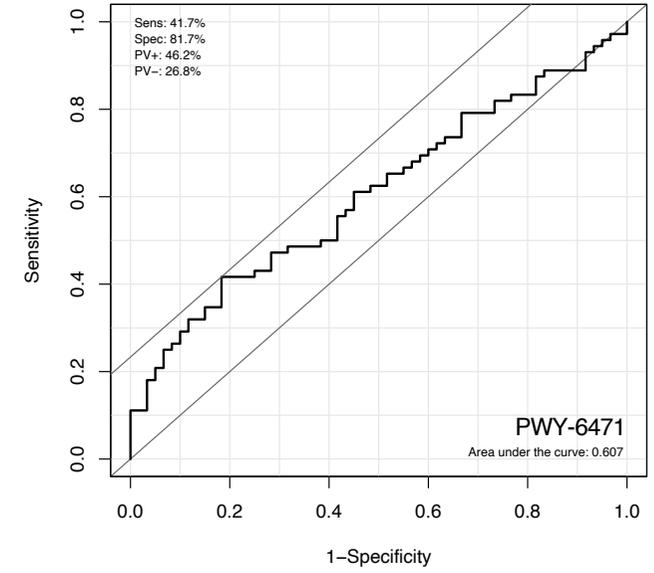
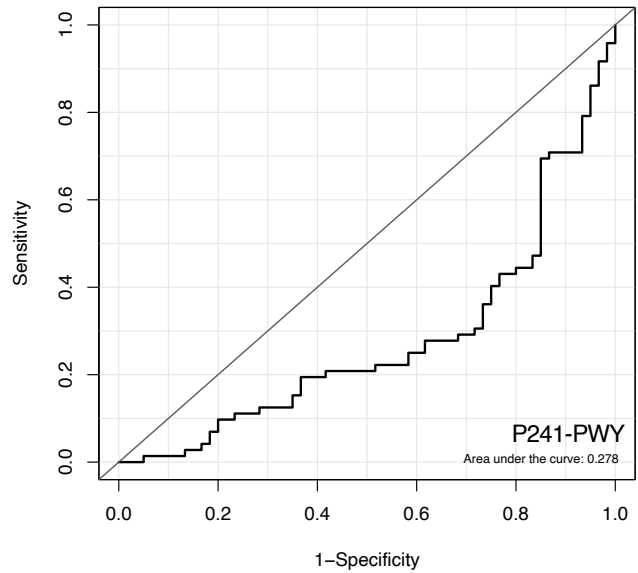
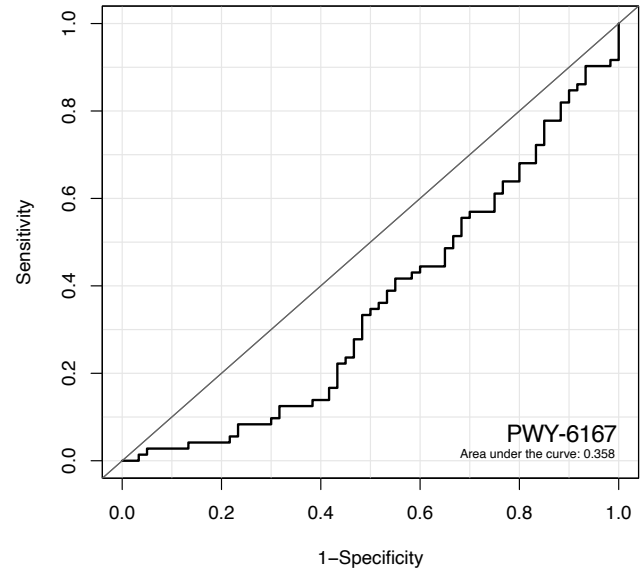

### Fig.S19

**Thresholds for the selected feature components.**
Receiver operating characteristic (ROC) curve analysis for PWY-2941, P241-PWY, PWY-4722, PWY-6167, and PWY-6471 is shown in Fig. 6b. The conditions for seagrass and non-seagrass were assigned as values of 1 and 0, respectively. On the basis of the two-dimensional surface for detection and false-detection rates of abrupt community-compositional changes (abruptness > 0.5), the area under the curve (AUC) was calculated. Abbreviations are as follows: Sens, sensitivity; Spec, specificity; PV+, positive predictive value; PV-, negative predictive value.

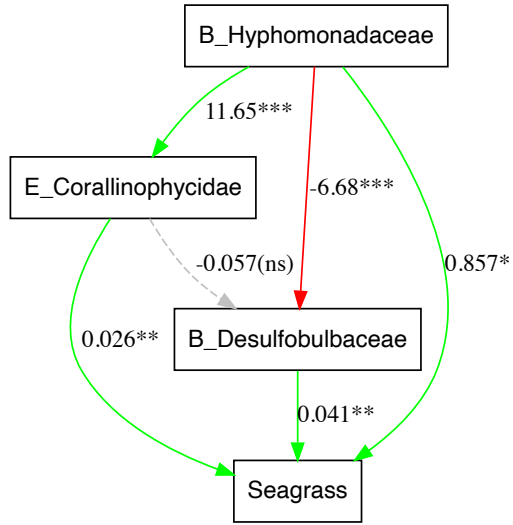
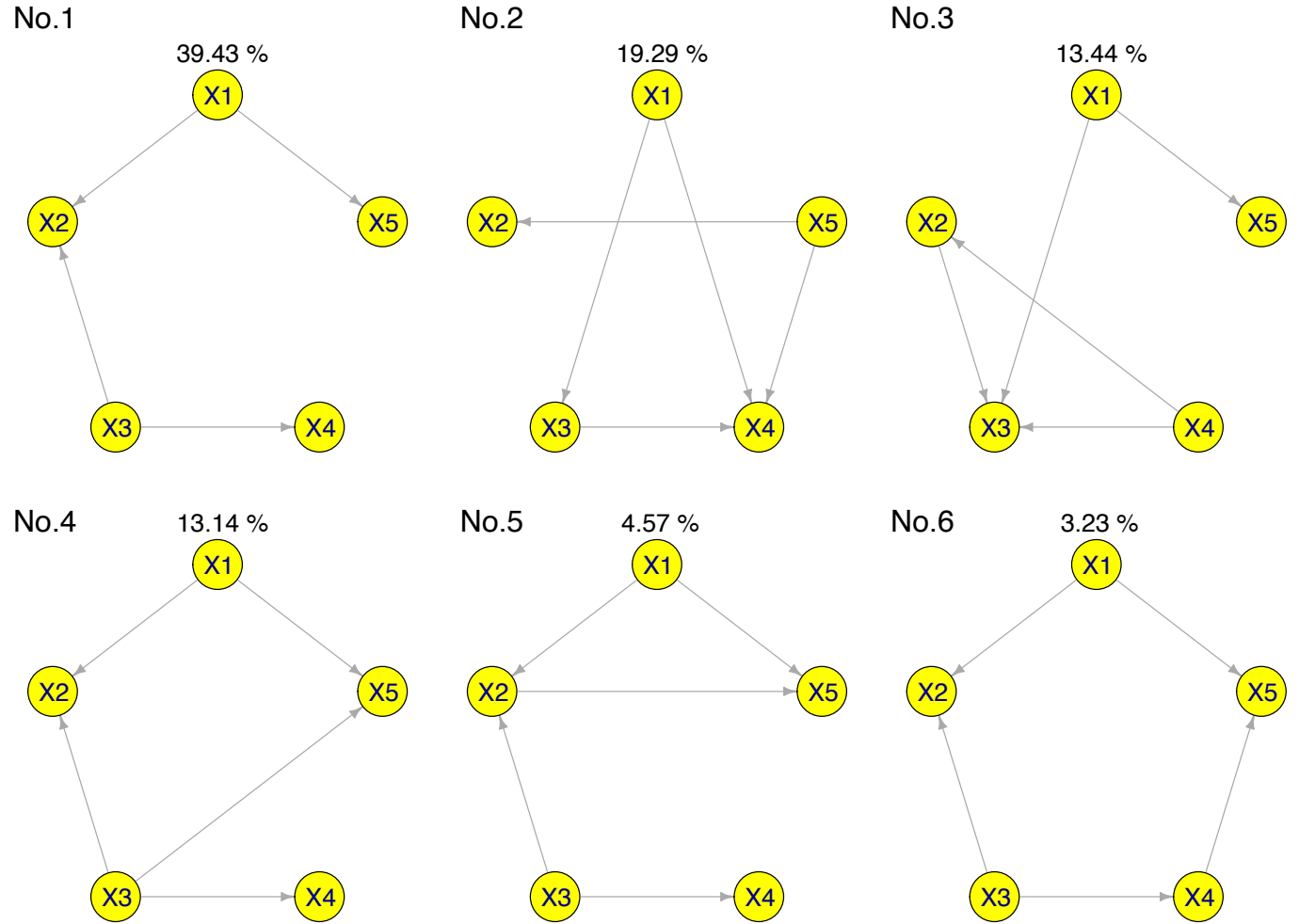

**Fig.S20**
**Causal inference for Fig. 5c.**
(a) A model combined with SEM (Table S5) and CMA (Table S6) visualized by the R library "DiagrammeR". The number on the line shows the value of ACME Estimate. A positive number of each DAG means a positive effect (green line), and a negative number means a negative effect (red line). The dotted DAG means a weak effect (p>0.05). (b) DAGs based on BayesLiNGAM. The components in Fig. 5c were computed. The upper percentage of each DAG indicates the calculated ratio. Abbreviations are as follows: ACME, average causal mediation effect; ns, not significant (gray line); *, $p<0.05$; **, $p<0.01$; ***, $p<0.001$; X1, Seagrass; X2, B_Desulfobulbaceae; X3, B_Hyphomonadaceae; X4, E_Corallinophycidae; and X5, E_Diatomea.

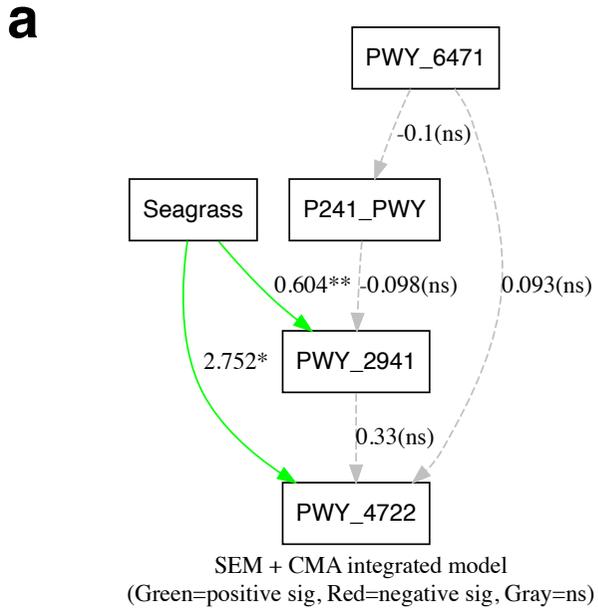
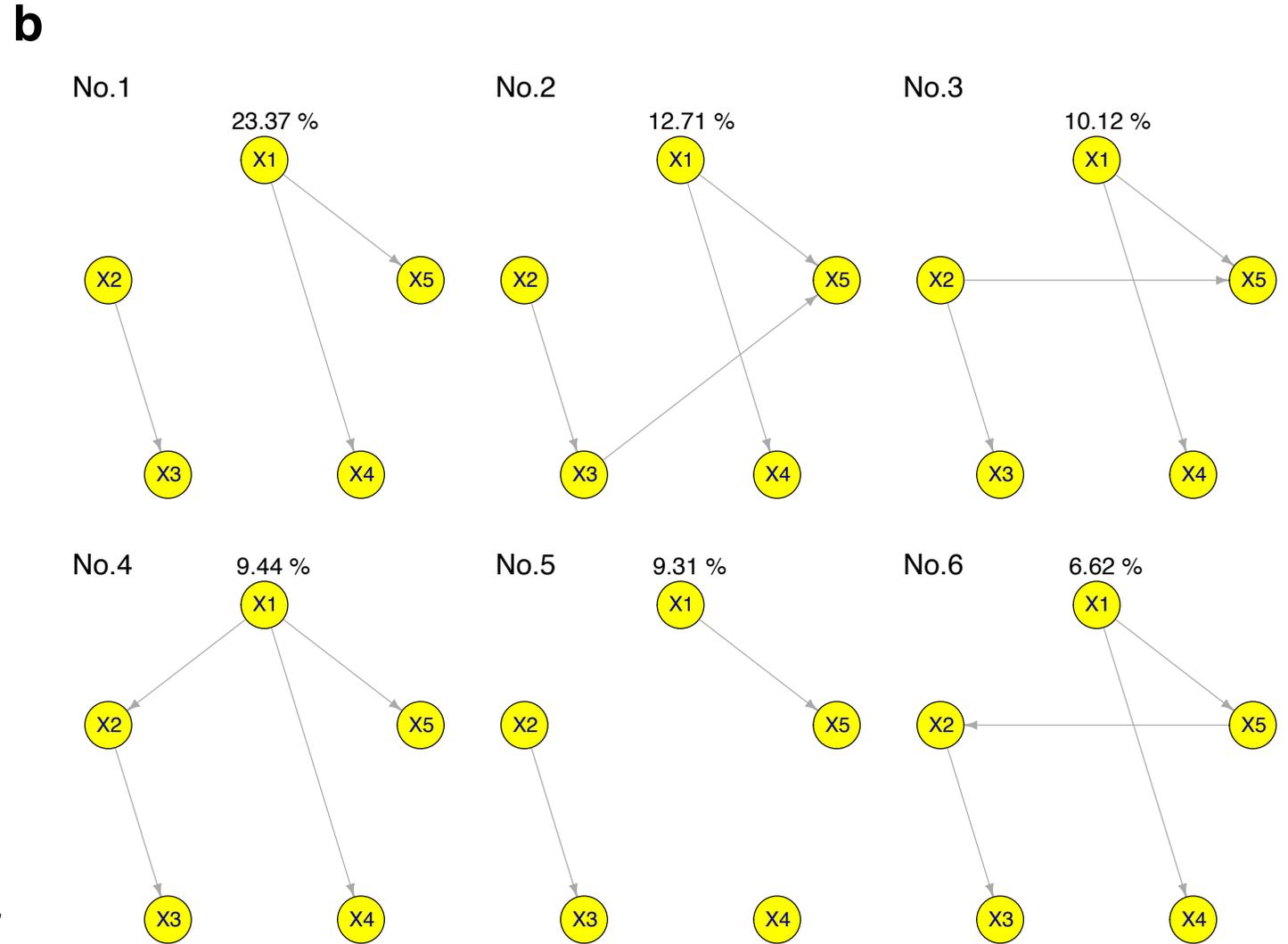

**Fig.S21**
**Causal inference for Fig. 6c.**
(a) A model combined with SEM and CMA (Table S7) visualized by the R library "DiagrammeR". The number on the line shows the value of ACME Estimate. A positive number of each DAG means a positive effect (green line). The dotted DAG means a weak effect ($p>0.05$). (b) DAGs based on BayesLiNGAM. The components in Fig. 6c were calculated. The upper percentage of each DAG indicates the calculated ratio. Abbreviations are as follows: ACME, average causal mediation effect; ns, not significant (gray line); *, $p<0.05$; **, $p<0.01$; (ns), not significant; X1, seagrass; X2, PWY-2941 (L-lysine biosynthesis II); X3, PWY-6471 (peptidoglycan biosynthesis IV); X4, PWY-4722 (creatinine degradation II); and X5, P241-PWY (coenzyme B biosynthesis).

**Table S1**
**Species of *Zostera* in the targeted regions.**

| Region | Target | Primer | Species |
|---|---|---|---|
| Hakata | ITS | | Zostera marina L. |
| Kitsuki | ITS | | Zostera japonica Ascheron et Graebner |
| | | | Zostera marina L. |
| Noto | ITS | | Zostera marina L. |
| Okayama_G | ITS | ITS1_F: CGTTCTTCATCGATGCGAGA | Zostera marina L. |
| Okayama_K | ITS | ITS1_R: CGTAACAAGGTTTCCGTAGG | Zostera marina L. |
| Saiki | ITS | | Zostera marina L. |
| Tokyo_HK | ITS | | Zostera marina L. |
| Tokyo_KC | ITS | | Zostera japonica |
| | | | Zostera marina L. |

# Table S2
**Statistical values of α-diversity for the targeted regions.**
The values for pairs of the target regions are presented in the table. The "*" symbol in the "sig" cell indicates statistical significance.

| Comparison | Bacteria | | | | | | Eukaryota | | | | | |
|---|---|---|---|---|---|---|---|---|---|---|---|---|
| | Shannon_Index | sig | Pielou_Evenness | sig | Simpson_Index | sig | Shannon_Index | sig | Pielou_Evenness | sig | Simpson_Index | sig |
| Kruskal-Wallis test | 4.17E-12 | * | 1.28E-10 | * | 1.44E-12 | * | 3.17E-06 | * | 8.76E-05 | * | 0.000509373 | * |
| Hakata vs Kitsuki | 4.57774E-08 | * | 0.890402596 | | 7.48908E-08 | * | 0.027947487 | * | 0.095016127 | | 0.100598681 | |
| Hakata vs Noto | 0.001186915 | * | 3.29748E-05 | * | 0.006576784 | * | 0.274052733 | | 0.251592801 | | 0.41531028 | |
| Hakata vs Okayama_G | 0.089959049 | | 0.090728787 | | 0.331032587 | | 0.146010688 | | 0.151670692 | | 0.207180696 | |
| Hakata vs Okayama_K | 0.030200614 | * | 0.643843447 | | 0.031171974 | * | 0.767523498 | | 0.987708512 | | 0.524451852 | |
| Hakata vs Saiki | 2.01126E-07 | * | 0.952456519 | | 4.06952E-07 | * | 0.000771739 | * | 0.348591537 | | 0.071881716 | |
| Hakata vs Tokyo_HK | 3.24449E-07 | * | 0.643843447 | | 4.07315E-07 | * | 0.944499595 | | 0.8326854 | | 0.41531028 | |
| Hakata vs Tokyo_KC | 3.24449E-07 | * | 0.00080893 | * | 7.96441E-05 | * | 0.146010688 | | 0.329688448 | | 0.184629679 | |
| Kitsuki vs Noto | 0.001813478 | * | 1.41335E-05 | * | 0.000295417 | * | 0.146010688 | | 0.436580359 | | 0.41531028 | |
| Kitsuki vs Okayama_G | 0.00018035 | * | 0.056892565 | | 6.55924E-06 | * | 0.445431324 | | 0.8326854 | | 0.881879618 | |
| Kitsuki vs Okayama_K | 0.00103428 | * | 0.583786068 | | 0.001065198 | * | 0.010244182 | * | 0.095016127 | | 0.02254447 | * |
| Kitsuki vs Saiki | 0.12041887 | | 0.796432732 | | 0.116682403 | | 0.58913864 | | 0.001873488 | * | 0.906500219 | |
| Kitsuki vs Tokyo_HK | 0.615574515 | | 0.583786068 | | 0.68364309 | | 0.033523637 | * | 0.060162512 | | 0.016169301 | * |
| Kitsuki vs Tokyo_KC | 0.615574515 | | 0.000408783 | * | 0.127870853 | | 0.445431324 | | 0.50213151 | | 0.906500219 | |
| Noto vs Okayama_G | 0.219536005 | | 0.046677406 | * | 0.116682403 | | 0.58913864 | | 0.641131031 | | 0.669042476 | |
| Noto vs Okayama_K | 0.458961233 | | 0.000408783 | * | 0.838857325 | | 0.146010688 | | 0.263786425 | | 0.064324607 | |
| Noto vs Saiki | 0.026674586 | * | 3.12575E-08 | * | 0.004567766 | * | 0.007104416 | * | 0.003106375 | * | 0.41531028 | |
| Noto vs Tokyo_HK | 0.010189131 | * | 0.000408783 | * | 0.002006882 | * | 0.326051525 | | 0.151670692 | | 0.039791298 | * |
| Noto vs Tokyo_KC | 0.010189131 | * | 0.796432732 | | 0.073372049 | | 0.58913864 | | 0.987708512 | | 0.612478681 | |
| Okayama_G vs Okayama_K | 0.666232362 | | 0.291458899 | | 0.236306262 | | 0.10337009 | | 0.151670692 | | 0.03760157 | * |
| Okayama_G vs Saiki | 0.001880313 | * | 0.046677406 | * | 7.96441E-05 | * | 0.146010688 | | 0.003106375 | * | 0.917355162 | |
| Okayama_G vs Tokyo_HK | 0.001150314 | * | 0.291458899 | | 5.2904E-05 | * | 0.163289124 | | 0.095016127 | | 0.028870832 | * |
| Okayama_G vs Tokyo_KC | 0.001150314 | * | 0.173136311 | | 0.003049275 | * | 0.982969034 | | 0.673948919 | | 0.928822413 | |
| Okayama_K vs Saiki | 0.010027153 | * | 0.643843447 | | 0.015251825 | * | 0.000205354 | * | 0.329688448 | | 0.012392377 | * |
| Okayama_K vs Tokyo_HK | 0.003799784 | * | 0.982969034 | | 0.004425526 | * | 0.706733896 | | 0.8326854 | | 0.917355162 | |
| Okayama_K vs Tokyo_KC | 0.003799784 | * | 0.006819408 | * | 0.086148677 | | 0.10337009 | | 0.329688448 | | 0.033000038 | * |
| Saiki vs Tokyo_HK | 0.372328672 | | 0.643843447 | | 0.287500759 | | 0.000914602 | * | 0.517689996 | | 0.010645863 | * |
| Saiki vs Tokyo_KC | 0.372328672 | | 7.31989E-05 | * | 0.793681602 | | 0.146010688 | | 0.029393508 | * | 0.928822413 | |
| Tokyo_HK vs Tokyo_KC | 1 | | 0.006819408 | * | 0.284886556 | | 0.163289124 | | 0.239080564 | | 0.026193668 | * |

## Table S3
**Statistical values of NMDS as β-diversity for the targeted regions.**
The values for pairs of the target regions are presented in the table. The "*" symbol in the "sig" cell indicates statistical significance.

| pairs | | | Bacteria | | | | Eukaryote | | | |
|---|---|---|---|---|---|---|---|---|---|---|
| | | | R2 | p.value | p.adjusted | sig | R2 | p.value | p.adjusted | sig |
| Saiki | vs | Noto | 0.2318438 | 0.001 | 0.028 | * | 0.14608693 | 0.001 | 0.028 | * |
| Saiki | vs | Tokyo_HK | 0.1228533 | 0.001 | 0.028 | * | 0.05901761 | 0.02 | 0.56 | |
| Saiki | vs | Tokyo_KC | 0.2132787 | 0.001 | 0.028 | * | 0.1250263 | 0.001 | 0.028 | * |
| Saiki | vs | Hakata | 0.1420916 | 0.001 | 0.028 | * | 0.11637207 | 0.001 | 0.028 | * |
| Saiki | vs | Okayama_K | 0.2923091 | 0.001 | 0.028 | * | 0.21664692 | 0.001 | 0.028 | * |
| Saiki | vs | Okayama_G | 0.250524 | 0.001 | 0.028 | * | 0.12867369 | 0.001 | 0.028 | * |
| Saiki | vs | Kitsuki | 0.2041308 | 0.001 | 0.028 | * | 0.14978927 | 0.001 | 0.028 | * |
| Noto | vs | Tokyo_HK | 0.2976199 | 0.001 | 0.028 | * | 0.26370303 | 0.001 | 0.028 | * |
| Noto | vs | Tokyo_KC | 0.3644753 | 0.001 | 0.028 | * | 0.0635102 | 0.066 | 1 | |
| Noto | vs | Hakata | 0.2731666 | 0.001 | 0.028 | * | 0.12603198 | 0.007 | 0.196 | |
| Noto | vs | Okayama_K | 0.3580615 | 0.001 | 0.028 | * | 0.33978379 | 0.001 | 0.028 | * |
| Noto | vs | Okayama_G | 0.3195594 | 0.001 | 0.028 | * | 0.17081676 | 0.001 | 0.028 | * |
| Noto | vs | Kitsuki | 0.3843862 | 0.001 | 0.028 | * | 0.16184296 | 0.001 | 0.028 | * |
| Tokyo_HK | vs | Tokyo_KC | 0.527653 | 0.001 | 0.028 | * | 0.37937341 | 0.001 | 0.028 | * |
| Tokyo_HK | vs | Hakata | 0.4738671 | 0.001 | 0.028 | * | 0.22754179 | 0.001 | 0.028 | * |
| Tokyo_HK | vs | Okayama_K | 0.6393581 | 0.001 | 0.028 | * | 0.44643243 | 0.001 | 0.028 | * |
| Tokyo_HK | vs | Okayama_G | 0.6138773 | 0.001 | 0.028 | * | 0.40791946 | 0.001 | 0.028 | * |
| Tokyo_HK | vs | Kitsuki | 0.410871 | 0.001 | 0.028 | * | 0.40117106 | 0.001 | 0.028 | * |
| Tokyo_KC | vs | Hakata | 0.5267474 | 0.001 | 0.028 | * | 0.23042336 | 0.002 | 0.056 | |
| Tokyo_KC | vs | Okayama_K | 0.6696437 | 0.001 | 0.028 | * | 0.53188392 | 0.001 | 0.028 | * |
| Tokyo_KC | vs | Okayama_G | 0.6365459 | 0.001 | 0.028 | * | 0.38942528 | 0.001 | 0.028 | * |
| Tokyo_KC | vs | Kitsuki | 0.2648103 | 0.001 | 0.028 | * | 0.11448211 | 0.014 | 0.392 | |
| Hakata | vs | Okayama_K | 0.5018314 | 0.001 | 0.028 | * | 0.29834493 | 0.001 | 0.028 | * |
| Hakata | vs | Okayama_G | 0.4612962 | 0.001 | 0.028 | * | 0.24767065 | 0.001 | 0.028 | * |
| Hakata | vs | Kitsuki | 0.4716457 | 0.001 | 0.028 | * | 0.30111662 | 0.001 | 0.028 | * |
| Okayama_K | vs | Okayama_G | 0.2804787 | 0.001 | 0.028 | * | 0.42008647 | 0.001 | 0.028 | * |
| Okayama_K | vs | Kitsuki | 0.6034175 | 0.001 | 0.028 | * | 0.50401443 | 0.001 | 0.028 | * |
| Okayama_G | vs | Kitsuki | 0.5759084 | 0.001 | 0.028 | * | 0.40049248 | 0.001 | 0.028 | * |

**Table S4**
**Statistical values for Fig. S16**.

| Components | KMO | Factor loadings | | | | EFA score | | | Anderson-Darling test | ROC analysis |
| --- | --- | --- | --- | --- | --- | --- | --- | --- | --- | --- |
| | | MR1 | MR2 | MR3 | MR4 | $h^2$ | $u^2$ | com | | AUC |
| | Overall MSA=0.68 | Cumulative explained variance=55.1% | | | | | | | | |
| Seagrass | 0.68 | -0.431 | 0.196 | 0.373 | | 0.33 | 0.67 | 2.5 | - | - |
| E_Corallinophycidae | 0.59 | -0.114 | -0.205 | 0.495 | | 0.22 | 0.78 | 1.5 | <0.001 | 0.645601852 |
| B_Hyphomonadaceae | 0.64 | | -0.164 | 0.688 | 0.299 | 0.68 | 0.32 | 1.5 | <0.001 | 0.609259259 |
| B_Desulfobulbaceae | 0.56 | -0.191 | 0.681 | -0.328 | | 0.55 | 0.45 | 1.6 | 8.00E-03 | 0.6125 |
| B_Kiloniellaceae | 0.74 | 0.773 | | | -0.166 | 0.59 | 0.41 | 1.1 | <0.001 | 0.33912037 |
| B_Flavobacteriaceae | 0.6 | | -0.144 | | 0.753 | 0.61 | 0.39 | 1.1 | <0.001 | 0.544444444 |
| B_Crocinitomicaceae | 0.67 | 0.361 | 0.273 | 0.268 | 0.565 | 0.74 | 0.26 | 2.7 | <0.001 | 0.552777778 |
| B_Marinilabiliaceae | 0.61 | -0.319 | -0.296 | -0.56 | 0.407 | 0.7 | 0.3 | 3.1 | <0.001 | 0.407175926 |
| B_BD7_8_other | 0.81 | 0.717 | | | 0.21 | 0.58 | 0.42 | 1.2 | <0.001 | 0.363657407 |
| B_bacterium_enrichment_culture_clone_22.2013 | 0.84 | -0.45 | | -0.24 | | 0.31 | 0.69 | 1.6 | <0.001 | 0.584027778 |
| E_Novel.Clade.Gran.1 | 0.67 | 0.666 | | -0.108 | | 0.43 | 0.57 | 1.1 | <0.001 | 0.42962963 |
| E_Diatomea | 0.7 | 0.86 | | -0.136 | | 0.72 | 0.28 | 1.1 | <0.001 | 0.344212963 |
| B_Unknown.Family | 0.54 | | 0.853 | -0.126 | | 0.73 | 0.27 | 1.1 | 0.014 | 0.551851852 |

The abbreviations are as follows: Seagrass, a condition in which the seagrasses grow; KMO, Kaiser-Meyer-Olkin; MSA, measure of sampling adequacy; factor loadings, the weights of variables on factors; cumulative explained variance, the total variance explained by the factors; $h^2$, communality score; $u^2$, uniqueness score; com, complexity, an information score that is generally related to uniqueness; ROC analysis, receiver operating characteristic (ROC) curve analysis ; and AUC, area under the curve.

## Table S5
**Statistical values of the final optimal structural equation models for bacteria and eukaryote in the marine sediment for seagrass flourished conditions in the targeted region.**

Column no. 1_a shows the best numerical structural equation model in Fig. 5c. Column no.1_b also shows the best numerical structural equation model. Column no.2 shows the models of the inferior numerical structural equations. The estimator used for these calculations was "MLR" (maximum likelihood robust).

| Category | Model | Fit indices | | |
|---|---|---|---|---|
| No.1_a | Seagrass ~ B_Desulfobulbaceae | chisq 0.113 | df 1.000 | *p*-value 0.737 |
| | B_Desulfobulbaceae + Seagrass ~ B_Hyphomonadaceae + E_Corallinophycidae + E_Diatomea | cfi.robust 1.000 | tli.robust 1.145 | rfi 0.986 |
| | E_Corallinophycidae ~ B_Hyphomonadaceae | nfi 0.998 | srmr 0.008 | AIC 1595.584 |
| | lavaan 0.6-19 ended normally after 1 iterations | rmsea 0.000 | gfi 0.999 | agfi 0.992 |
| | Number of successful bootstrap draws    1000 | | | |
| No.1_b | B_Desulfobulbaceae ~ Seagrass | chisq 0.113 | df 1.000 | *p*-value 0.737 |
| | B_Desulfobulbaceae + Seagrass ~ B_Hyphomonadaceae + E_Corallinophycidae + E_Diatomea | cfi.robust 1.000 | tli.robust 1.026 | rfi 0.986 |
| | E_Corallinophycidae ~ B_Hyphomonadaceae | nfi 0.998 | srmr 0.008 | AIC 1595.584 |
| | lavaan 0.6-19 ended normally after 1iterations | rmsea 0.000 | gfi 0.999 | agfi 0.992 |
| | Number of successful bootstrap draws    1000 | | | |
| No.2 | B_Desulfobulbaceae ~ Seagrass | chisq 0.220 | df 3.000 | *p*-value 0.639 |
| | B_Desulfobulbaceae + Seagrass ~ B_Hyphomonadaceae + E_Corallinophycidae + E_Diatomea | cfi.robust 1.000 | tli.robust 1.172 | rfi 0.973 |
| | B_Hyphomonadaceae ~ E_Corallinophycida | nfi 0.997 | srmr 0.011 | AIC 631.723 |
| | lavaan 0.6-19 ended normally after 1 iterations | rmsea 0.000 | gfi 0.999 | agfi 0.984 |
| | Number of successful bootstrap draws    1000 | | | |

Abbreviations are as follows: chisq, chi-square χ2; df, degrees of freedom; *p*-value, *p* value from the chi-square test; cfi.robust, robust CFI (comparative fit index); tli.robust, robust TLI (Tucker–Lewis Index); nfi, normed fit index (NFI); rfi, relative fit index (RFI); srmr, standardized root mean residuals (SRMR); AIC, Akaike information criterion; rmsea, root mean square error of approximation (RMSEA); gfi, goodness-of-fit index (GFI); and agfi, adjusted goodness-of-fit index (AGFI). The underlines are shown as values inferior to those of other models with goodness-of-fit indices.

## Table S6
**Causal mediation analysis for Fig. 5c and their statistical values.**
The calculated data based on the regression models are shown. The a) to d) indicate a pattern of treat, mediator, and outcome.

| CMA results based the regression models | (I) Seagrass ~ B_Desulfobulbaceae |
|---|---|
| | (II) Desulfobulbaceae + Seagrass ~ B_Hyphomonadaceae + E_Corallinophycidae + E_Diatomea |
| | (III) E_Corallinophycidae ~ B_Hyphomonadaceae |

Nonparametric Bootstrap Confidence Intervals with the Percentile Method

a) Treat: B_Hyphomonadaceae  Mediator: E_Corallinophycidae  Outcome: B_Desulfobulbaceae

|  | Estimate | 95% CI Lower | 95% CI Upper | p-value |  |
|---|---|---|---|---|---|
| ACME | -0.666185 | -1.978671 | 0.124822 | 0.092 | # |
| ADE | -6.678638 | -9.968599 | -4.206752 | <2e-16 | *** |
| Total Effect | -7.344823 | -10.587965 | -5.150853 | <2e-16 | *** |
| Prop. Mediated | 0.090701 | -0.015325 | 0.274627 | 0.092 | # |

Sample Size Used: 132     Simulations: 1000

c) Treat: E_Corallinophycidae  Mediator: B_Desulfobulbaceae  Outcome: Seagrass

|  | Estimate | 95% CI Lower | 95% CI Upper | p-value |  |
|---|---|---|---|---|---|
| ACME | -0.00445927 | -0.00910059 | -0.00080757 | 0.02 | * |
| ADE | 0.02601637 | 0.01975401 | 0.04580236 | <2e-16 | *** |
| Total Effect | 0.0215571 | 0.0168459 | 0.04033042 | <2e-16 | *** |
| Prop. Mediated | -0.20685846 | -0.41516468 | -0.02970372 | 0.02 | * |

Sample Size Used: 132     Simulations: 1000

b) Treat: B_Hyphomonadaceae  Mediator: B_Desulfobulbaceae  Outcome: Seagrass

|  | Estimate | 95% CI Lower | 95% CI Upper | p-value |  |
|---|---|---|---|---|---|
| ACME | -0.39451 | -0.74955 | -0.15357 | 0.002 | ** |
| ADE | 1.4233 | 1.01679 | 2.04854 | <2e-16 | *** |
| Total Effect | 1.0288 | 0.69709 | 1.53431 | <2e-16 | *** |
| Prop. Mediated | -0.38346 | -0.80025 | -0.14855 | 0.002 | ** |

Sample Size Used: 132     Simulations: 1000

d) Treat: B_Hyphomonadaceae  Mediator: E_Corallinophycidae  Outcome: Seagrass

|  | Estimate | 95% CI Lower | 95% CI Upper | p-value |  |
|---|---|---|---|---|---|
| ACME | 0.172778 | 0.048857 | 0.385838 | 0.002 | ** |
| ADE | 0.856021 | 0.526281 | 1.340517 | <2e-16 | *** |
| Total Effect | 1.028799 | 0.717591 | 1.547292 | <2e-16 | *** |
| Prop. Mediated | 0.167941 | 0.05326 | 0.368287 | 0.002 | ** |

Sample Size Used: 132     Simulations: 1000

Abbreviations are as follows: Estimate, estimate of the effect; 95% CI Lower, lower limit of the 95% confidence interval; 95% CI Upper, upper limit of the 95% confidence interval; p-value, probability of observing the result under the null hypothesis; ACME, average causal mediation effect; and ADE, average direct effect; Total Effect, combined (total) effect of the independent variable on the dependent variable; Prop. Mediated, proportion of the total effect; Sample Size Used, number of observations included in the analysis; Simulations, number of bootstrap replications used to estimate confidence intervals; #, $p<0.1$; *, $p<0.05$; **, $p<0.01$; and ***, $p<0.001$.

## Table S7
**Statistical values of the final optimal structural equation models for bacterial metabolic pathways in the marine sediment under seagrass growth conditions in the targeted region.**

The values in the category of the optimal model show the fit indices of the best numerical structural equation model in Fig. 6c. The estimator used for these calculations was "MLR" (maximum likelihood robust). The category of the linear regression models show the results calculated by mediation analysis. The a) to c) indicate a pattern of treat, mediator, and outcome.

| Category | Contents | | | | | Fit indices | | |
|---|---|---|---|---|---|---|---|---|
| Optimal model | (I) PWY_2941 + PWY_4722 ~ Seagrass | | | | | chisq 0.163 | df 1.000 | *p*-value 0.686 |
| | (II) PWY_4722 + PWY_2941 ~ PWY_6471 + P241_PWY | | | | | cfi.robust 1.000 | tli.robust 1.296 | rfi .957 |
| | | | | | | nfi 0.994 | srmr 0.008 | AIC 1313.659 |
| | lavaan 0.6-19 ended normally after 1 iterations | | | | | rmsea 0.000 | gfi 0.999 | agfi 0.976 |
| | Number of successful bootstrap draws: 1000 | | | | | | | |
| CMA | a) Treat: Seagrass  Mediator: PWY_2941  Outcome: PWY_4722 | | | | | Sample Size Used: 132 | | Simulations: 1000 |
| | | Estimate | 95% CI Lower | 95% CI Upper | p-value | | | |
| | ACME | 0.199472 | -0.316526 | 0.809081 | 0.448 | | | |
| | ADE | 2.752318 | 0.358479 | 5.125727 | 0.022 | * | | |
| | Total Effect | 2.951791 | 0.714333 | 5.198854 | 0.006 | ** | | |
| | Prop. Mediated | 0.067577 | -0.111305 | 0.49733 | 0.454 | | | |
| | b) Treat: PWY_6471  Mediator: P241_PWY  Outcome: PWY_2941 | | | | | Sample Size Used: 132 | | Simulations: 1000 |
| | | Estimate | 95% CI Lower | 95% CI Upper | p-value | | | |
| | ACME | 0.0036513 | -0.0015757 | 0.0139662 | 0.226 | | | |
| | ADE | 0.0437431 | 0.0095187 | 0.0761118 | 0.006 | ** | | |
| | Total Effect | 0.0473944 | 0.0134908 | 0.0798397 | 0.006 | ** | | |
| | Prop. Mediated | 0.077041 | -0.0391495 | 0.3594037 | 0.232 | | | |
| | c) Treat: PWY_6471  Mediator: P241_PWY  Outcome: PWY_4722 | | | | | Sample Size Used: 132 | | Simulations: 1000 |
| | not significant (ACME Estimate=0.0098351, p-value=0.328;  ADE Estimate=0.0935015, p-value=0.282) | | | | | | | |

Abbreviations are as follows: chisq, chi-square χ2; df, degrees of freedom; *p*-value, *p* value from the chi-square test; cfi.robust, robust CFI (comparative fit index); tli.robust, robust TLI (Tucker–Lewis Index);  nfi, normed fit index (NFI); rfi, relative fit index (RFI); srmr, standardized root mean residuals (SRMR); AIC, Akaike information criterion; rmsea, root mean square error of approximation (RMSEA); gfi, goodness-of-fit index (GFI); and agfi, adjusted goodness-of-fit index (AGFI); Estimate, estimate of the effect; 95% CI Lower, lower limit of the 95% confidence interval; 95% CI Upper, upper limit of the 95% confidence interval; p-value, probability of observing the result under the null hypothesis; ACME, average causal mediation effect; and ADE, average direct effect; Total Effect, combined (total) effect of the independent variable on the dependent variable; Prop. Mediated, proportion of the total effect; Sample Size Used, number of observations included in the analysis; Simulations, number of bootstrap replications used to estimate confidence intervals; *, p<0.05; and **, p<0.01.